\documentclass[pdflatex,sn-mathphys-num]{ZoeRNPONeutral}

\usepackage{graphicx}%
\usepackage{multirow}%
\usepackage{amsmath,amssymb,amsfonts}%
\usepackage{amsthm}%
\usepackage{mathrsfs}%
\usepackage[title]{appendix}%
\usepackage{xcolor}%
\usepackage{textcomp}%
\usepackage{manyfoot}%
\usepackage{booktabs}%
\usepackage{algorithm}%
\usepackage{algorithmicx}%
\usepackage{algpseudocode}%
\usepackage{listings}%
\usepackage[caption=false]{subfig}
\usepackage{tabularx,ragged2e}
\newcolumntype{L}{>{\RaggedRight}X}
\newcolumntype{C}{>{\centering}X}

\DeclareMathOperator{\sn}{sn}
\DeclareMathOperator{\cn}{cn}

\begin{document}

\title[Article Title]{Periodic orbits of neutral test particles
in Reissner-Nordstr\"{o}m naked singularities}

\author*[1]{\fnm{Zoe C S} \sur{Chan}}\email{ bravychancs@gmail.com}

\author[2]{\fnm{Yen-Kheng} \sur{Lim}}\email{yenkheng.lim@gmail.com, yenkheng.lim@xmu.edu.my}

\affil[1,2]{\normalsize{\textit{Department of Physics, Xiamen University Malaysia, 43900 Sepang, Malaysia}}}

\abstract{We conduct studies on Levin's taxonomy of periodic orbits for neutral test particles around a 
Reissner-Nordstr\"{o}m naked singularity. It was known that naked singularities could harbor two 
distinct regions of time-like bound orbits and thus we expect periodic orbits to appear in both regions.
It is possible for a pair of periodic orbits from both regions to possess the exact same angular 
momentum $L$ and energy $E$ values. We chart the sets of periodic orbits in $(L,E)$-parameter space 
and highlight the general distribution pattern of these sets for three possible scenarios. Regions 
within $(L,E)$-space can be partitioned into multiple domains $\dom_k$ based on the roots configuration 
of the quartic polynomial $P(u)$ where $u$ is the inverse radial coordinate. Consequently, each domain 
and interestingly enough, portions of certain periodic orbits sets that lie in different $\dom_k$ require 
different analytical solutions to plot the resulting orbit. Furthermore, we uncover physical properties of some 
hypothetical circular orbits residing in the inner region from analysing the $(L,E)$-space.}


\maketitle

\section{Introduction} \label{sec_intro}

The Reissner-Nordstr\"{o}m (RN) metric is an exact solution to Einstein-Maxwell field equations
which describe a static, asymptotically flat spacetime outside a charged, non-rotating, spherically
symmetric compact body of mass $M$. It was determined independently by Reissner \cite{reissner} in 
1916 and Nordstr\"{o}m \cite{nordstrom} in 1918. This solution was generally deemed unrealisitc since 
the compact body is unlikely able to acquire a relatively large charge-to-mass ratio $Q/M$. This was
exemplify by Zaja\v{c}ek et. al. \cite{zajacekSgrAQ} where they find constraints of the the Milky 
Way's central black hole (Sgr A*) charge to be $Q/M\simeq0$. Nevertheless, studying the geodesics of 
test particles in the vicinity of a charged compact body are still important in relativistic astrophysics 
as they provide approximate models to astrophysical processes such as gravitational waves detection 
\cite{gwave1,gwave2} or direct imaging of black holes shadows \cite{EHT1,EHT2,KerrLens}.

Earlier, Levin et.al.classified the anatomy of periodic, zoom-whirl orbits \cite{Glampedakis:2002} 
in a series of papers \cite{levinpg08,levindyn1,levindyn2,levinhomo1,levinhomo2,levinzw09,levinel09,levinmisra10,levin11}.
These periodic orbits were indexed by three non-negative integers $(z,w,v)$ based on
the geometric and topological features of the orbits structure. $z$ counts the number of 
ellipses in an orbit, $w$ is the number of near-center whirls and $v$ is the order in which the ellipses 
are traced out. Then, every periodic orbits can be parametrized by a dimensionless rational number $q$, given 
by $q=w+\frac{v}{z}$. If $q$ is irrational, it corresponds to a quasiperiodic orbit. Initially, the taxonomy was meant 
to aid calculations in gravitional-waves detection, specifically for extreme mass ratio inspirals that exhibit zoom-whirl behaviour. 
Lately, these periodic orbits descriptions were applied to other theoretical spacetime settings or alternate gravity theories. 
\cite{babaryklim,liuKerrSen,2020xmumcollab,zhangbbrn,wangQuinPO,TuLQGPO,dengBWPO,dengQCBH,dengEpicyclicPO,liPOcubic}.

Naked singularities are hypothetical singularities where there is no event horizon surrounding
the singularity, so any distant observer could observe the singularity in principle.
For RN case, this occurs when the compact body's charge exceeds its mass, i.e. $Q/M>1$. 
Penrose conjectured that naked singularities are impossible to observe in nature and must 
always be hidden behind an event horizon via his infamous cosmic censhorship hypothesis 
\cite{penrose69WeakCCC}. On the flip side, there is an ongoing trend to detect the existence of  
naked singularities based on its exotic property of allowing multiple regions of bound orbits
\cite{patilNSShellAccel,vieraNSCircles,vieiraNSCloak,MishraNSTori,shaikhNSshadow}.
Some newly proposed naked singularity kinds such as sub-solar mass \cite{SubSolarMassNS} 
or primordial \cite{PrimordialNS} have been put forward too.

Recently, one of the authors contribute more theoretical descriptions to Levin's taxonomy
for Schwarzschild spacetime \cite{yklimzcy24}. Several analytical solutions were parametrized 
in terms of parameters characterizing the geometry of an ellipse, the eccentricity $e$ and 
latus rectum $\lr$. This enable the complete interpolation of $q$-distribution of sets 
containing values of the conserved quantities $L$ and $E$ for $0\leq e<1$. Each of these sets 
appear as a `branch' emanating from the stable circular orbit segment (which acts as the 
zero-eccentricity limit). These $q$-branches distribute as a discrete line spectrum, resembling 
the emission spectrum of chemcial elements. An advantage of performing analysis from the $(L,E)$-parameter 
space is that it uncover features of certain types of orbits that may not be present in other graphs.

Gathering the motivations mentioned above, for this paper, we will like to extend the works of \cite{yklimzcy24} 
by considering only neutral time-like test particles orbiting a Reissner-Nordstr\"{o}m naked singularity.  
For black holes, time-like circular orbits were proven to exist only in a single region outside the event horizon 
\cite{levinmisra10,pugliese11n,vieraNSCircles}. So, the circular orbits curves of RN black holes have similiar structures 
as the Schwarzschild case. The main difference is that larger black holes charge decrease the lower bound of 
circular orbits radius and in return, increase the range of $L$ and $E$ values. Accordingly, the circular orbits curve of RN black holes with 
larger $Q/M$ shift towards the $L$ and $E$-axes in $(L,E)$-space. 

Key results for circular orbits of RN naked singularity by Pugliese at.al. \cite{pugliese11n} show that for a 
small range of $Q/M$, there exist a second smaller region that allow time-like stable circular orbits. As such, 
we expect the circular orbits curve in $(L,E)$-space to have another segment representing the stable circular orbits 
of this smaller region and thus, new sets of $q$-branches could be distributed along this segment. There are also 
three distinct scenarios for naked singularity circular orbits. Each scenario occur in the following charge range: 
$M<Q<\sqrt{\frac{9}{8}}\,M$, $\sqrt{\frac{9}{8}}\,M \leq Q<\frac{\sqrt{5}}{2}\,M$ and 
$Q\geq\frac{\sqrt{5}}{2}\,M$. From here on, we will refer to these scenarios, in order, as Case 1, 2 and 3.

The rest of the paper is organized as follows. In Sec. \ref{sec_EoM}, we derive the relevant 
equations of motion from the RN metric and also review key techniques from \cite{yklimzcy24} 
to chart the periodic orbits distribution in $(L,E)$-space. In Sec. \ref{sec_3}, we relate how
the roots configuration of the quartic polynomial $P(u)$ at different points in $(L,E)$-space 
determine the orbit types and then show which analytical solution in terms of $P(u)$ roots to use 
at each point. Sec. \ref{sec_4} generalize the domains in $(L,E)$-space where we can locate periodic 
orbits and highlight $q$-branch distribution pattern for all three naked singularity scenarios. 

We inform the readers that we work in Lorentzian metric signature $(-,+,+,+)$ alongside
geometrized units $c=G=1$. We adopt the extended periodic orbit taxonomy notation, 
$(z,w,v;e)$ introduced in \cite{yklimzcy24} and notations of some special kinds of circular orbits 
from \cite{pugliese11n}. We generally colored the text and labels in the figures 
as \textcolor{blue}{blue} ro represent periodic orbits residing in the outer region and \textcolor{red}{red} 
for the inner ones. All numerical values in the figures are displayed up to 5 significant figures 
unless stated otherwise.

\section{Equations of motions and the \texorpdfstring{$(L,E)$}{TEXT} parameter space} \label{sec_EoM} 
Here, we review the derivation of the geodesics for neutral test particle
via Lagrangian formalism in Sec. \ref{sec_2.1}. In Sec. \ref{sec_2.2}, we show how
the analytical expressions for the particle's conserved energy $E$ and angular momenta $L$ 
in terms of circular orbits radius $\rc$ and geometric parameters $e$ and $\lr$ describe 
periodic orbits graphically on $(L,E)$-parameter space. Derivation of $e,\lr$ parametric expressions 
is given in Appendix \hyperref[app_A]{A}. In Sec. \ref{sec_2.3}, we review techniques 
from \cite{yklimzcy24} for determining $\lr$ numerically while fixing every other parameters 
$Q,e,z,w$ and $v$ via the relation between $P(u)$ and the elliptic integral. This enable us to map 
the whole distribution of periodic orbits on $(L,E)$-space, taking a RN black hole as an example. 

\subsection{Geodesics from Lagrangian formalism}\label{sec_2.1}
The Reissner-Nordstr\"{o}m spacetime is described by the metric line element and horizon function
\begin{subequations}
    \begin{align}
      ds^2&=-f\,(r)\,dt^2+f^{-1}\,(r)\,dr^2+r^2\,(d\theta^2+\sin^2{\theta}\,d\phi^2),\label{rnmetric}\\
  f\,(r)&=1-\dfrac{2M}{r}+\dfrac{Q^2}{r^2},  
    \end{align}
    \end{subequations}
with associated electromagnetic gauge potential and field
\begin{equation}\label{gauge} 
    A=\dfrac{Q}{r}\,dt,\qquad F=dA=-\dfrac{Q}{r^2}\,dt\wedge dr, 
\end{equation}    
The central body is a black hole if $0\leq Q\leq M$ with horizons at $r_{\pm}=M\pm\sqrt{M^2-Q^2}$ and is
a horizonless naked singularity if $Q>M$. Geodesics of neutral time-like particles are 
described by the parametrised curve $x^{\mu}(\tau)$, where $\tau$ is the proper time parameter. The 
test particle's Lagrangian $\Lagr(x,\dot{x})=\half\,g_{\mu\nu}\dot{x}^{\mu}\dot{x}^{\nu}$ is explicitly
\begin{equation}\label{lagr}
\Lagr=\half\brac{-f\dot{t}^2+f^{-1}\,\dot{r}^2+r^2\,\dot{\theta}^2+r^2\sin^2{\theta}\,\dot{\phi}^2},
\end{equation}
where over-dots denote derivatives with respect to $\tau$.

The canonical momenta are $p_{\mu}=\pdv{\Lagr}{\dot{x}^{\mu}}$. The isometries of 
spacetime are generated by the Killing vector fields $\xi=\xi^{\mu}\partial_{\mu}$, that are
associated with quantities conserved throughout the geodesics. In this case, 
$\partial_{t}$ and $\partial_{\phi}$ are the Killing vectors and their corresponding momenta 
$p_t$ and $p_{\phi}$ are conserved. Referring to components of the Lagrangian (\ref{lagr}), the 
constants of motion associated with these fields are
\begin{equation}\label{killV}
    E\equiv-\xi^{t}p_{t}=f\dot{t},\quad 
    L\equiv\xi^{\phi}p_{\phi}=r^2\sin^2{\theta}\,\dot{\phi}\,,
\end{equation}
where $E$ and $L$ are the energy and angular momentum of the test particle respectively.
Then, we can represent two first integrals in terms of these conserved quantities. Rearranging 
Eqs.(\ref{killV}) gives\footnote{Notice these first integrals are similiar to the Schwarzschild ones, 
with only the metric function $f$ having an additional $\frac{Q^2}{r^2}$ term. An alternative formalism 
using Hamilton-Jacobi equations can be found in Ref.\cite{yklimzcy24}.}
\begin{equation}\label{1stintA}
    \dot{t}=\dfrac{E}{f},\quad \dot{\phi}=\dfrac{L}{r^2\sin^2{\theta}}
\end{equation}

The normalization condition for time-like particles gives another first integral,
\begin{equation}\label{guv1stint}
    g_{\mu\nu}\dot{x}^{\mu}\dot{x}^{\nu}=-1,
\end{equation}
rearranging Eq.(\ref{rnmetric}) and doing relevant substitutions with Eqs.(\ref{1stintA}) leads to
\begin{equation}\label{1stintB}
    \dot{r}^2+r^2f\dot{\theta}^2=E^2-\brac{1+\dfrac{L^2}{r^2\sin^2{\theta}}}f
\end{equation}
Due to the geometry being spherically symmetric, we can restrict all trajectories to lie on the 
equatorial plane $\theta=\frac{\pi}{2}$, without loss of generality. In the end, the essential 
equations are the energy equation upon rearranging Eq.(\ref{1stintB}),
\begin{subequations}\label{energyEq}
    \begin{align}
    &\dot{r}^2+\veff=E^2,\\
    &\veff=\brac{1+\dfrac{L^2}{r^2}}\brac{1-\dfrac{2M}{r}+\dfrac{Q^2}{r^2}},
    \end{align}
\end{subequations}
and the quartic radial polynomial from taking $\frac{\dot{r}}{\dot{\phi}}=\frac{dr}{d\phi}$. 
As in classical central force problems, we change the variable of this differential equation 
to the inverse radial coordinate $u=\frac{1}{r}$, giving us the $P(u)$ polynomial which can be 
written out in two forms, 
\begin{subequations}\label{Pudu}
    \begin{align}
     \dfrac{du}{d\phi}&=\pm\sqrt{P(u)},\label{Pu0}\\
                 P(u)&=-Q^2u^4+2Mu^3-\brac{1+\dfrac{Q^2}{L^2}}u^2
                      +\dfrac{2M}{L^2}u-\dfrac{1-E^2}{L^2}\label{Pu1}\\
                     &\equiv Q^2(a-u)(b-u)(c-u)(u-d)\,, \label{Pu2}
    \end{align}
\end{subequations}
where $a,b,c,d$ are the roots of $P(u)$ that could take on real or complex values. 

In essence, $P(u)$ determines the geometry of orbits from its roots disposition 
where each root represent the turning point ($\dot{r}=0$) of the orbit \cite{ChandrashekarBook}.
This also correlate with the number of bound orbits that can exist in a given spacetime. 
From the presence of the square root in (\ref{Pu0}), it is clear that physcially allowed solutions 
only occur for $P(u)\geq 0$. We will explain more on $P(u)$ and its applications in later sections.

\subsection{Solutions in terms of circular orbits and \texorpdfstring{$e,\lr$}{TEXT} parameters}\label{sec_2.2}
In the context of the periodic orbits taxonomy \cite{levinpg08,yklimzcy24}, circular orbits were 
identified to be the zero-eccentricity limit. Expressions of $L$ and $E$ in terms of the circular 
orbits radius $\rc$ for neutral test particle in RN spacetime have appeared in the literature 
\cite{pugliese11n,levinmisra10}. A detailed derivation method can be found in Ref.\cite{levinmisra10}.  
Basically, the conditions to obtain circular orbits are $\dot{r}=0$ and $V_{\mathrm{eff}}'=0$. 
The expressions are given by,
\begin{equation}\label{LEcircular}
    L(r_{c},Q)=\frac{r_c\sqrt{Mr_c-Q^2}}{\sqrt{r^{2}_{c}-3Mr_c+2Q^2}}\,,\quad
    E(r_{c},Q)=\frac{r^{2}_{c}-2Mr_c+Q^2}{r_c\sqrt{r^{2}_{c}-3Mr_c+2Q^2}}
\end{equation}
Then, we can map the set of $\rc$ values with Eqs.(\ref{LEcircular}) for some fix $Q$ in $(L,E)$-space. 
An additional condition $V_{\mathrm{eff}}''$ checks the stability of orbits of each segment of the 
curve (Fig.\hyperref[Fig:1]{1}). Orbits are stable if this second derivative is positive and unstable if negative. 
$V_{\mathrm{eff}}''=0$ occur at the critical points, which can be interpretated as the point where the set of $\rc$ 
values change stability upon crossing it. The critical point is normally comprise of the \textit{innermost stable circular orbit} 
(ISCO) but a second one, sometimes referred to as the \textit{outermost stable circular orbit} (OSCO) \cite{schroeven,Zhang2022} 
of the inner region appear in Case 1 and 2 naked singularity (Fig.\hyperref[Fig:2]{2}).  

Geometrical bound and unbound orbits are related to the eccentricity $e$ of a conic section. 
Bound orbits have $0\leq e<1$, resulting in the orbits having circular or elliptic-like shapes 
whereas $e\geq 1$ describes unbound orbits that are hyperbolic trajectories. Periodic orbits are trajectories 
that return exactly to its initial condition after a fixed orbital period and some finite amount of 
precession. Thus, all periodic orbits must be bound geometrically. The non-relativistic limit 
occurs at $\rc\rightarrow\infty$. On the unstable circular orbit segment, $\rc$ approach the other limit, 
the photon sphere at $r_{\gamma^{+}}$ as both $L$ and $E$ diverge 
(Fig.\hyperref[Fig:1]{1}).

\begin{figure}[h]
    \centering
    \includegraphics[width=.85\linewidth]{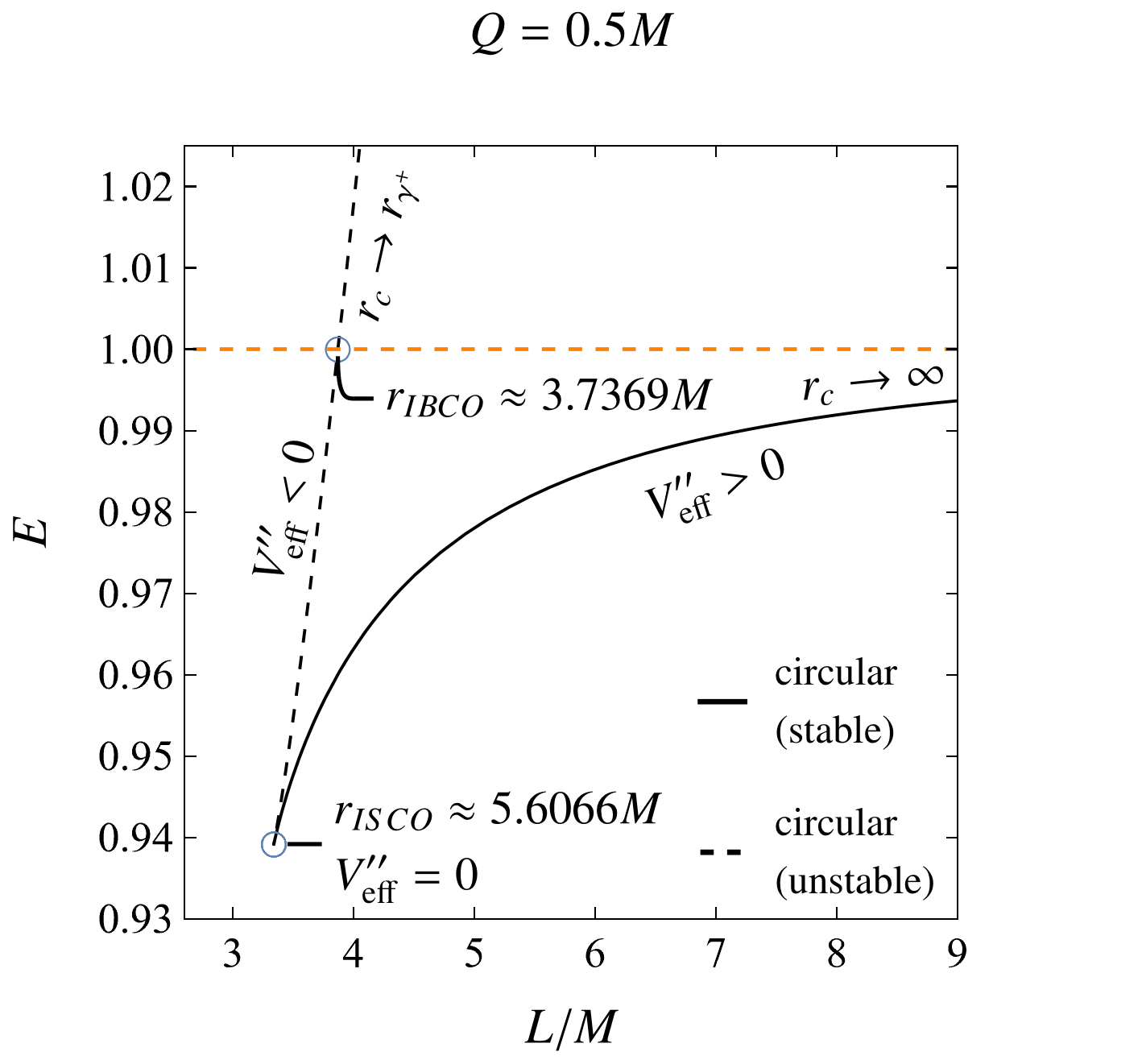}
\caption{A time-like circular orbits parametric curve for RN black holes in $(L,E)$-space. 
The orange dashed line at $E=1$ acts as the separator between regions containing orbits that are 
energetically bound region $(E<1)$ and unbound $(E\geq1)$.}
\end{figure}\label{Fig:1}

Now, we summarize key features of some special types of circular orbits that are exclusive to naked
singularity as covered in \cite{pugliese11n}. $r_{\gamma^{-}}$ represent the null circular orbit in 
the inner region and is stable unlike the outer counterpart. Time-like circular orbits are forbidden 
within the range $r_{\gamma^{-}}\leq r_c \leq r_{\gamma^{+}}$. Both null circular orbits 
only appear in Case 1 since both of their radii merge at $r=1.5M$ when $Q=\sqrt{\frac{9}{8}}\,M$. 
$r_*$ is the radius of the smallest possible time-like circular orbit surrounding a naked singularity where 
particles possess zero angular momentum and appear static to distant observers. Alternatively, we could 
describe $r_*$ as the spherical interface between the gravitational attraction region $r>r_*$ and 
the repulsive `anti-gravity' region $r<r_*$. \cite{pugliese11n,cohen79} gave justificiations on how $r_*$ exists.

There are also $\rc$ values located exactly on the $E=1$ line in $(L,E)$-space  
(see Fig.\hyperref[Fig:1]{1} and \hyperref[Fig:2]{2}). These act like transition points
between energetically bound $(E<1)$ and energetically unbound $(E\geq1)$ circular orbits. The one 
at the unstable $\rc$ segment is normally known as the \textit{innermost bound circular orbit} (IBCO), 
although it is technically an energetically unbound circular orbit already. For Cases 1 and 2 naked singularity, 
there is another $E=1$ point at the inner stable $\rc$ segment which we shall label it as the 
\textit{innermost unbound circular orbit} (IUCO)\footnote{Sec. \hyperref[sec_3.3]{3.3} 
explains the role and naming convention of this circular orbits better when we get into root 
configuration analysis.}. It is indeed the smallest possible energetically unbound circular orbit and also contain 
information of the innermost distance a hyperbolic trajectory could reach in a naked singularity background.

On a side note, the singular point of the curves where different stability segments meet are 
known as \textit{cusps} (Figs.\hyperref[Fig:1]{1}, \hyperref[Fig:2]{2}). 
The usual way of deriving the rather long analytical solution for critical points 
involve solving the conditions $\dot{r}=0,\,V_{\mathrm{eff}}'=0$ and $V_{\mathrm{eff}}''=0$ together \cite{pugliese11n,levinmisra10}. 
However, in our case, we can use the fact that the cusp is also a point on the circular orbits parametric curve
where the first derivative with respect to $\rc$ of both Eqs.(\ref{LEcircular}) are zero. 
So we could easily get the numerical value of critical points by solving
\begin{equation}\label{cuspcond}
    \pdv{}{\rc}L(\rc,Q)=0\en\mbox{or}\quad\pdv{}{\rc}E(\rc,Q)=0
\end{equation} 

\begin{figure}[ht!]
    \centering
    \begin{minipage}{\linewidth}
    \centering
    {\includegraphics[scale=1,height=6.2cm]{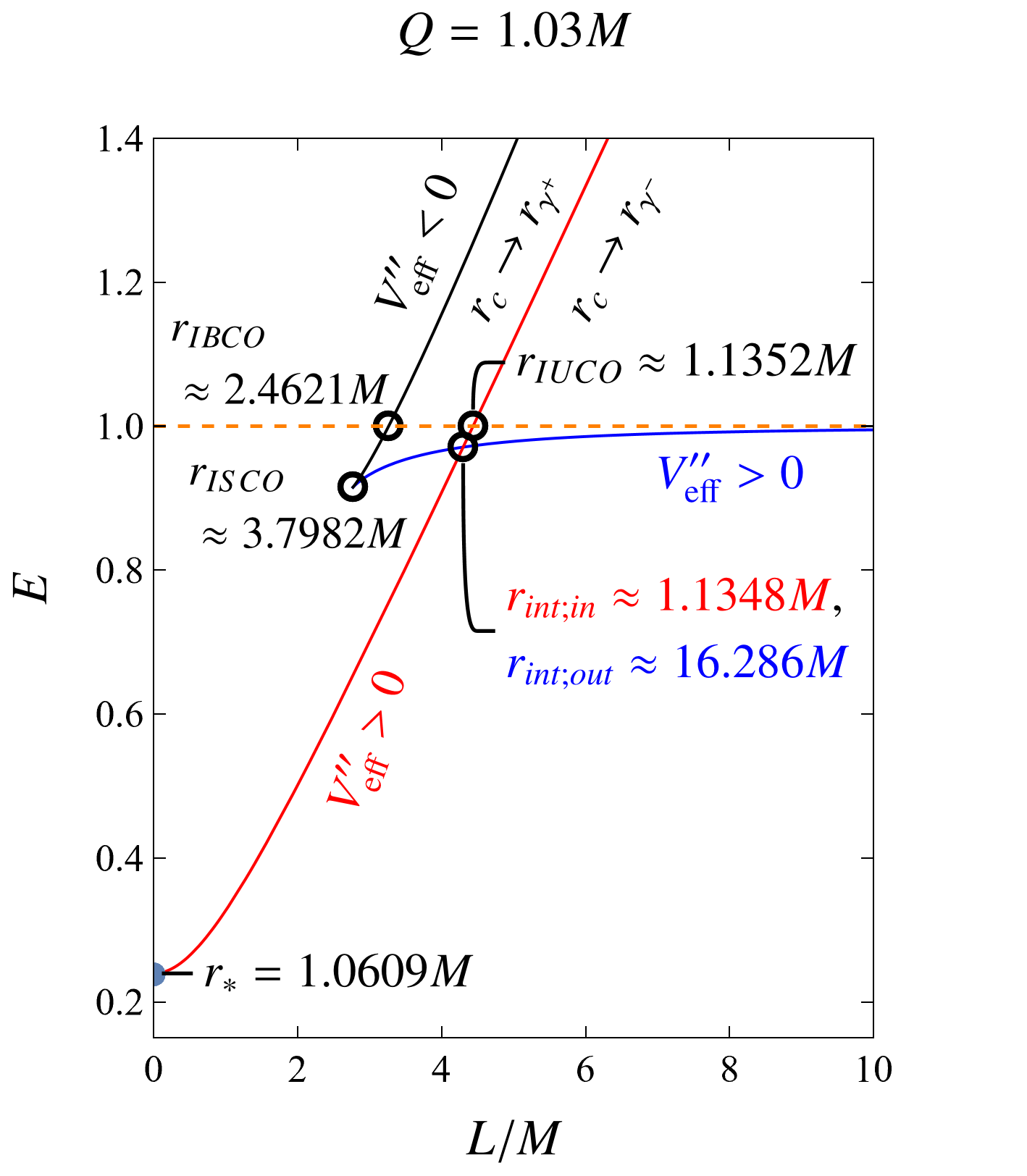}}
    \enspace
    {\includegraphics[scale=1,height=6.2cm]{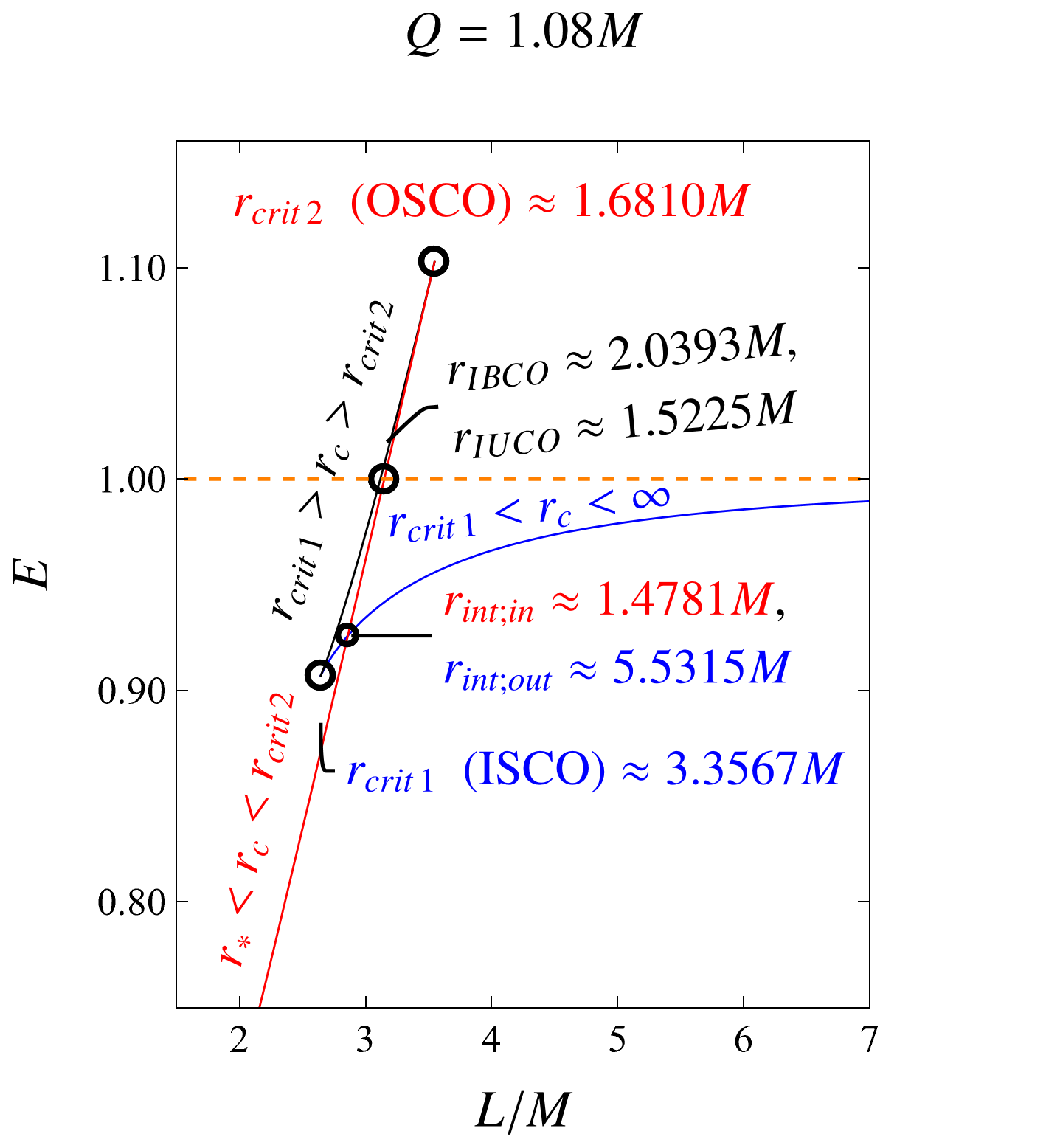}}
    \end{minipage}\par\smallskip
    \begin{minipage}{\linewidth}
    \centering
    {\includegraphics[scale=1,height=5.8cm]{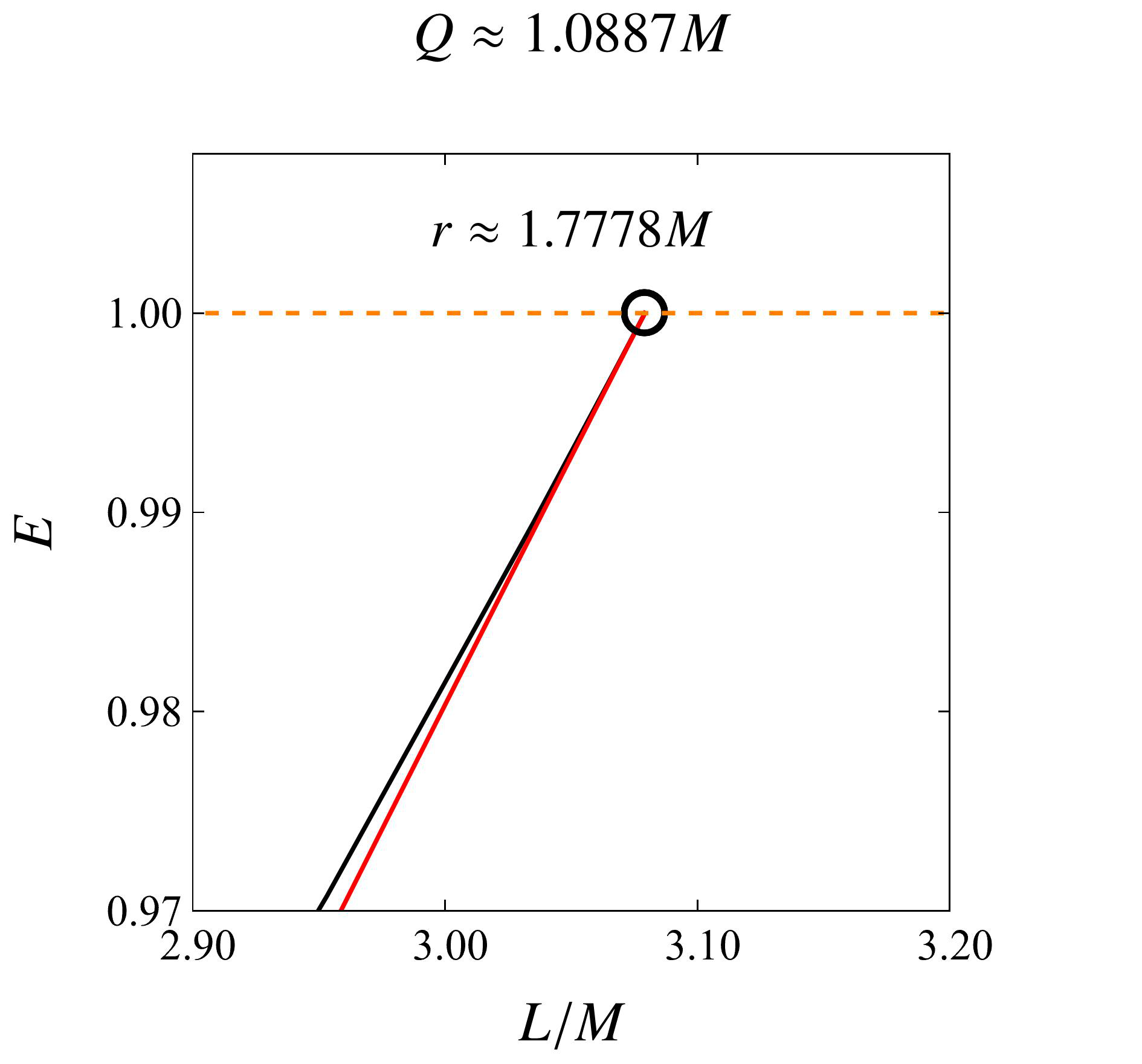}}
    \enspace
    {\includegraphics[scale=1,height=5.8cm]{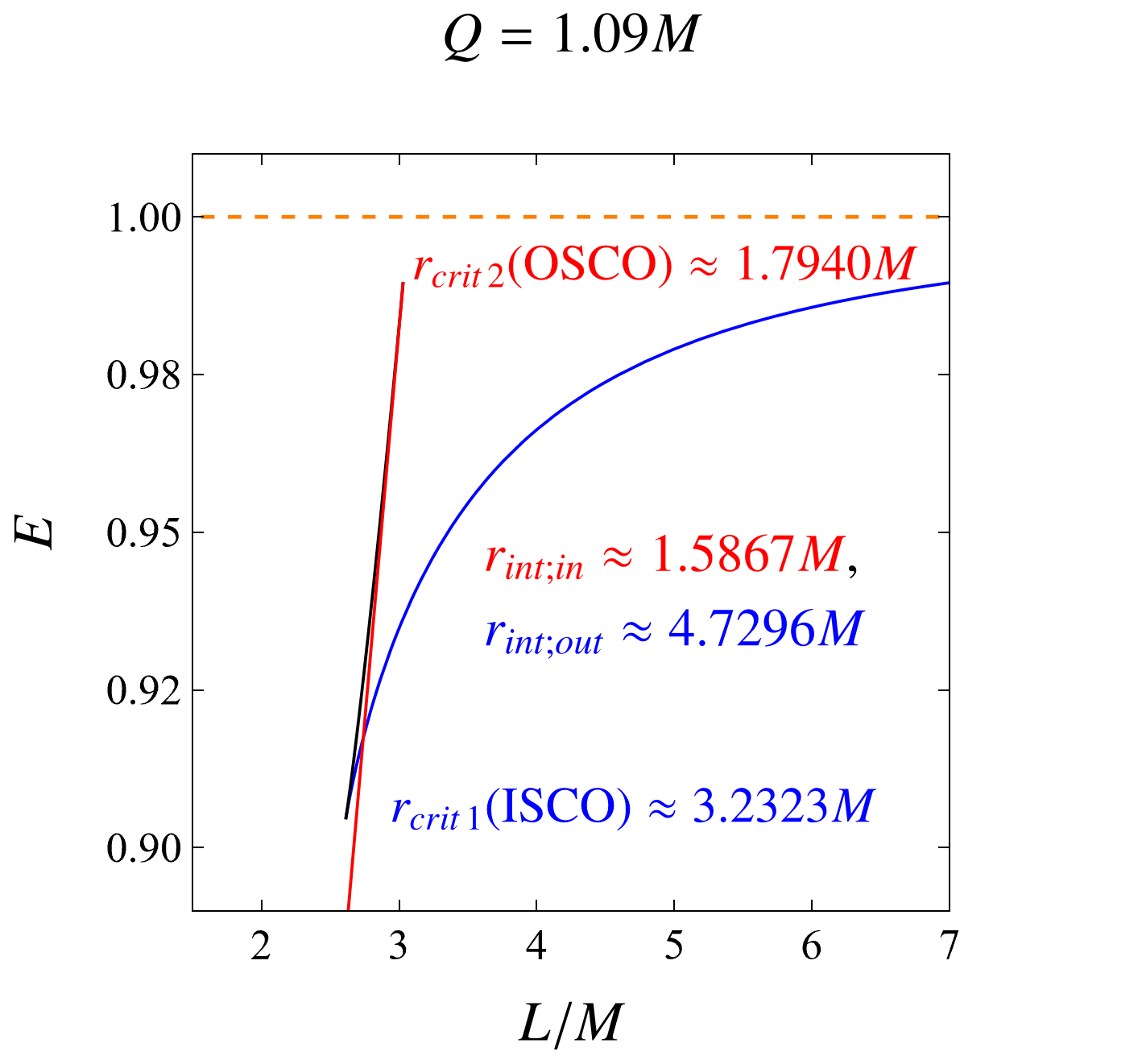}}
    \end{minipage}\par\smallskip
    \begin{minipage}{\linewidth}
    \centering
    {\includegraphics[scale=1,height=5.8cm]{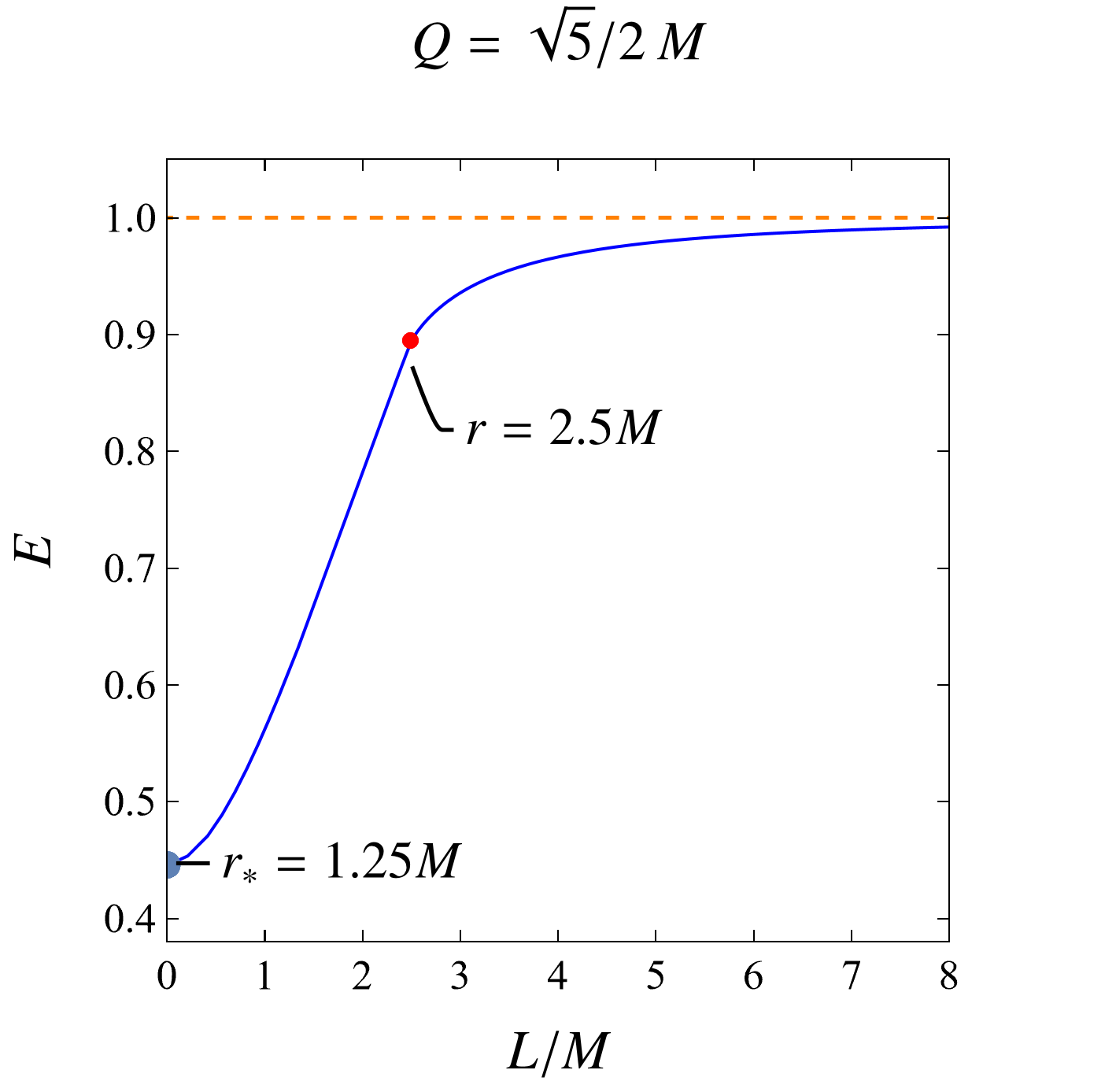}}
    \enspace
    {\includegraphics[scale=1,height=5.8cm]{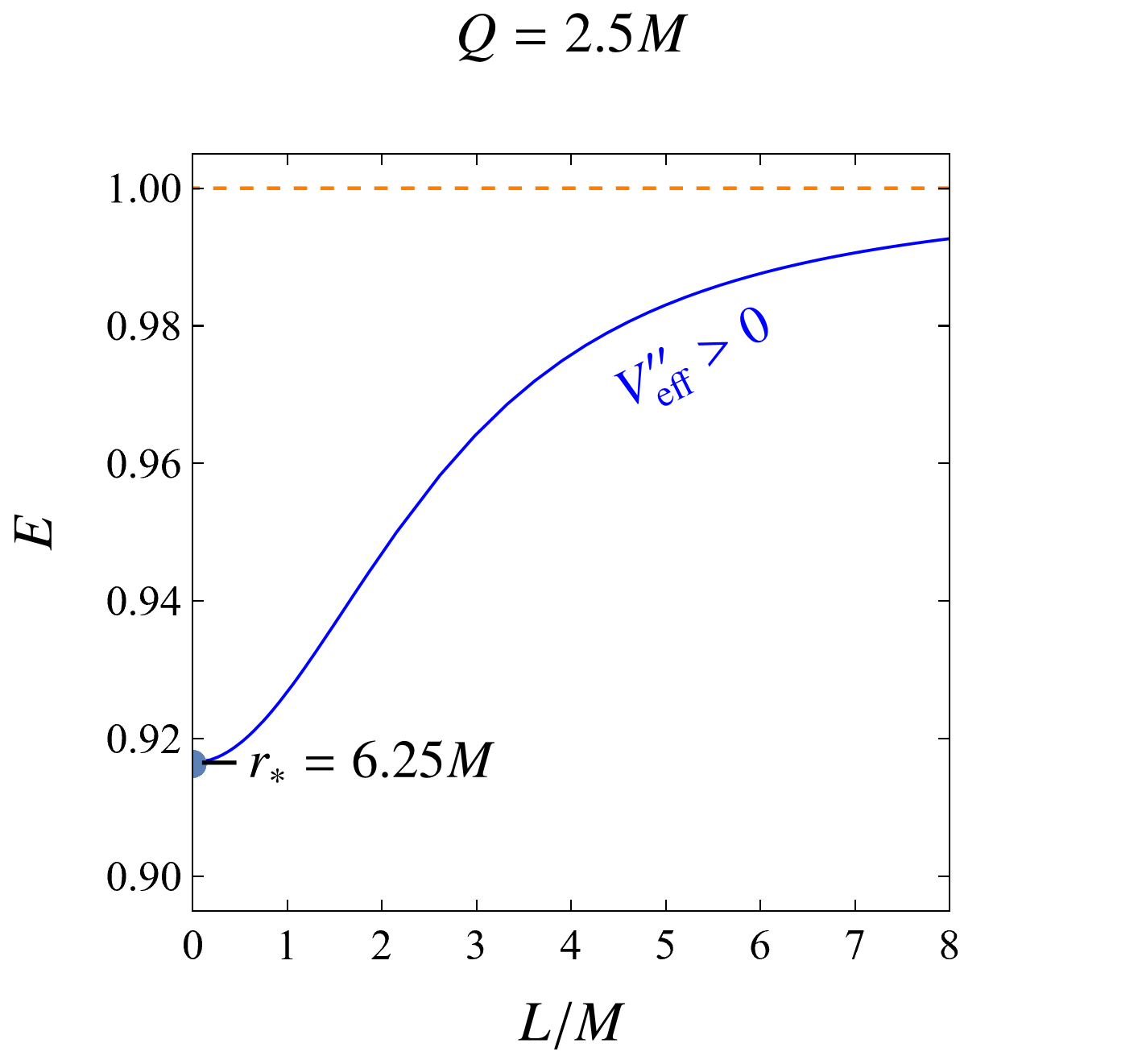}}
    \end{minipage}%
    \caption{The evolution of circular orbits curves for naked singularities from 
     Case 1 (top left) to Case 2A (top right and center left), Case 2B 
     (center right) and finally Case 3 in the bottom row. }
\end{figure}\label{Fig:2}

There will be three solutions as $V_{\mathrm{eff}}''=0$ yields a cubic equation.
For black holes, the only real solution correspond to the ISCO. For naked singularities,
there are three real solutions where ISCO is the largest solution, the Case 2 OSCO is 
the middle value and the smallest solution is unphysical since it lie in $\rc<r_*$. 
Alternatively, analytical expressions for the critical points are provided in 
Appendix \hyperref[app_B]{B}; (\hyperref[B2]{B2}) for the ISCO and (\hyperref[B1]{B1}) 
for the OSCO. Case 2 can be subdivided into two variations where Case 2A and Case 2B 
fall in $Q\in\sbrac{\sqrt{9/8}\,M,1.0887M}$ and $Q\in\brac{1.0887M,\sqrt{5/2}M}$ respectively. 
We obtain the maximum charge $Q\approx1.0887M$ for Case 2A by solving the conditions 
$E(\rc,Q)=1$ (\ref{LEcircular}) and $\pdv{}{\rc}E(\rc,Q)=0$. In Case 3, the two critical points 
coalesece at $r=2.5M$ of $Q=\frac{\sqrt{5}}{2}\,M$ and thereby, only a single region of stable 
bound orbits remain.

To see the full picture of periodic orbits map in $(L,E)$-space, we still
need expressions in terms of $e$ and $\lr$. This can be done 
by first reparametrizing the $u$-roots with these parameters \cite{yklimzcy24,ChandrashekarBook}. 
Refer to Appendix \hyperref[app_A]{A} for derivation guide. 
We label the four roots of $P(u)$ as
\begin{subequations}\label{rootpaireqs}
\begin{align}
a=&\dfrac{-(Q^2-M\lr)^2-H(Q,e,\lr)}{Q^2\lr(Q^2-M\lr)}\,,\phantom{=}b=
  \dfrac{-(Q^2-M\lr)^2+H(Q,e,\lr)}{Q^2\lr(Q^2-M\lr)}\,,\label{rootpaireqs1}\\
c&=\frac{1+e}{\lr}, \phantom{=}d=\frac{1-e}{\lr},\label{rootpaireqs2}\\
H(Q,e,\lr)=&\sqrt{(Q^2-M\lr)(e^2Q^6+M\lr(Q^4-M^2\lr^2+Q^2\lr(\lr-M)))}\label{He}
\end{align}
\end{subequations}
As in \cite{ChandrashekarBook,yklimzcy24}, the roots $c$ and $d$ represent the $u$-value of the 
periastron $u_p$ and the apastron $u_a$ of the orbits respectively 
and always take real values. $e$ and $\lr$ can be easily checked with \cite{levinhomo1},
\begin{equation}\label{elruroot}
   e=\dfrac{u_p-u_a}{u_p+u_a},\quad \lr=\dfrac{2}{u_p+u_a}
\end{equation}
The $a,b$ pair (\ref{rootpaireqs1}) is a consequence from the polynomial being quartic and have 
no analogue in Schwarzschild metric (or any spacetime with a degree 3 radial polynomial). This pair
could take complex values due to the presence of the square root function $H(Q,e,\lr)$ (\ref{He}).  
The reparametrized $L,E$ expressions are 
\begin{subequations}
\begin{align}    
L(e,\lr,Q)=&\sqrt{\dfrac{(M\lr^3-Q^2\lr^2)}{g(e,\lr,Q)}}\,,\quad \label{LEellipse}
E(e,\lr,Q)=\sqrt{\dfrac{f(e,\lr,Q)}{\lr^2\,g(e,\lr,Q)}}\,,\\
f(e,\lr,Q)=&\,\lr^4-4M\lr^3+(2(e^2+1)Q^2-4(e^2-1)M^2)\lr^2\\
          &+4(e^2-1)MQ^2\lr+(e^2-1)^2Q^4, \nonumber\\
g(e,\lr,Q)=&\,\lr^2-(e^2+3)M\lr+2(e^2+1)Q^2
\end{align}
\end{subequations}
Eqs.(\ref{LEellipse}) allow us to interpolate curves that encode the set of periodic orbits 
$(L,E)$ values for any $q$ within $0\leq e<1$. Visually, we observe these curves emanating 
as `branches' from the stable $\rc$ segments which are the $e=0$ limit and hence why we 
refer to them as $q$-branches (see Figs.\hyperref[Fig:3]{3} and \hyperref[BHqbranch]{4}). 
The upper limit of $e$ for any $q$-branch is clearly $e=1$ since this represent the parabola,
a type of hyperbolic trajectory.

\begin{figure}[ht]
\centering
    \begin{minipage}{\linewidth}
    \centering
    \subfloat{\includegraphics[scale=.82]{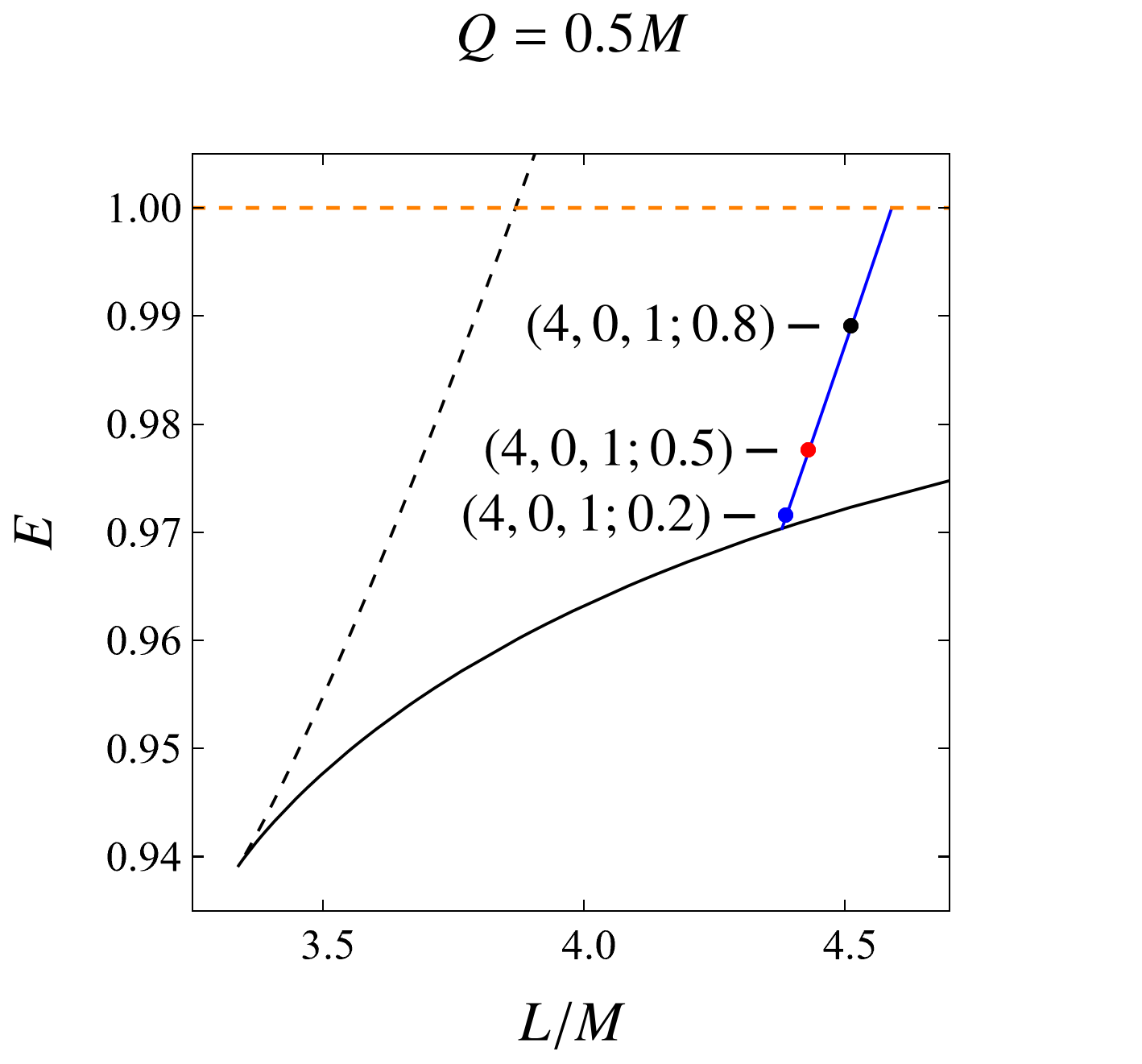}}
    \end{minipage}\par\smallskip
    \begin{minipage}{\linewidth}
    \centering
    \subfloat{\includegraphics[scale=.44]{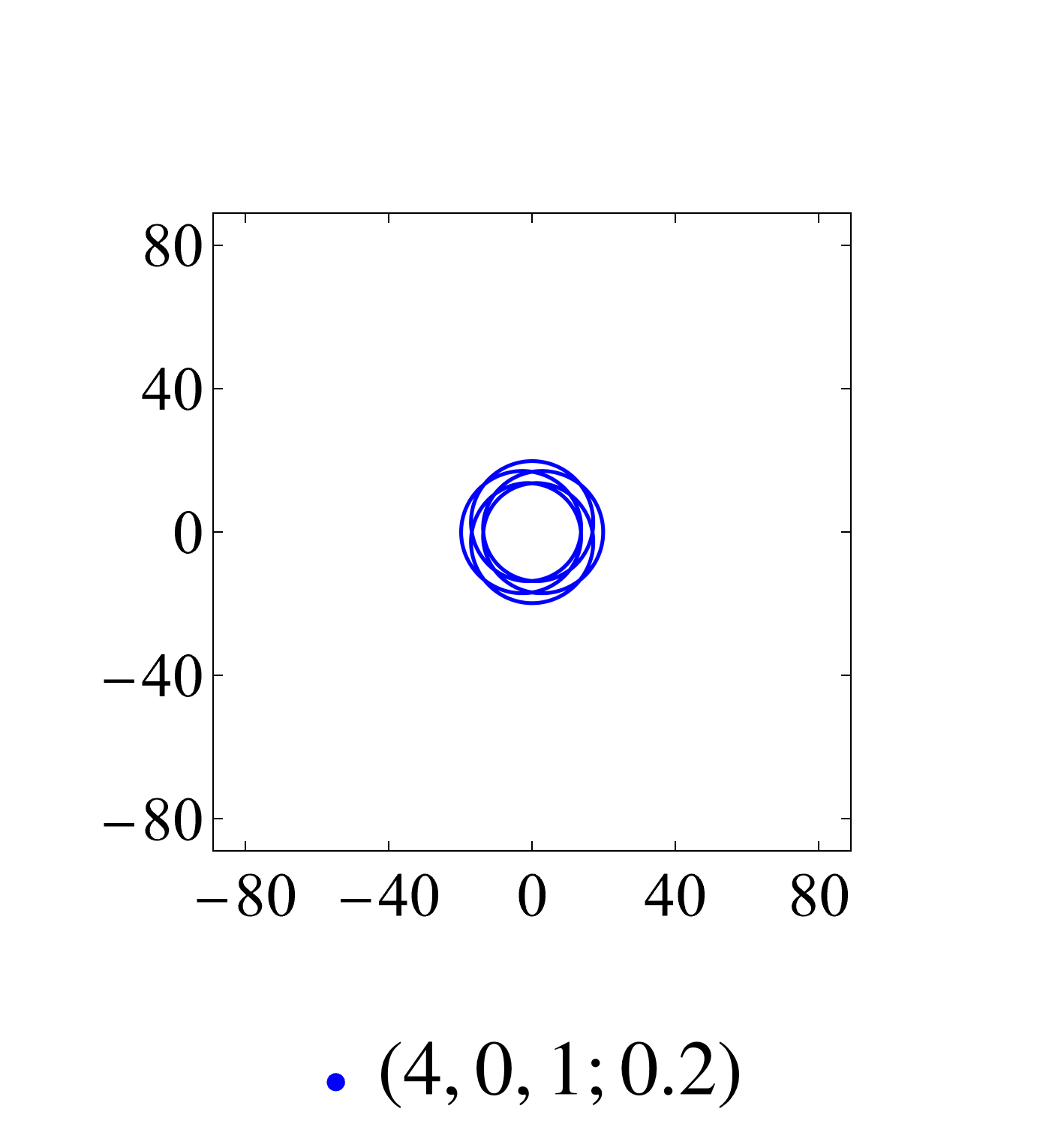}}
    \enspace
    \subfloat{\includegraphics[scale=.44]{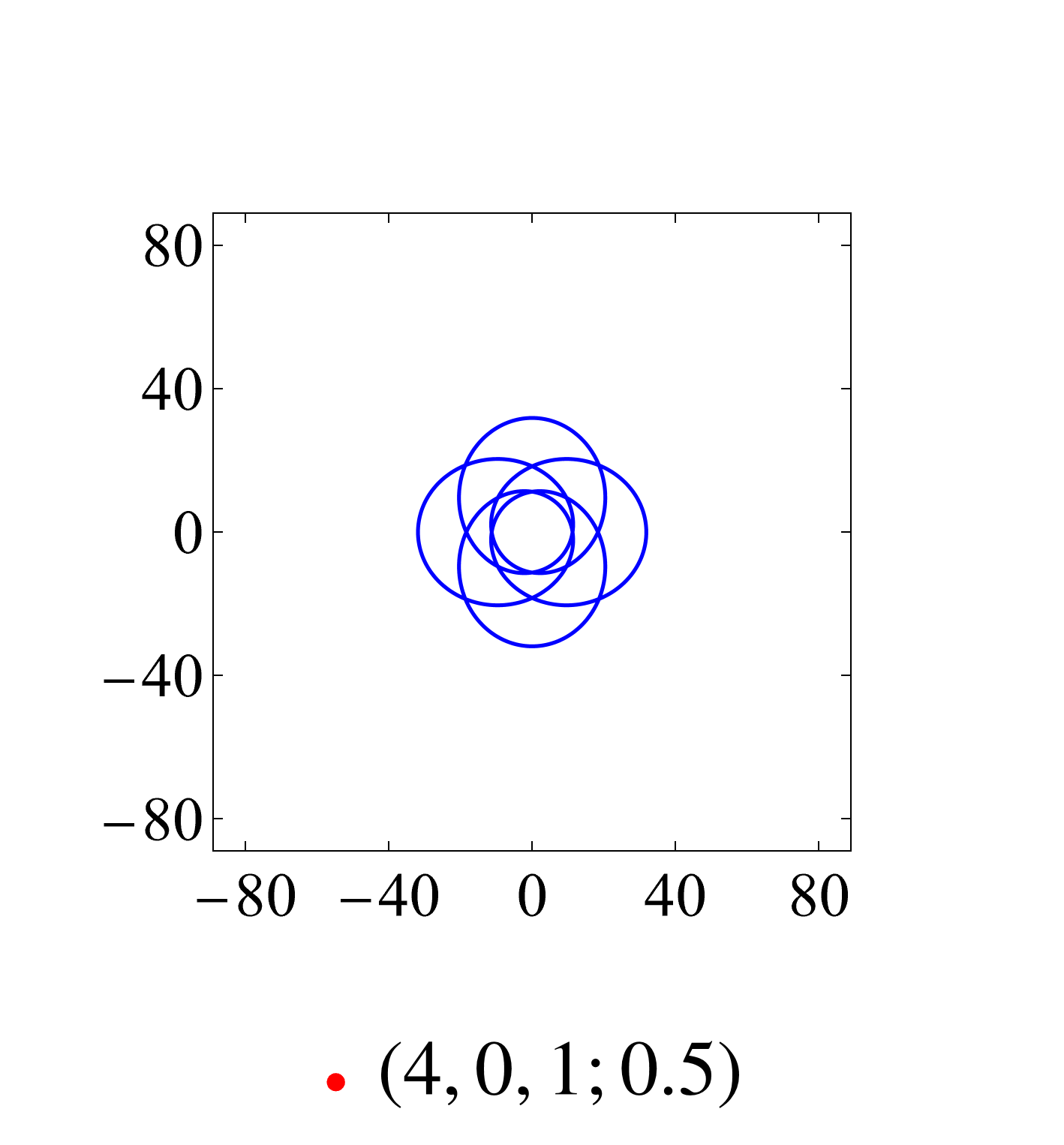}}
    \enspace
    \subfloat{\includegraphics[scale=.44]{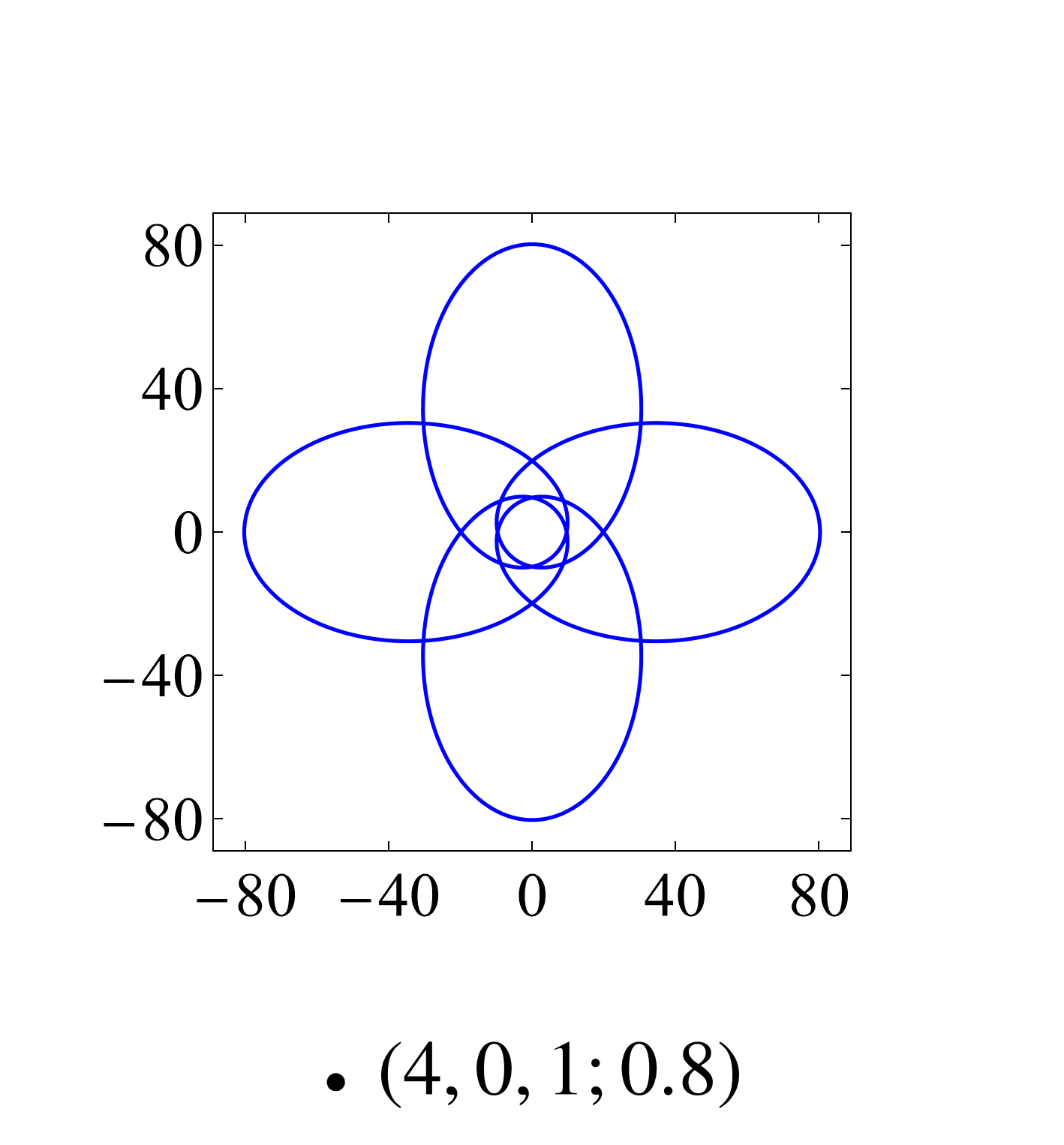}}
    \end{minipage}%
\caption{Top figure: The interpolated blue solid line is the $q$-branch containing the set of 
$(L,E)$ values of the $(4,0,1;e)$ orbit for $e=0.2,\,e=0.5$ and $e=0.8$.}
\end{figure}\label{Fig:3}

\subsection{Black holes \texorpdfstring{$q$}{TEXT}-branch distribution in \texorpdfstring{$(L,E)$}{TEXT}-space}\label{sec_2.3}
Now, we will show the procedures that allow us to chart out the entire rational 
$q$-branch distribution for different configurations of $z,w,v$ and $Q\,$. 
With the $\rc$ parameter, we can derive an equation that compute the discrete sets of 
the $q$-branches \textit{emanation points} from the stable circular orbits segments. 
First, we obtain the the second-order differential equation in $r$ from solving the 
Euler-Lagrange equation $\frac{d}{d\tau}\pdv{\Lagr}{\dot{r}}=\pdv{\Lagr}{r}$ of (\ref{lagr}),
\begin{equation}\label{ELE}
    \ddot{r}=\dfrac{f'\dot{r}^2}{2f}-\dfrac{f'E^2}{2f}+\dfrac{L^2f}{r^3}
\end{equation} 
Then, we linearize Eq.(\ref{ELE}) in the vicinity of $\rc$ by small perturbation 
approach, 
\begin{equation}\label{pertb1stord}
    r(\tau)=\rc+\varepsilon\,\delta r(\tau),
\end{equation}
Substituting Eqs.(\ref{LEcircular}) and (\ref{pertb1stord}) into Eq.(\ref{ELE}), 
up to linear order in $\varepsilon$, we get
\begin{equation}
\delta\ddot{r}=-\Omega^2\delta r,\quad\Omega=\dfrac{1}{\rc^2}
                 \sqrt{\dfrac{M\rc^3-6M\rc^2+9MQ^2\rc-4Q^4}{\rc^2-3M\rc+2Q^2}}
\end{equation}
Choosing suitable initial conditions \cite{yklimzcy24}, we relate the particle's $\phi$ increment to
its orbital period given by $\tau=\frac{2\pi}{\Omega}$ and the $q$-parameter \cite{levinpg08},
\begin{equation}\label{Omega}
q+1=\dfrac{\Delta\phi}{2\pi}\simeq\dfrac{L_c}{\rc^2\Omega}=
    \rc\,\sqrt{\dfrac{M\rc-Q^2}{M\rc^3-6M\rc^2+9MQ^2\rc-4Q^4}}
\end{equation}
Then rearrange Eq.(\ref{Omega}) to obtain a cubic equation in $\rc$,
\begin{equation}\label{rcq}
M((q+1)^2-1)\,\rc^3+(Q^2-6M^2(q+1)^2)\,\rc^2+9MQ^2(q+1)^2\,\rc-4Q^4(q+1)^2=0    
\end{equation}
The emanation points are the solutions of Eq.(\ref{rcq}). Similar to (\ref{cuspcond}), there are three real 
$\rc$ solutions for $Q>M$. For Case 1 and 2, we take the larger two solutions and
ignore the smallest solution whose value is always less than $r_*$. But for Case 3, we only need 
the smallest solution instead since that is the only $\rc$ value where the $q$-branches could 
emanate from continuously for $e\simeq0$. We will remind the readers of these choices again in 
later sections.

Coming back to the $P(u)$ polynomial, we will first separate the differential equation 
(\ref{Pu0}) and then take the integrals on both sides. This gives the general form
of an \textit{elliptic integral},
\begin{equation}\label{EI}
\phi(u)=\int_{u_1}^{u}\dfrac{du}{\sqrt{P(u)}}
\end{equation}
with $u_1$ as the initial condition. As taught in \cite{brydEI,gradshteyn2014table}, 
by algebraic reduction, the polynomial $P(u)$ can be written in its factored form (\ref{Pu2}). 
Furthermore, (\ref{EI}) can be expressed in the Jacobi form of the first kind, 
that is $g\,\mbox{F}(\psi,k)$ where $g$ is a constant, $\psi$ is the
amplitude and $k$ is the elliptic modulus which could be complex. The function $\mbox{F}$ is 
the incomplete elliptic integral of the first kind. Values of $g,\psi$ and $k$ are 
dependant on the roots $a,b,c,d$ (\ref{rootpaireqs}).

We have two usages for this elliptic integral. Here, we use it to set up the relation to determine $\lr$ 
numerically for any fixed $(z,w,v;e)$ sets. This enable us to chart $(L,E)$ values for any $e>0$ to 
complement with Eq.(\ref{rcq}). Following the procedure in Ref.\cite{yklimzcy24}, we choose the initial condition to 
be at the apastron $d$ and roots ordering $a>b>c\geq u>d$, as given in 252.00, pg.103 \cite{brydEI},
\begin{align}\label{EIouteraphe}
     \phi(u)&=\int_{d}^{u}\dfrac{du}{\sqrt{Q^2(a-u)(b-u)(c-u)(u-d)}}\nonumber\\
            &=\dfrac{2}{Q\sqrt{(a-c)(b-d)}}\,
              \mbox{F}\brac{\arcsin{\sqrt{\dfrac{(a-c)(u-d)}{(c-d)(a-u)}}},
              \sqrt{\dfrac{(a-b)(c-d)}{(a-c)(b-d)}}},   
\end{align}
where, $g=\dfrac{2}{Q\sqrt{(a-c)(b-d)}}$, 
$\psi=\arcsin{\sqrt{\dfrac{(a-c)(u-d)}{(c-d)(a-u)}}}$
and $k^2=\dfrac{(a-b)(c-d)}{(a-c)(b-d)}$.

Then, the accumulated angle $\Delta\phi_r$ between successive periastrons per orbital period is 
\begin{equation}\label{K(k)relation}
    \Delta\phi_r=2\phi(u_p)=2\phi(c)=g\,\mbox{F}\brac{\frac{\pi}{2},k}=g\,K(k),
\end{equation}
where $K(k)$ is the complete elliptic integral of the first kind. Relating it with $q$ again,
\begin{equation}\label{LRnumerical}
    \dfrac{\Delta\phi_r}{2\pi}=\dfrac{g\,K(k)}{2\pi}=q+1=w+\dfrac{v}{z}+1,
\end{equation}
then substitute Eqs.(\ref{rootpaireqs}) first into  
(\ref{EIouteraphe}) then (\ref{LRnumerical}). Upon rearrangement, we have 
\begin{equation}\label{LRnumericalrel}
\dfrac{\pi}{2}\sqrt{\dfrac{J(Q,e,\lr)}{(M\lr^3-Q^2\lr^2)}}\brac{w+\dfrac{v}{z}+1}
= K\brac{\sqrt{\dfrac{4eH(Q,e,\lr)}{J(Q,e,\lr)}}}
\end{equation}
where $J(Q,e,\lr)=2eH(Q,e,\lr)+M\lr^3-6M^2\lr^2-(e^2-9)MQ^2\lr+2(e^2-2)Q^4$ and $H(Q,e,\lr)$ is
as given in (\ref{He}). Eq.(\ref{LRnumericalrel}) coupled with relevant solutions of Eq.(\ref{rcq}) 
reveal the complete distribution of periodic orbits for any fixed $Q$. It can be tricky to find the right $\lr$, 
especially for periodic orbits from the inner region since the polynomials in (\ref{LRnumerical})
typically yield multiple real solutions. Our practice is to take the closest monotonically increasing 
$(e,\lr)$ values starting from $e=0$ and avoid nonsensical large jumps between values. 
Subsequently, it is checked that the solutions return rational $q$ and produce plots of orbits that closes on itself.       

\begin{figure}[h]
\centering
{\includegraphics[width=.56\textwidth]{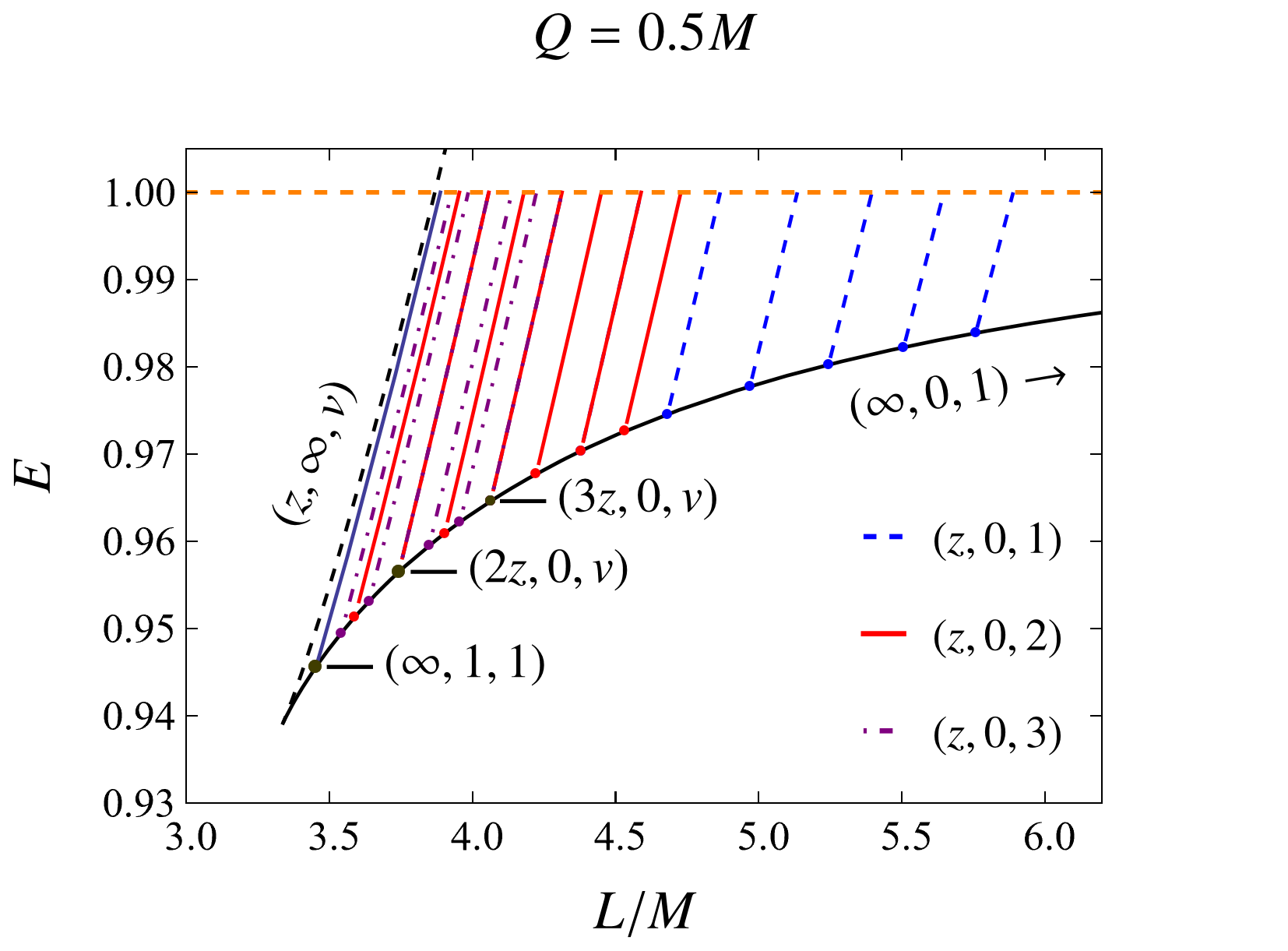}}
\enspace
{\includegraphics[width=.41\textwidth]{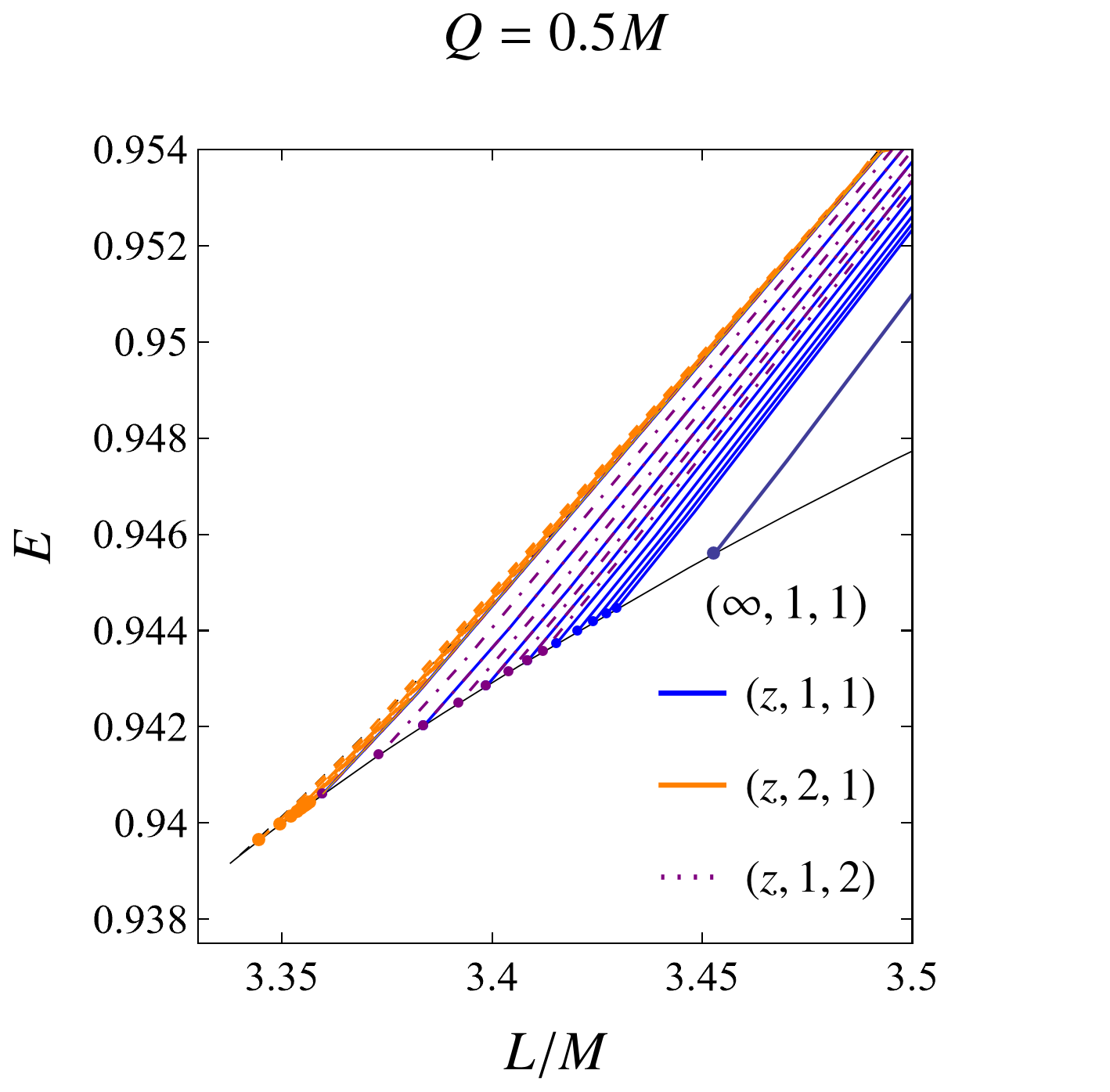}}
\caption{The $q$-branch general distribution for RN black holes with $0\leq Q\leq M$. 
Dots on the stable $\rc$ segment are the emanation points. Here, $z$ values are from 
$1$ to $9$.}
\end{figure}\label{BHqbranch}

The general $q$-branch distribution pattern for black holes in $(L,E)$-space 
was first discovered in \cite{yklimzcy24} for the Schwarzschild ($Q=0$) case.
To summarize the distribution pattern, the sequence of branch of increasing $z$ always 
goes from left to right whereas sequence of increasing $v$ shift all branches of fixed $z,w$ 
to the left. Increasing $w$ drastically shrink the range of $L$ values and 
visually, all $w\geq2$ sequences are indistinguishable from one another (Fig.\hyperref[BHqbranch]{4}).
 
Because of the coprime condition of $z$ and $v$ \cite{levinpg08}, increasing unit value of $v$ 
require removing all integers of $z\leq v$. Taking an example, if the largest $z$ value is $9$ and  
$v$ is $2$, then the smallest $z$ value must be $3$. We certainly notice that branches where 
$z$ and $v$ are not coprime all emanate from the same $\rc$ point and completely overlap one 
another, like the $(2,0,1)-(4,0,2)-(6,0,3)$ triplets in Fig.\hyperref[BHqbranch]{4}

The limit branches are also important, especially when 
dealing with domains in $(L,E)$-space involving complex $u$-roots later on. As before, increasing $w$ cause 
the $(z,w,v)$ sets to converge towards the unstable circular orbit segment. Thus, the unstable circular 
orbits segment is equivalent to the $w=\infty$ branch and is the overall left-most boundary of the 
distribution \cite{yklimzcy24}. $w=0$ are the right boundary for $w$ sub-divisions. By the coprime 
condition again, the $(z,w,1)$ and $(z,w,v)$, for $v=z-1,\,z>2$ are the right and left boundaries for 
$v$ sub-divisions respectively. The $z\rightarrow\infty$ limit branch can be approximated with  
\begin{equation}\label{limqbranch}
\underset{z \to \infty}{\lim}(z,w,1)\simeq(1,w,0),
\end{equation}
and is the right boundary for $z$ sub-division. This limit branch 
acts as a separator between different $w$ sub-divisions since it is impossible for any
$(z,w,v)$ branches to cross over to its left and likewise for the $(z,w+1,v)$ ones to 
cross over to its right. By (\ref{limqbranch}), the infinite $z$ limit recover the 
$(1,0,0)$ Keplerian orbits in agreement with the non-relativistic limit.

\section{Roots configuration and analytical solutions} \label{sec_3}
We will first explain the significance of the $P(u)$ polynomial and how its root configuration 
constitute periodic orbits. To rephrase our earlier statement about the roots disposition \cite{ChandrashekarBook}, 
a geometrical bound orbit is defined by a closed domain in $P(u)$ where values between two distinct positive 
real roots contain the set of all possible radial distance in that orbit. Within this context, a periodic orbit 
radial motion can be translated as the particle's initial $u$-value starting from one of the root and then 
oscillate to and fro between the other root of the same domain. 

No oscillations occur when the domain have a root that is zero or negative. This 
correspond to the unbound \footnote{Another type of unbound trajectory is the plunging orbit, an irreversible 
journey into the central singularity. In $P(u)$ context, the $u$-value make a `one-way trip' from the periastron root 
to positive infinity. This is not possible for RN as radial polynomials with even number terms always exhibit an 
inifinte potential barrier in $\veff$ that prevent anything from reaching the singularity.} 
escaping orbits. We can intepret it as the $u$-value coming from negative infinity approaching the positive 
periastron root, then bounce off this root before going back to the same direction it came from. Circular orbits 
roots $\uc$ are \textit{degenerate}, meaning the root value is repeated at least once and thus reduce the total 
number of distinct roots. These points are illustrated by $P(u)$ plots in Fig.\hyperref[p(u)108(3,1,1)]{5}.
The periastron root always lie to the right of the apastron's one since $u_p>u_a$. As $e\rightarrow0$, the gap 
between roots of a domain shrink and eventually coalesce at $\uc$ value.
We further extend the periodic orbit notation as $(z,w,v;e)_{outer/inner}$ to indicate the region
outside a naked singularity where the periodic orbit reside.

\begin{figure}[t]
    \centering
        \begin{minipage}{\linewidth}
        \centering
        {\includegraphics[scale=.53]{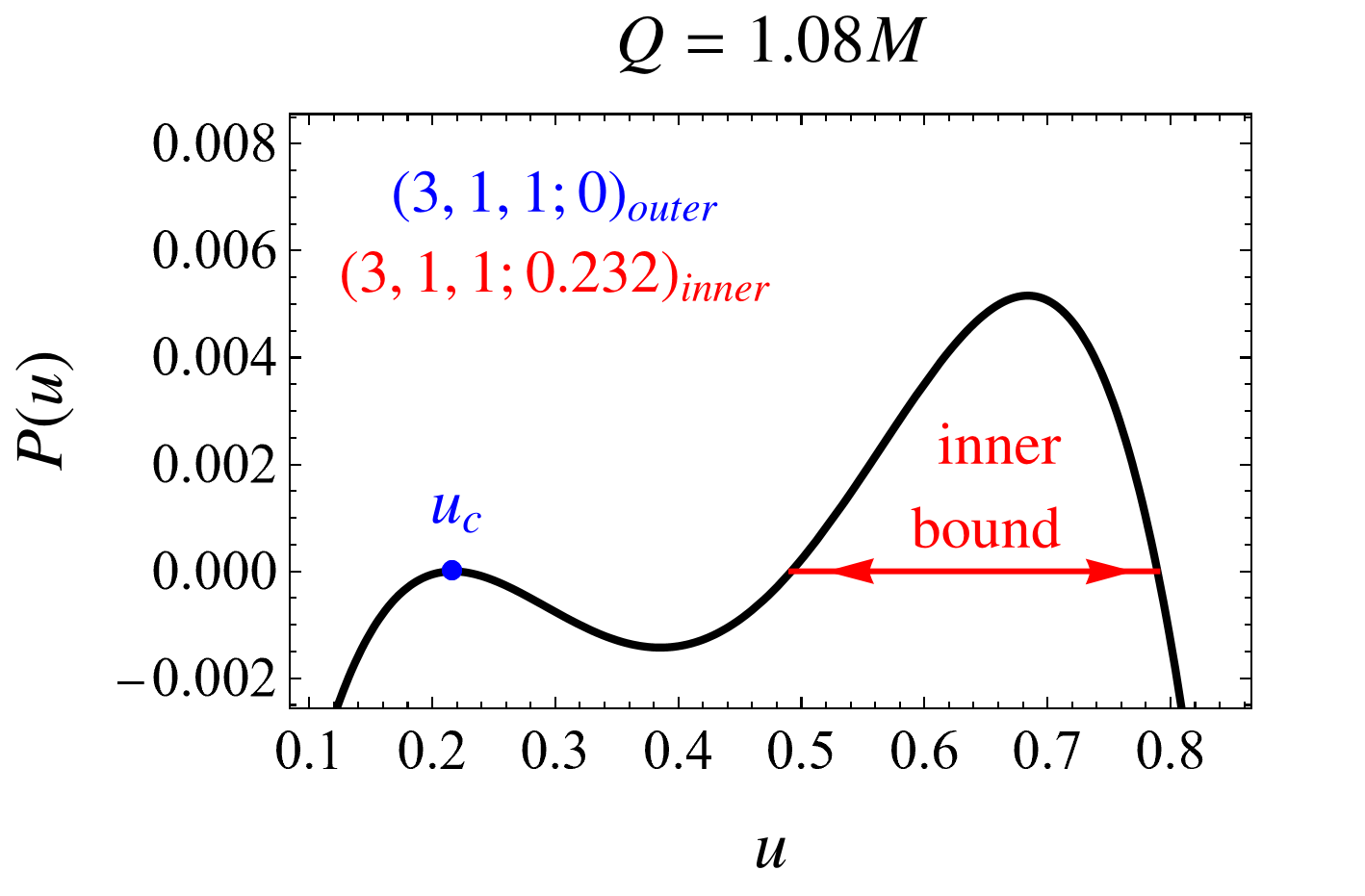}}
        \enspace
       {\includegraphics[scale=.53]{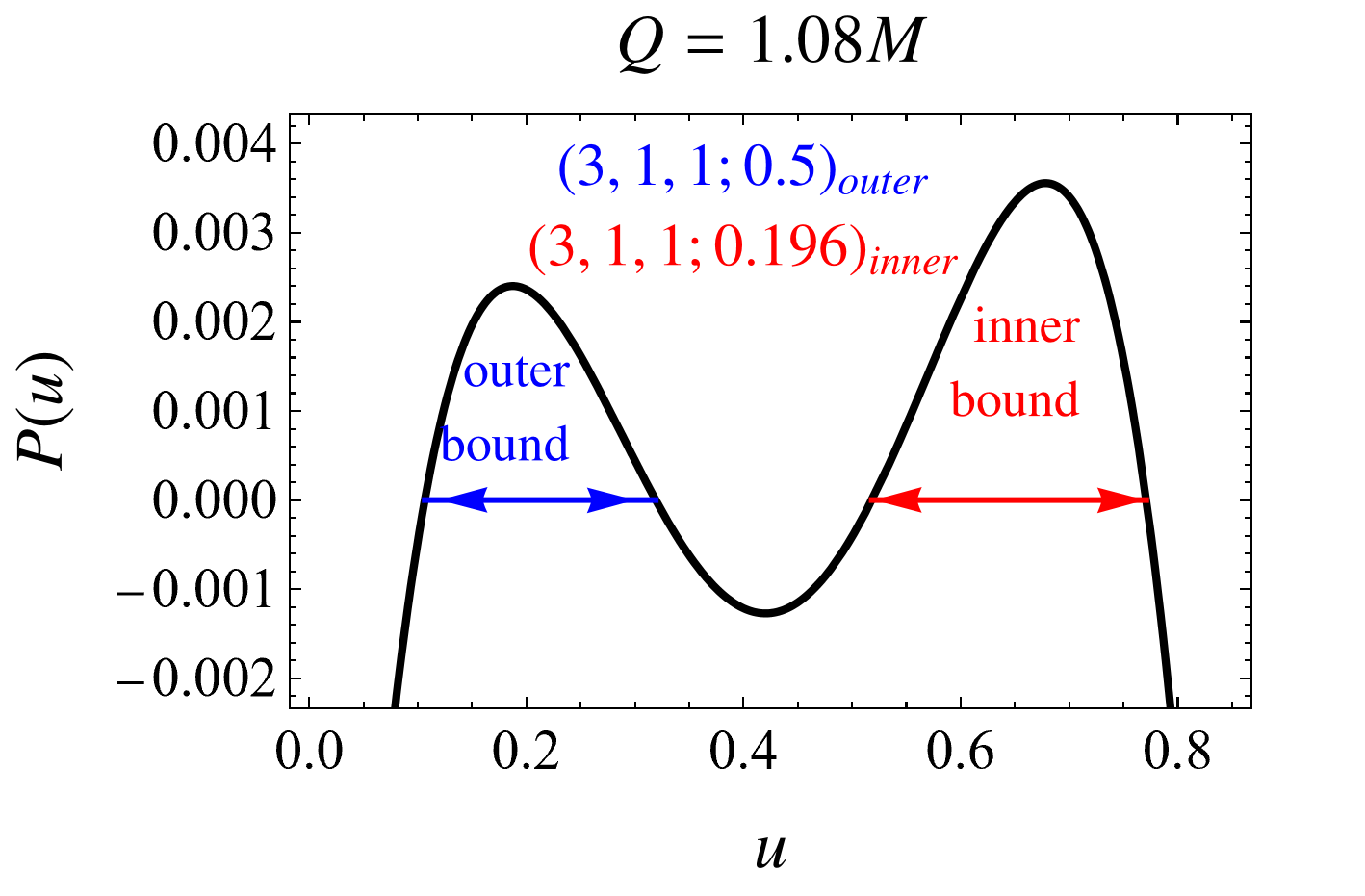}}
        \end{minipage}\par\smallskip
        \begin{minipage}{\linewidth}
        \centering
        {\includegraphics[scale=.53]{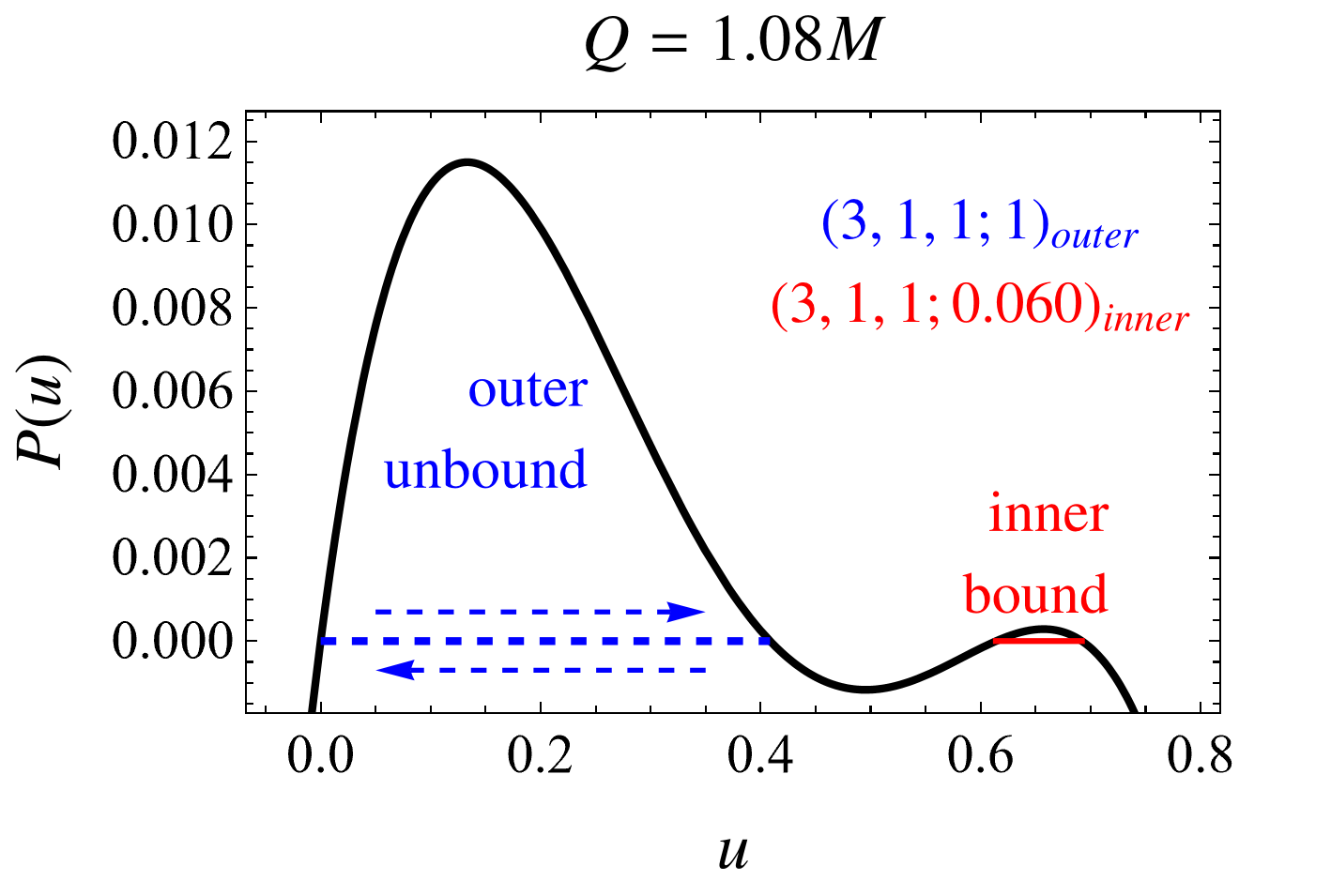}}
        \enspace
        {\includegraphics[scale=.53]{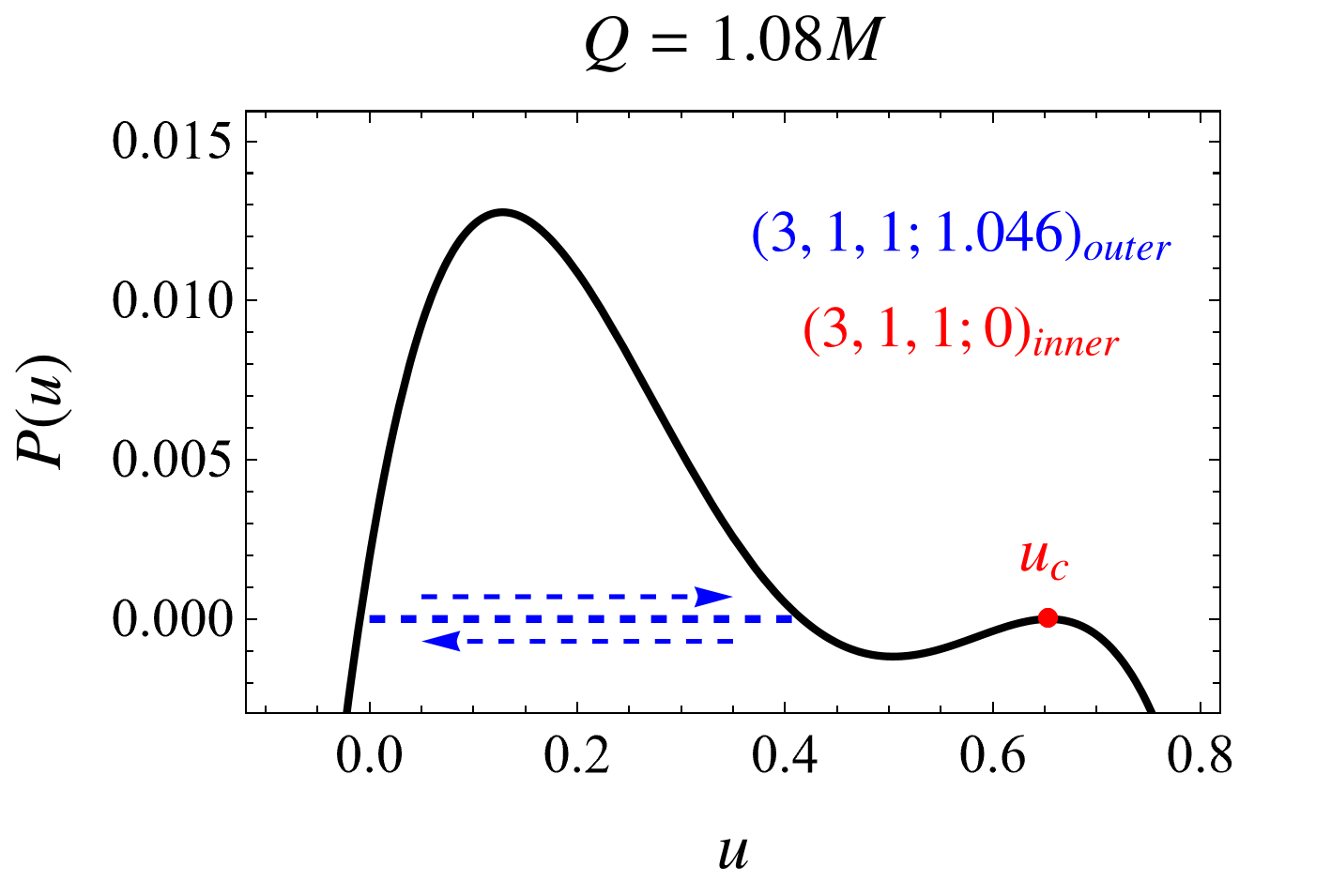}}
        \end{minipage}%
\caption{Evolution of $P(u)$ moving upwards along the slope of the $Q=1.08M,\,(3,1,1)$-branch, 
from the emanation point of the outer orbit up to that of the inner orbit. Each plots correspond to values of 
(\ref{Pu2}), (\ref{eabcdvalues}), (\ref{eabcdvalues2}) and (\ref{(3,1,1;0)in}). Dashed lines represent the 
`one-way' motion of unbound escaping orbits.}
\end{figure} \label{p(u)108(3,1,1)}

Since the polynomial for RN spacetime is quartic, there are up to two domains corresponding to 
bound orbits which further solidify the possibility of a periodic orbit pair. 
Referencing previous works \cite{ChandrashekarBook,szelakova,grunauRN}, 
the smaller value domain correspond to orbits of the first kind, the relativistic analogue of 
Keplerian orbits whilst the larger value domain correspond to orbits of the second kind, a purely 
relativistic effect. This feature is also present in RN black holes, but it have a more complex global 
causal strucutre where the Killing vector fields in the region between the two horizons, 
$r_-<r<r_+$ switch causality. On top of that, we need to maximally extend the manifold 
at both horizons in order to describe a complete geodesic \cite{soltani}. We will not pursue this 
case further. Solutions of unbound trajectories will 
not be presented but readers can find it in \cite{szelakova}.

\subsection{Four real roots}\label{sec_3.1}
We start by analysing the $Q=1.08M,\,(3,1,1)$ orbit. We determine $\lr$ unique to some $e$ from 
Eq.(\ref{LRnumericalrel}) and then plug these into Eqs.(\ref{rootpaireqs}) to obtain the 
numerical value of all four roots $a,b,c,d$ as follows;
\begin{subequations}\label{eabcdvalues}
\begin{align}
e&=0,  & \lr&=4.6133M, & a&=0.78938, & b&=0.49177, & c&=d=0.21677,\\
e&=0.5,& \lr&=4.7004M, & a&=0.77102, & b&=0.51816, & c&=0.31912,\en d=0.10637,\\
e&=1,  & \lr&=4.8945M, & a&=0.69212, & b&=0.61394, & c&=0.40862,\en d=0
\end{align}
\end{subequations} 
Observe that a non-zero $e$ causes all four roots to take real distinct positive values but $e=0$
have a value repeated for $c$ and $d$. If we pair up $a$ with $b$ and $c$ with $d$ and then inspect 
the corresponding $e$ and $\lr$ values for each pairing with Eqs.(\ref{elruroot}), the
$c,d$ pair return the initial value whilst the corresponding $a,b$ pair output
smaller values. The exception to this is the $c,d$ pair in (\hyperref[eabcdvalues]{26a}). Here, 
the $a,b$ pair yield a non-zero $e$. So, we deduce that if any $(z,w,v;e)$ sets 
produce four real roots, then there should be an associative pair of periodic orbits possessing 
the same $(L,E)$ values, each residing in different regions surrounding the naked 
singularity. The corresponding values of $L$ and $E$ for each set in (\ref{eabcdvalues}) 
where the $c,d$ pair is \textcolor{blue}{blue} (left) and $a,b$ pair is 
\textcolor{red}{red} (right) are;  
\begin{subequations}\label{eabcdvalues2}
\begin{align}
&(26\mbox{a})\Rightarrow L=2.7394, & E&=0.91670, & \color{blue}(0,4.6133M)&\color{black}
                                         \Leftrightarrow\color{red}(0.232,1.5611M),\label{eq32a}\\
&(26\mbox{b})\Rightarrow L=2.8322, & E&=0.93440, & \color{blue}(0.5,4.7004M)&\color{black}
                                         \Leftrightarrow\color{red}(0.196,1.5514M),\label{eq32b}\\
&(26\mbox{c})\Rightarrow L=3.1425, & E&=1, & \color{blue}(1,4.8945M)&\color{black}
                                        \Leftrightarrow\color{red}(0.0599, 1.5313M)
\end{align}
\end{subequations}
Clearly, the pairs $c,d$ and $a,b$ correspond to the outer and inner region respectively. 
Then, when we interpolate $q$-branch for the outer periodic orbit, values of both $L$ and $E$ increase 
monotonically together with $e$. So, the outer $q$-branch are straight lines branching 
upwards diagonally right from the outer stable $\rc$ segment and end just below the $E=1$ line as in 
Fig.\hyperref[Fig:3]{3} and Fig.\hyperref[BHqbranch]{4}. Based on the values in (\ref{eabcdvalues2}), the 
$e$ values for the inner region decrease in contrast to increasing $L$ and $E$. So,
the inner $q$-branch could even extend above the $E=1$ line as $e\rightarrow0$ and thus emanate from some 
point on the $E\geq1$ portion of the inner stable $\rc$ segment. The values at the emanation point 
of the inner $(3,1,1)$ orbit are  
\begin{align}\label{(3,1,1;0)in}
L=3.1861,\en E=&1.0098,\,a=b=0.65382,\,c=0.41637,\,d=-0.009333\nonumber\\
&\Rightarrow\color{blue}(1.046,4.9136M)\color{black}
 \Leftrightarrow\color{red}(0,1.5295M)
\end{align}
Emanation points of inner periodic orbits cannot originate from $r_{\gamma^-}\leq\rc\leq r_{\gamma^+}$, 
where time-like circular orbits are forbidden in Case 1. This imply that there is a limited number 
of inner periodic orbit that could exist for certain $Q$ and even fewer associative pairs of periodic orbits
(Fig.\hyperref[overlap]{6}). Also, any inner $q$-branch from the four real roots configuration must specifically emanate from points 
above $r_{int}$, excluding the critical point in Case 2. Thus, the emanation range\footnote{This is the range 
where the middle value solution of Eq.(\ref{rcq}) representing the $e=0$ limit of inner periodic orbits must take. 
Obviously, the associated outer periodic orbit correspond to the largest solution.} of an inner periodic 
orbit $q$-branch containing four real roots is $r_{int;\,in}<\rc<r_{\gamma^-}$ for Case 1 and 
$r_{int;\,in}<\rc<r_{OSCO}$ for Case 2. A small eccentricity range for inner periodic orbits from the emanation point 
down to the $e$ value at the $E=1$ line have an associated outer escaping orbit. 
As expected, we find that an inner periodic orbit is always `contained' within its outer counterpart since 
the apastron of the former do not overlap the periastron of the outer orbit (Fig.\hyperref[PO4R]{7}).

\begin{figure}[]
    \centering
    \subfloat{\includegraphics[width=.53\textwidth]{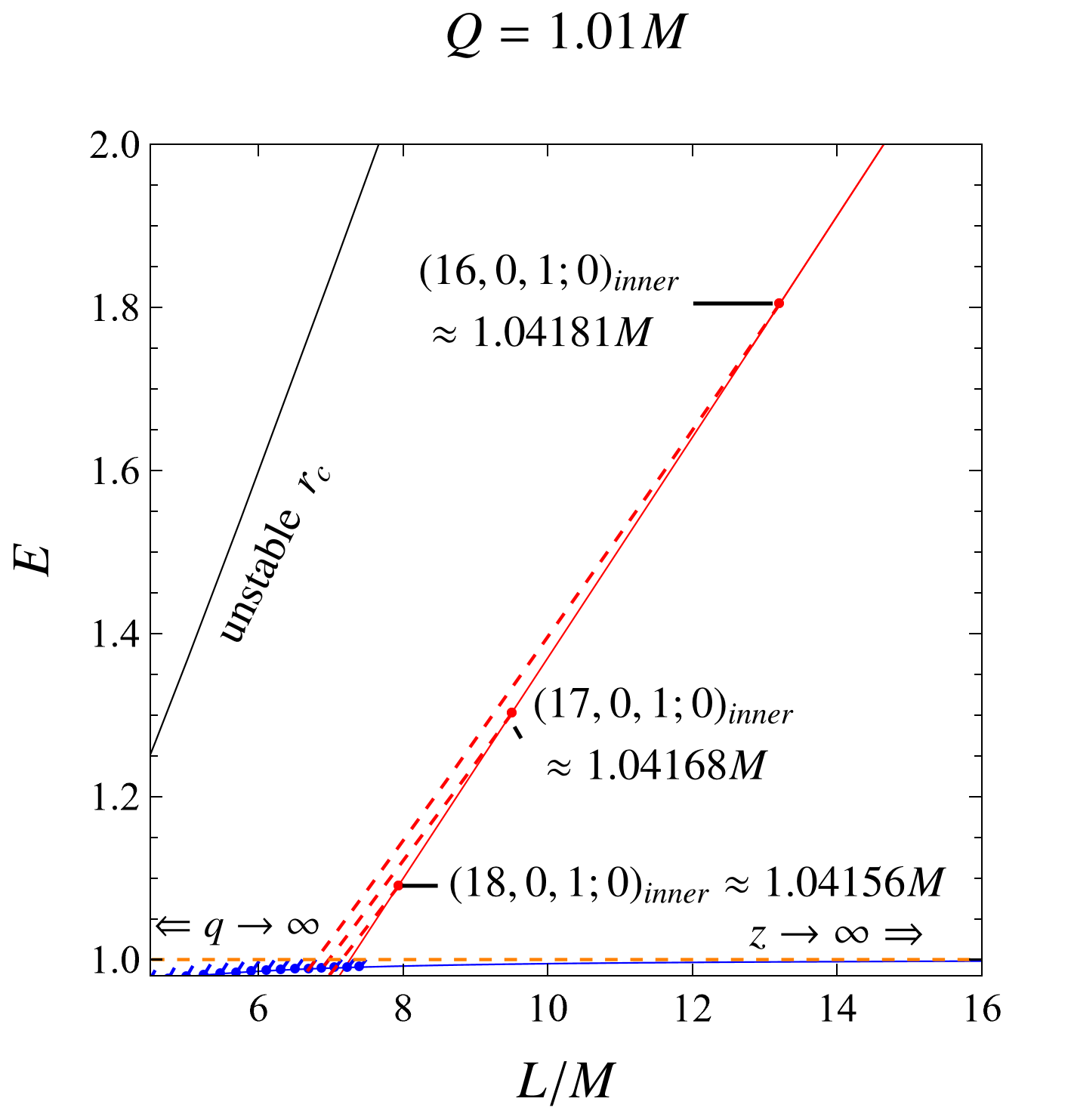}}
    \enspace
    \subfloat{\includegraphics[width=.55\textwidth]{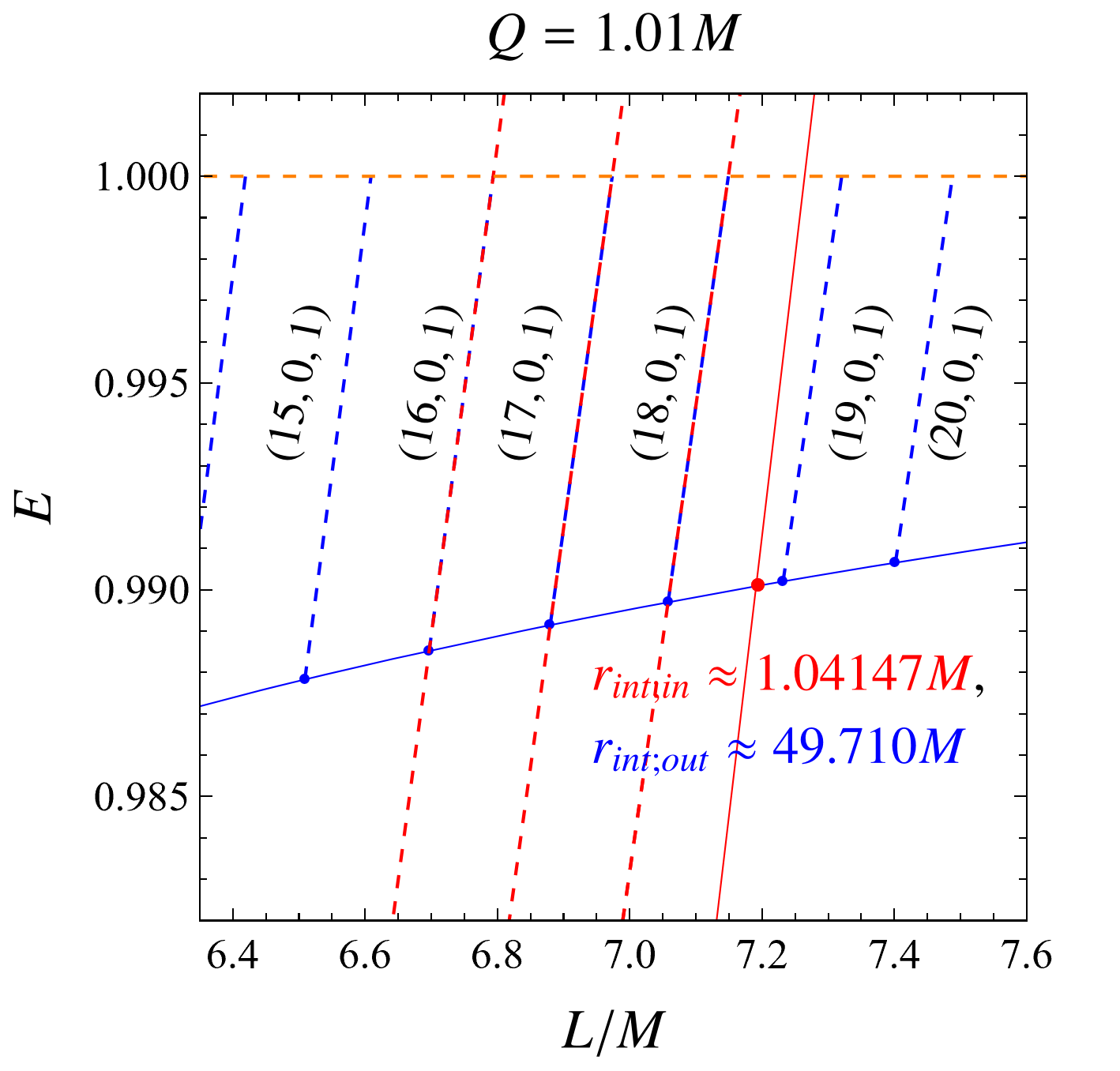}}
    \caption{Portions of the inner periodic orbits $q$-branch with four real roots.
    An extra significant figure was added to distinguish the emanation points radial values 
    and $r_{init;in}$. Right figure: $q$-branches of the outer and inner orbits completely 
    overlap one another.}
    \end{figure}\label{overlap}

To plot periodic orbits, we seek expressions of the radial distance as a function of $\phi$ in terms of the roots 
values of $P(u)$. So this is where the second usage of elliptic integral comes in. The procedure is 
to first invert the Jacobi form of the integral, $\phi(u)=g\,\mbox{F}(\psi,k)$ to $u(\phi)$, then take the
reciprocal to get $r(\phi)$. When all four roots have real values (including negative reals), 
the general form of $r$ as in 250.04, pg 97 of \cite{brydEI} is 
\begin{equation}\label{EF4realgen}
    r(\phi)=\dfrac{A_1+A_2\,\sn^2(g^{-1}\,\phi,k)}{A_3+A_4\,\sn^2(g^{-1}\,\phi,k)}
\end{equation}
where $\sn$ is the Jacobi elliptic sine function and $A_1,A_2,A_3,A_4$ are functions of roots $a,b,c,d$. 
Arrangement of roots in these functions depends on $\psi$ which in turn, depend on the initial conditions. 
Because of our choice of root ordering, $g^{-1}$ and $k$ for the general form in 
(\ref{EF4realgen}) will always be (see pg.275 \cite{gradshteyn2014table}),
\begin{equation}\label{ginvk}
    g^{-1}=\dfrac{Q}{2}\sqrt{(a-c)(b-d)}\en,\quad k^2=\dfrac{(a-b)(c-d)}{(a-c)(b-d)}\,,
\end{equation}

Choosing the initial condition to be at the apastron of each orbit, $d$ for the 
outer region and $b$ for the inner region couple with (\ref{ginvk}), we have
\begin{subequations}\label{EF4realboth}
\begin{align}
&\mbox{outer}:\en r(\phi)_{\,\RN{1}}=\dfrac{(a-c)+(c-d)\,\sn^2\,(g^{-1}\,\phi,k)}
                       {(a-c)\,d+a\,(c-d)\,\sn^2\,(g^{-1}\,\phi,k)}\label{EF4realoutap}\,,\\ 
&\mbox{inner}:\en r(\phi)_{\,\RN{2}}=\dfrac{(c-a)+(a-b)\,\sn^2\,(g^{-1}\,\phi,k)}
                       {(c-a)\,b+c\,(a-b)\,\sn^2\,(g^{-1}\,\phi,k)}\label{EF4realinap}
\end{align}
\end{subequations}
We plot all orbits in Cartesian coordinates $(r(\phi)\cos{\phi}\,,\,r(\phi)\sin{\phi})$. 
Orbits in Fig.\hyperref[Fig:3]{3} and Fig.\hyperref[PO4R]{7} are plotted with Eqs.(\ref{EF4realboth}). 
We could use analytical solutions expressed with Weierstrass elliptic functions \cite{grunauRN} but it require
converting the radial polynomial from quartic to cubic. We will just stick to the Jacobi version as 
it allow a more intuitive visual representation of $P(u)$ plots. 

\begin{figure}
\centering
{\includegraphics[width=.31\textwidth]{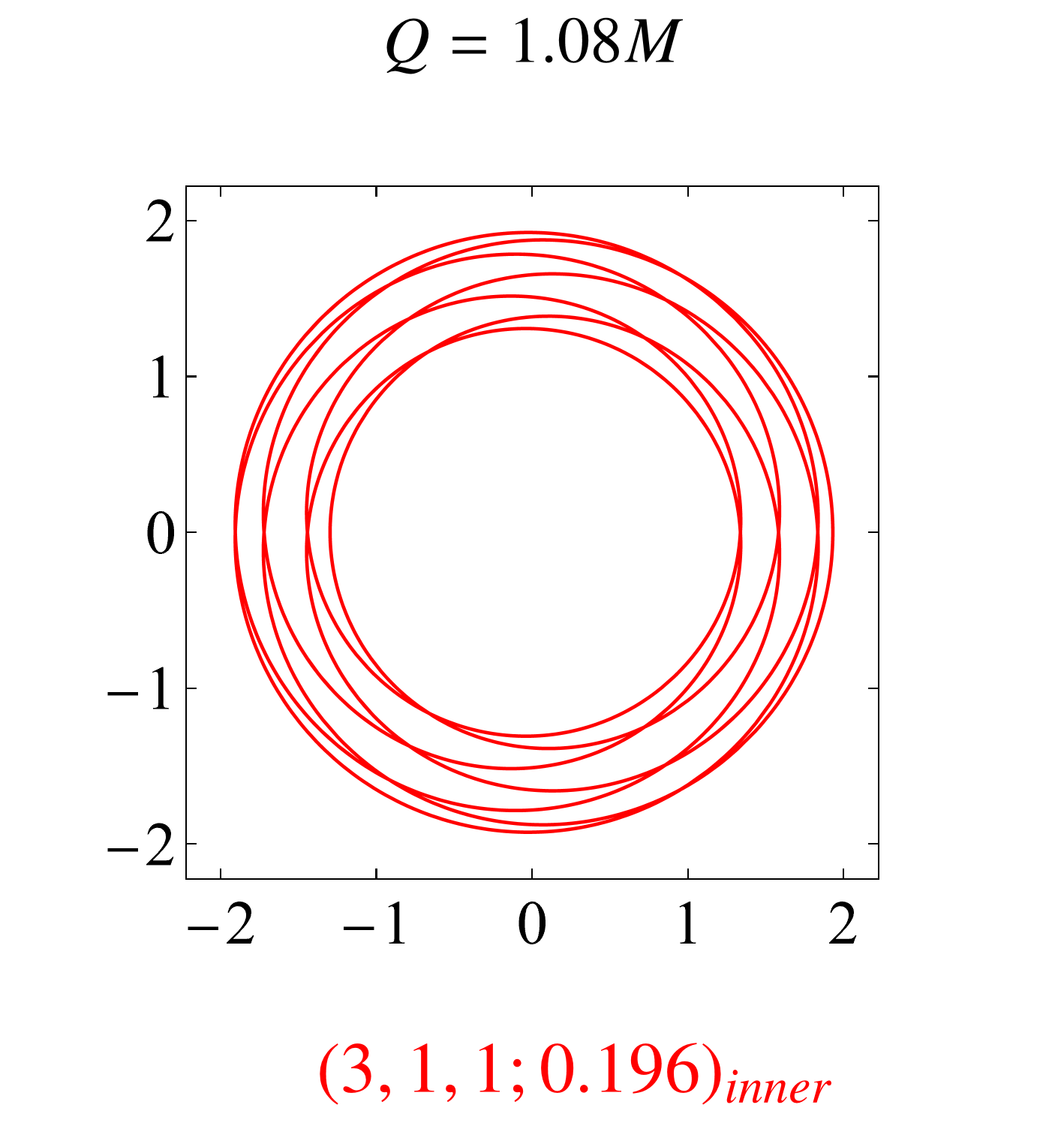}}
\enspace
{\includegraphics[width=.31\textwidth]{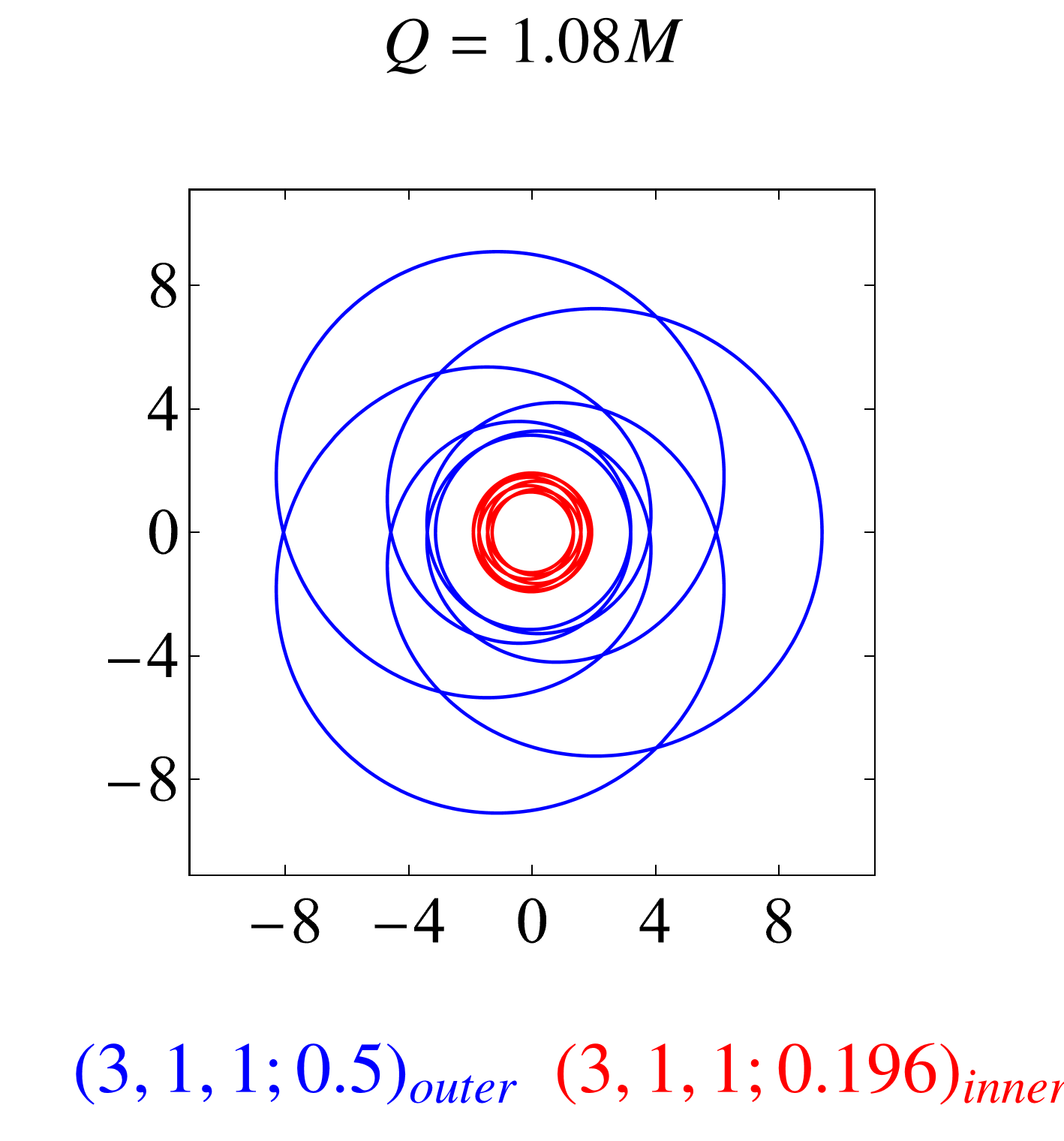}}
\enspace
{\includegraphics[width=.31\textwidth]{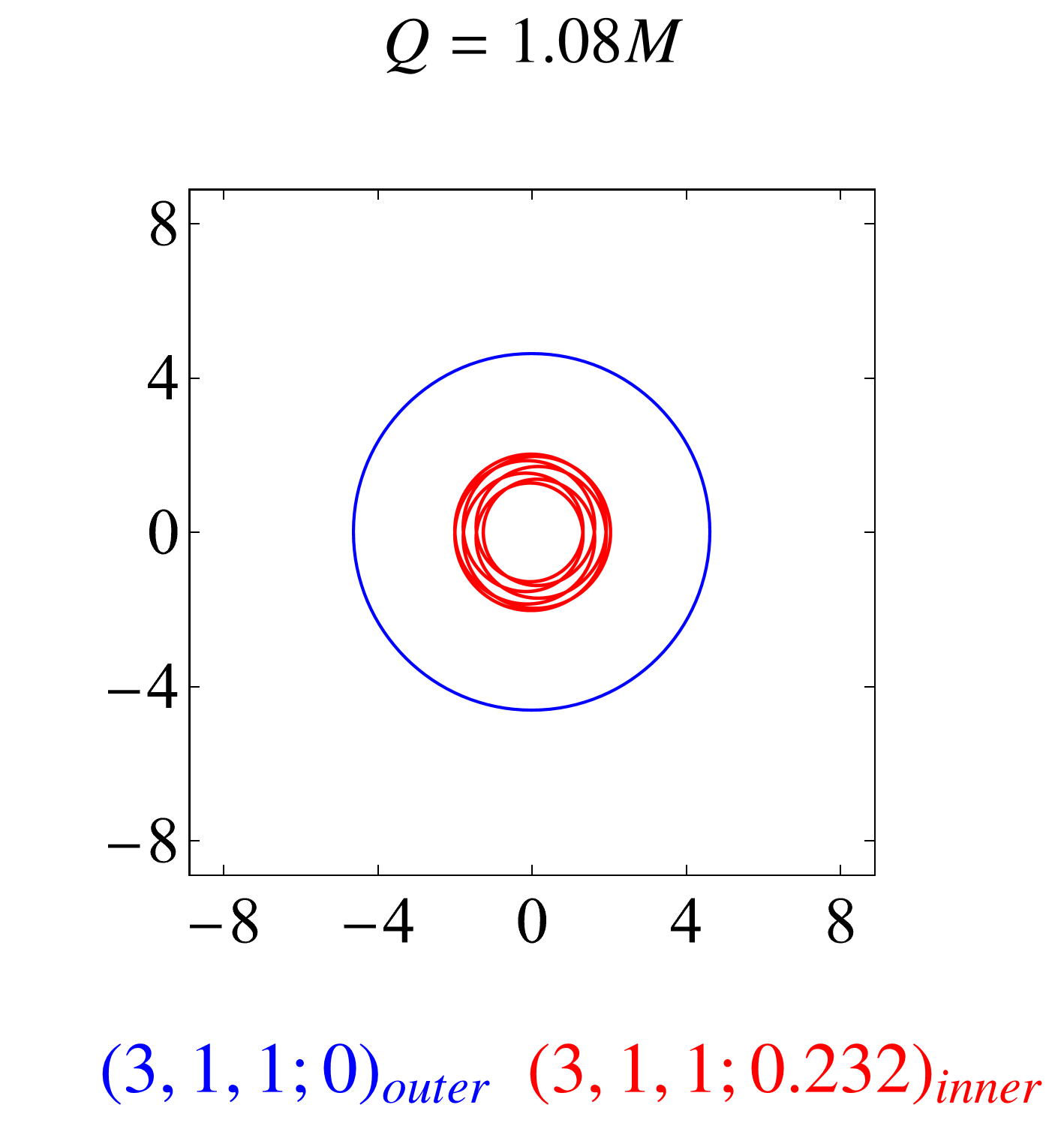}}
\caption{Plots of the associative pair of $Q=1.08M,\,(3,1,1)$ orbits. 
Both orbits in the middle and right figures possess the exact same $L$ and $E$ values.}
\end{figure}\label{PO4R}

\subsection{Two real and two complex roots}\label{sec_3.2}
If we now increase the eccentricity of an inner periodic orbit and extend the $q$-branch below the outer 
stable circular orbits segment, the values of $a$ and $b$ turn complex and are conjugates of each other. The 
shape of the inner $q$-branch starts to curve and reach an energy minima at some moderate eccentricity, 
roughly $0.1\lesssim e\lesssim0.5$, before extending upwards left of the minima and terminate at $E=1$. 
A rule of thumb is that a larger $q$ generates larger $e$ corresponding to the minimum $E$. This means that  
it is now possible for two different eccentricities of a single inner orbit set to possess the same $E$ value
(Fig.\hyperref[Q101branchB]{8}). Like the outer $q$-branches \cite{yklimzcy24}, 
it is not possible for neighbouring inner $q$-branches to cross or intersect one another. 

All other outer and inner periodic orbits that emanate from 
$\rc>r_{int;\,out}$ and $r_*<\rc<r_{int;\,in}$ respectively also display this root configuration. 
The $q$-branch shape for each respective regions is the same as described previously except for 
inner $q$-branches that emanate from points closer to $r_*$. These particular branches resemble half of a parabola with 
$e=0$ being the minimum point and increasing $e$ always extend the branches upwards. Hence, 
$E$ values increase monotonically with $e$ but $L$ values not necessarily so as the branch may start swerving 
towards the right, usually around the half way mark, $e\gtrsim0.5$ (Fig.\hyperref[Case2qbranch]{15}). Most of Case 3 
$q$-branches take this shape but could only emanate from the left side 
\footnote{Emanation points are smallest solutions of Eq.(\ref{rcq}).} of the circular orbits curve 
(Fig.\hyperref[Case3qbranch]{16}). The energetically unbound region to the left and right of the top part 
of both stable $\rc$ segments also contain two complex roots (Fig.\hyperref[Q101branchB]{8}). 
This can be surveyed by calculating the root values (\ref{rootpaireqs}) from any arbitary points from those regions. 

\begin{figure}[t!]
    \centering
    {\includegraphics[width=.52\textwidth]{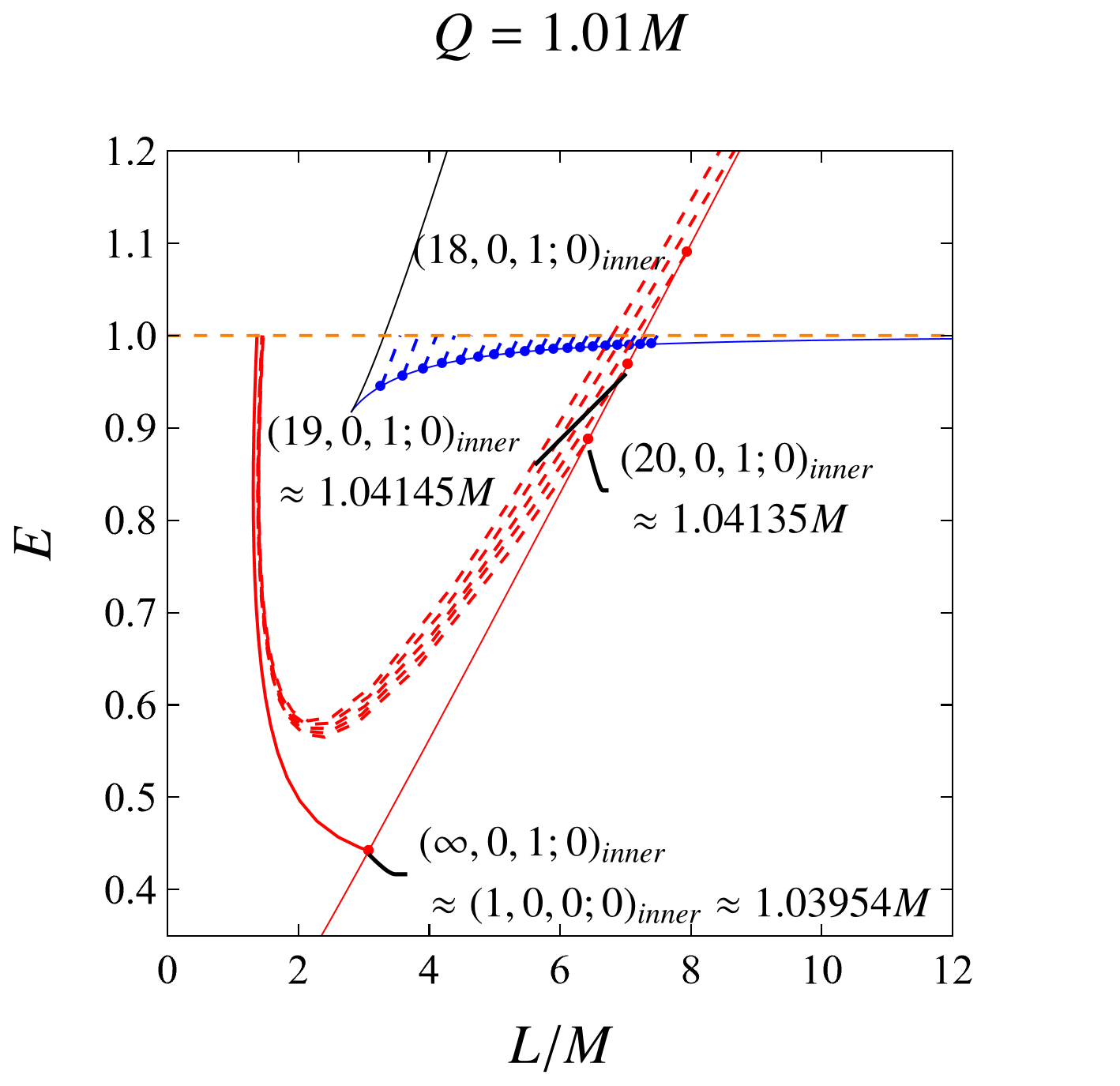}}
    \enspace
    {\includegraphics[width=.45\textwidth]{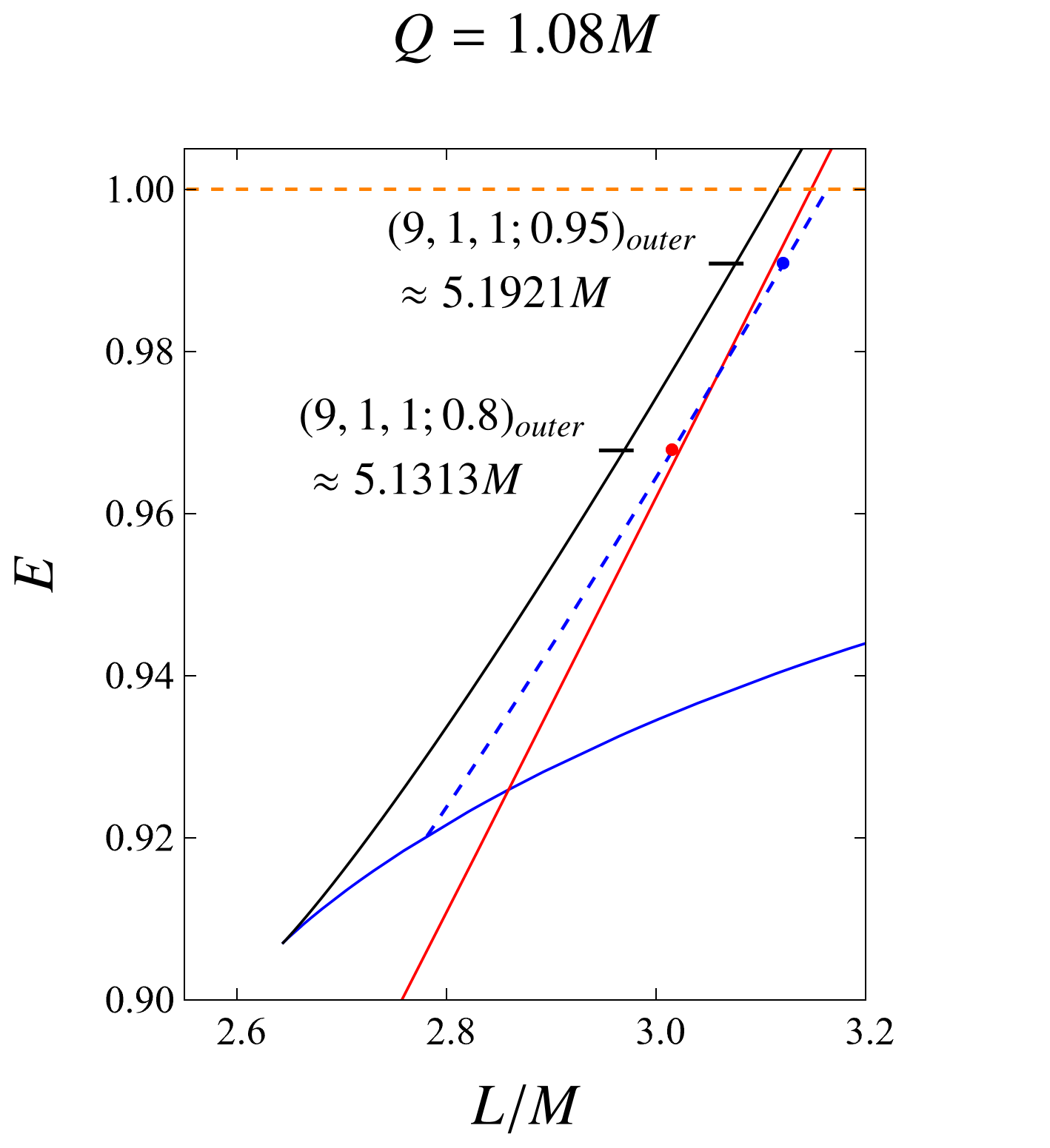}}
    \caption{Left figure: The limit branch for the inner Keplerian orbit  
    is the solid red half parabola shape curve. Right figure: 
    This outer $q$-branch resides in domains with four real roots at $0<e\leq0.864$ 
    and two complex roots at $0.864<e<1$.}
\end{figure}\label{Q101branchB}

Eqs.(\ref{EF4realboth}) could not be used when complex value roots are present. 
Instead, we need the form given in 250.05, pg 97 of \cite{brydEI},
\begin{equation}\label{EF4complxgen}
    r(\phi)_{\,\RN{3}}=\dfrac{\alpha_1+\alpha_2\,\cn\,(g^{-1}\,\phi,k)}
                            {\alpha_3+\alpha_4\,\cn\,(g^{-1}\,\phi,k)}\,,
\end{equation}
where $\cn$ is the Jacobi elliptic cosine function and the functions $\alpha_1,\alpha_2,\alpha_3,\alpha_4$ 
are all dependant on the roots values but $\alpha$-values must be real. Eq.(\ref{EF4complxgen}) 
work exactly the same way for both regions if we define the initial condition to start at the apastron like 
before. In this case, we shall relabel the roots $c,d$ to $u_p,u_a$ and $a,b$ to $z,\bar{z}$.
The rest of the parameters for this configuration are given by 259.00, pg. 133 of \cite{brydEI}:
\begin{align*}
    a^2_1&=-\dfrac{(z-\bar{z})^2}{4}, & b_1&=\dfrac{z+\bar{z}}{2},\\
    A^2&=(u_p-b_1)^2+a^2_1, & B^2&=(u_a-b_1)^2+a^2_1,\\
    g&=\dfrac{1}{Q\sqrt{AB}}, & k^2&=\dfrac{(u_p-u_a)^2-(A-B)^2}{4AB}, 
\end{align*}\\
Or we could express them, including $\alpha$, explicitly involving all four roots $u_p,u_a,z,\bar{z}$, 
\begin{align}\label{cmplxplot}
\alpha_{1,2}&=A\pm B=\sqrt{(u_p-z)(u_p-\bar{z})}\pm\sqrt{(u_a-z)(u_a-\bar{z})}\,,\\ 
\alpha_{3,4}&=Au_a\pm u_pB=u_a\sqrt{(u_p-z)(u_p-\bar{z})}\pm u_p\sqrt{(u_a-z)(u_a-\bar{z})}\,,\\ 
g^{-1}&= Q\sqrt{AB}=Q\sqrt{\sqrt{(u_p-z)(u_p-\bar{z})}\sqrt{(u_a-z)(u_a-\bar{z})}}\,,\\
   k^2&=\dfrac{1}{4}\brac{2+\dfrac{u_p(z+\bar{z}-2u_a)+u_a(z-\bar{z})-2z\bar{z}}
            {\sqrt{(u_p-z)(u_p-\bar{z})}\sqrt{(u_a-z)(u_a-\bar{z})}}}
\end{align}
Plots of periodic orbits with Eqs.(\hyperref[cmplxplot]{33-36}) are shown in Fig.\hyperref[complexdomPO]{9}.
Obtaining a sensible real $\lr$ via (\ref{LRnumericalrel}) is done by first considering the $e=0$ points
along the allowed emanation range $\rc>r_{int;\,out}$ or $r_*<\rc<r_{int;\,in}$. For inner periodic orbits,
the lower limit of $\lr$ would be the emanation point of the inner $(\infty,0,1)$ limit branch.
Then, seek the closest monotonically increasing $\lr$ values for $e>0$ as before.

\begin{figure}[h]
\centering
\begin{minipage}{\linewidth}
\centering
\subfloat{\includegraphics[scale=.62]{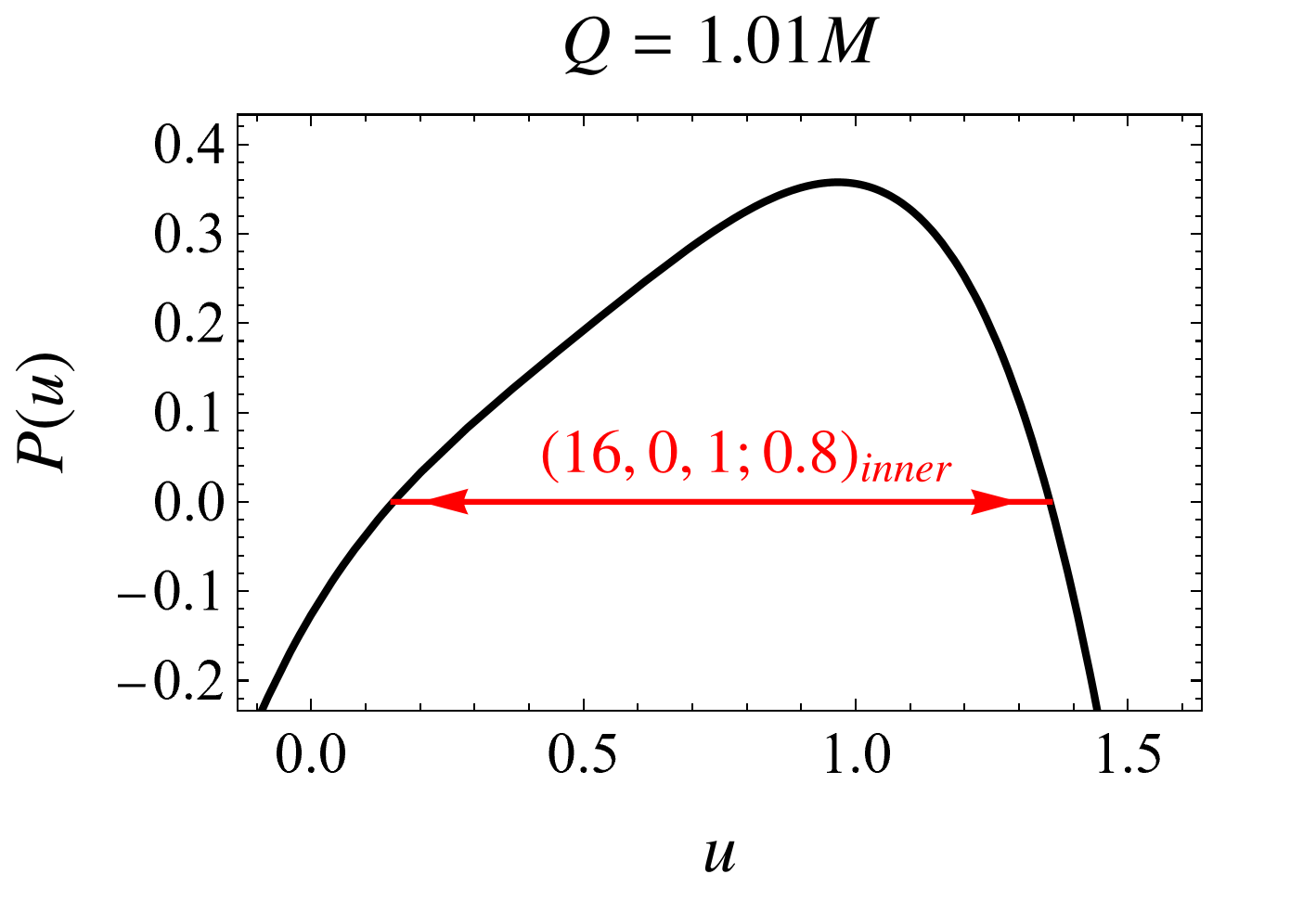}}
\enspace
\subfloat{\includegraphics[scale=.64]{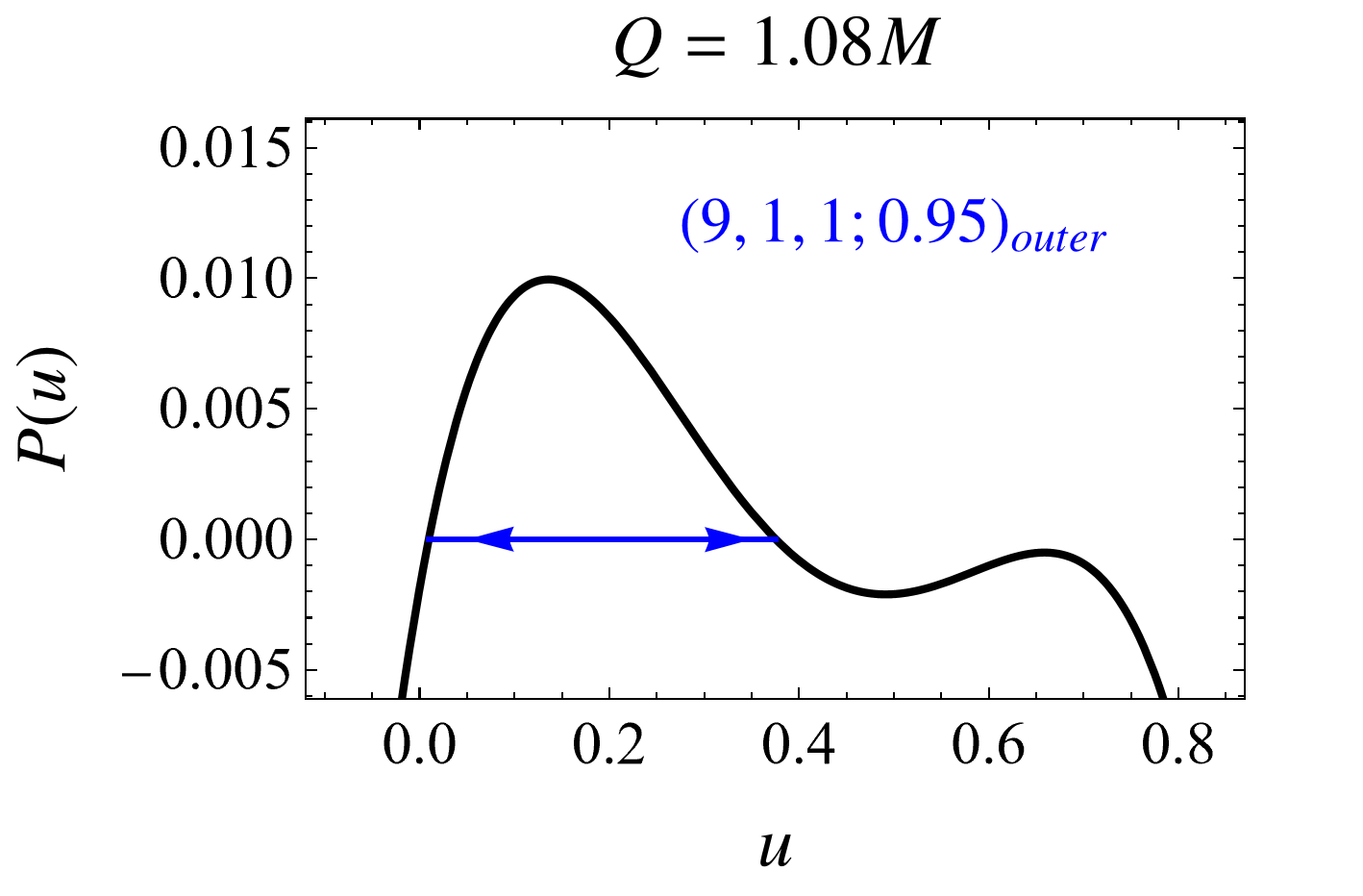}}
\end{minipage}\par\smallskip
\begin{minipage}{\linewidth}
\centering
\subfloat{\includegraphics[scale=.4]{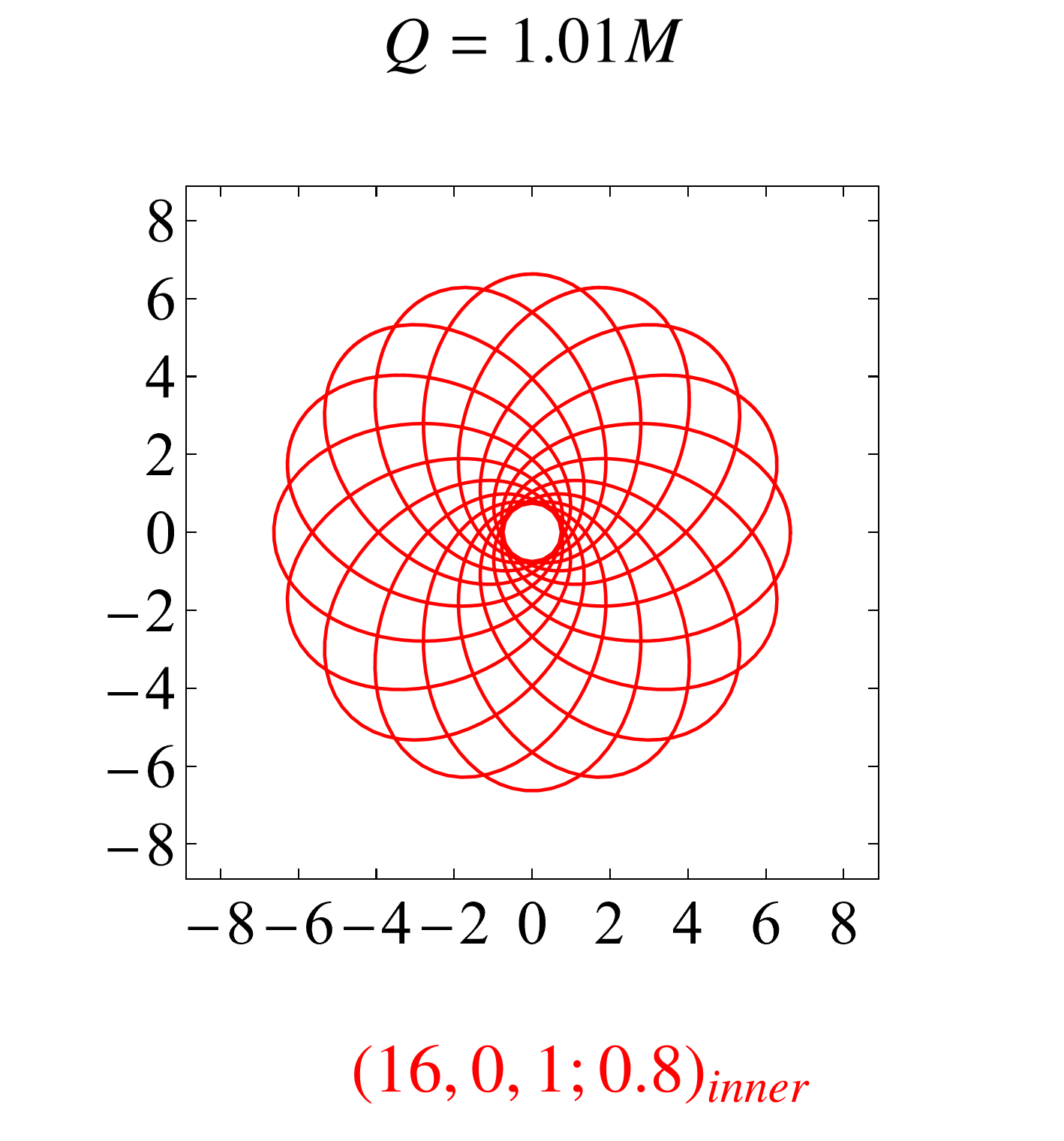}}
\enspace
\subfloat{\includegraphics[scale=.41]{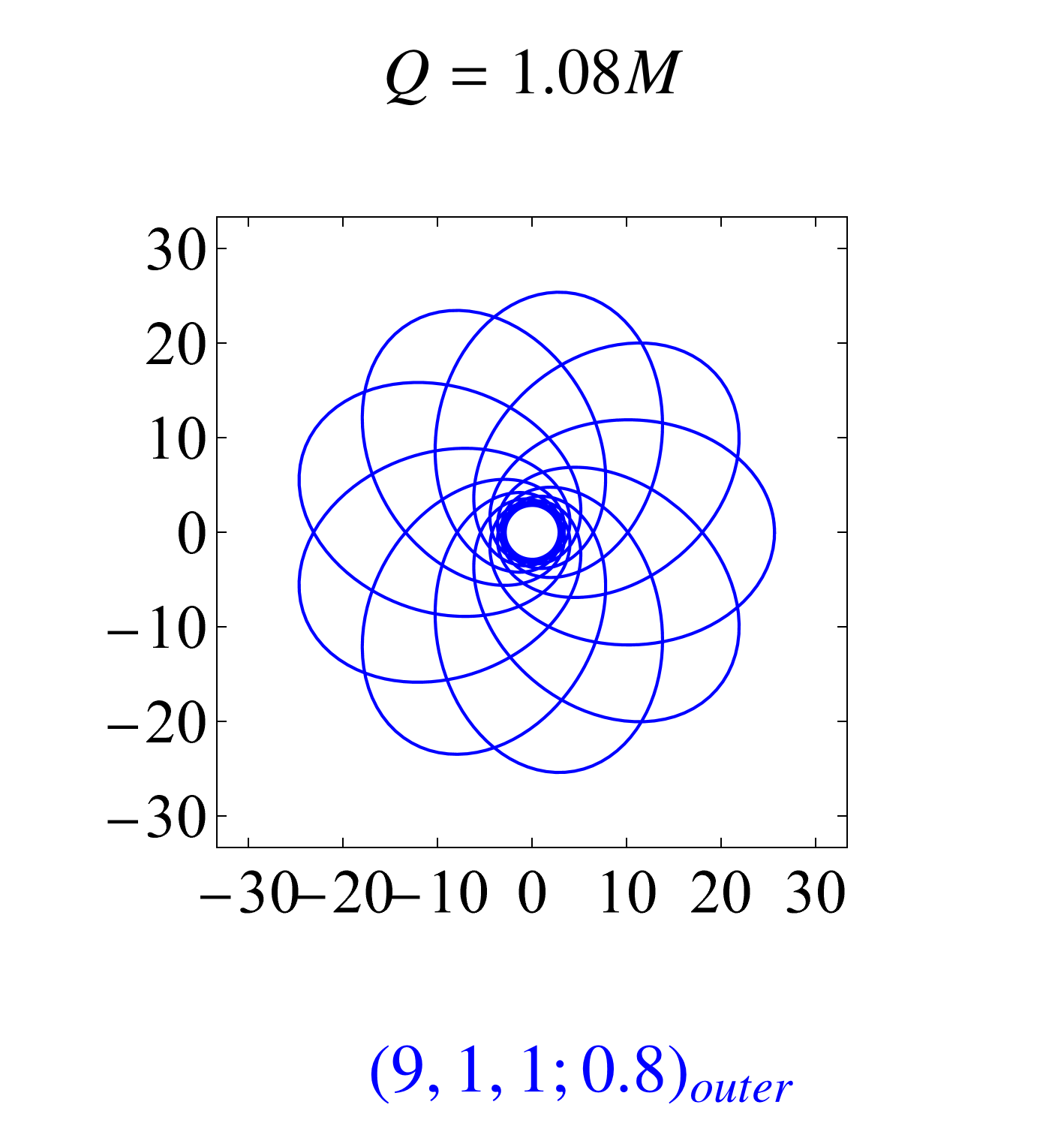}}
\enspace
\subfloat{\includegraphics[scale=.42]{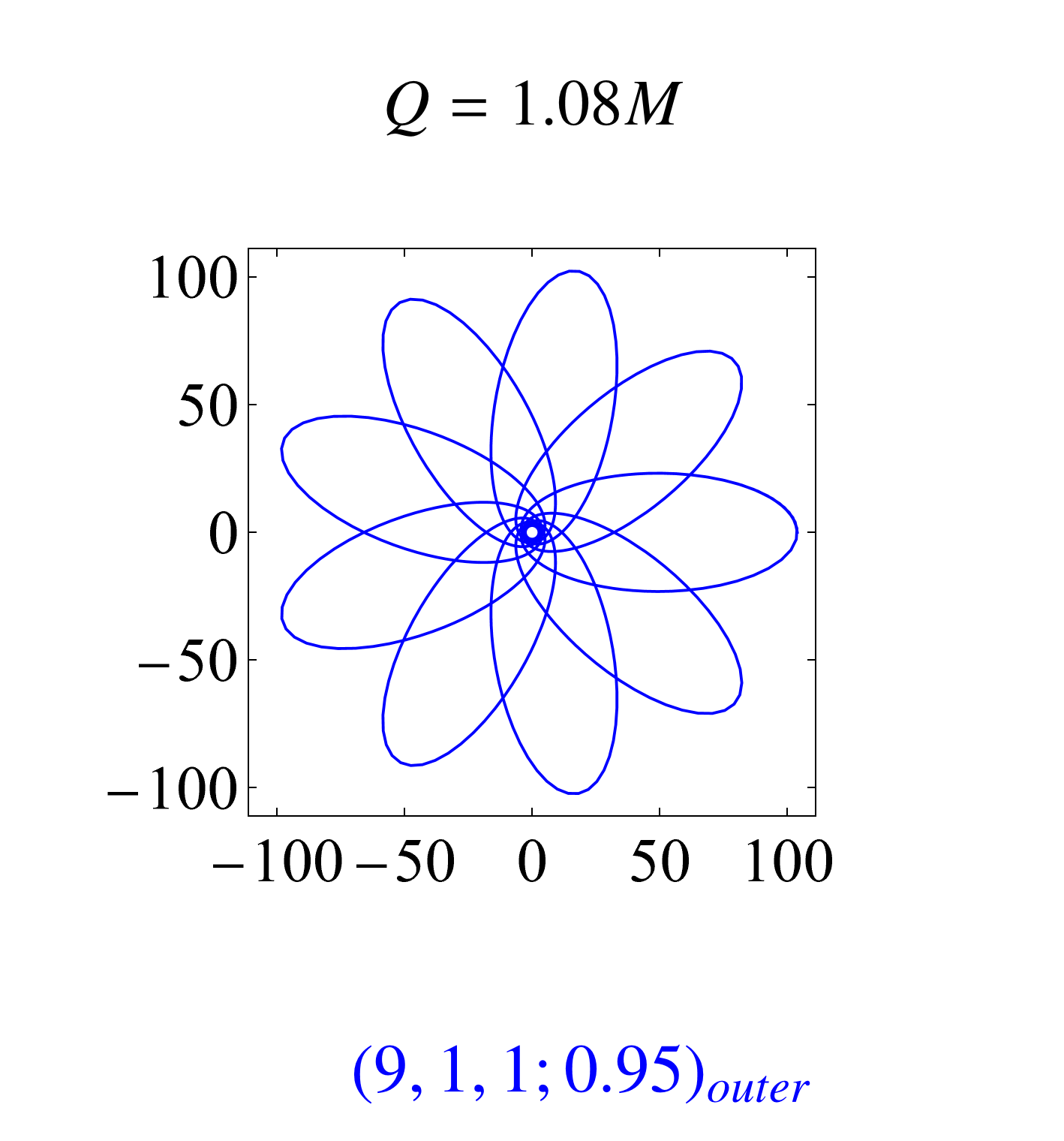}}
\end{minipage}%
\caption{Top row: $P(u)$ plots for $q$-branches that typically emanate from 
$\rc>r_{int;\,out}$ or $r_*<\rc<r_{int;\,in}$. Bottom row: The middle and right orbits 
are based of the two points of the $q$-branch in Fig.\hyperref[Q101branchB]{8}. They should be plotted 
with Eq.(\ref{EF4realoutap}) and Eq.(\ref{cmplxplot}) accordingly.}
\end{figure}\label{complexdomPO}

\subsection{Special cases of degenerate roots}\label{sec_3.3}
Here, we will mention some intriguing parts on the circular orbits curve with degenerate real roots and 
reveal some striking properties. On any point on the unstable circular orbits segment, we encounter another 
case of non-oscillatory motion within domains of $P(u)$. This happen for the infinite-whirl limit homoclinic orbit 
\cite{levinpg08,levinhomo1,yklimzcy24}. In physical space, this orbit occur when the trajectory start from some 
arbitary location before asymptotically approach the radius of an unstable circular orbit. It then remained as a circular 
orbit for an infinite time unless it is perturbed. In the naked singularity background and with $P(u)$ being quartic,
there exist a pair of homoclinic orbits for a single $\rc$ (Fig.\hyperref[p(u)108homo]{10}). 
In $P(u)$ plots, we view it as the $u$-value making a `one-way' trip from either the periastron or 
apastron root to the degenerate root $\uc$, which is now the middle value in the 
order: $a>\,\uc=b=c\,>d$ (Fig.\hyperref[p(u)108homo]{10}).

Studies on analytical solutions for the homoclinic orbits in axially-symmetric spacetimes 
can be found in \cite{levinhomo1,liKNhomo}. For spherically-symmetric case,
we can easily plot the outer homoclinic orbits via Eq.(\ref{EF4realoutap}) by inserting 
the roots $a,\uc,d$. However, we need a different initial condition to plot the homoclinic orbits 
arising from the inner region. This will be 257.00, pg.124 \cite{brydEI}. It describes trajectory that 
starts from the periastron, giving $\psi=\arcsin{\sqrt{\dfrac{(b-d)(a-u)}{(a-b)(u-d)}}}$. 

Together with $g^-1$ and $k$ (\ref{ginvk}), we have
\begin{equation}\label{EFinnerhomo}
r(\phi)_{\,\RN{4}}=\dfrac{(b-d)+(a-b)\,\sn^2\,(g^{-1}\,\phi,k)}
                             {(b-d)\,a+d\,(a-b)\,\sn^2\,(g^{-1}\,\phi,k)}
\end{equation}

\begin{figure}[t!]
    \centering
\begin{minipage}{\linewidth}
\centering
\subfloat{\includegraphics[scale=.64]{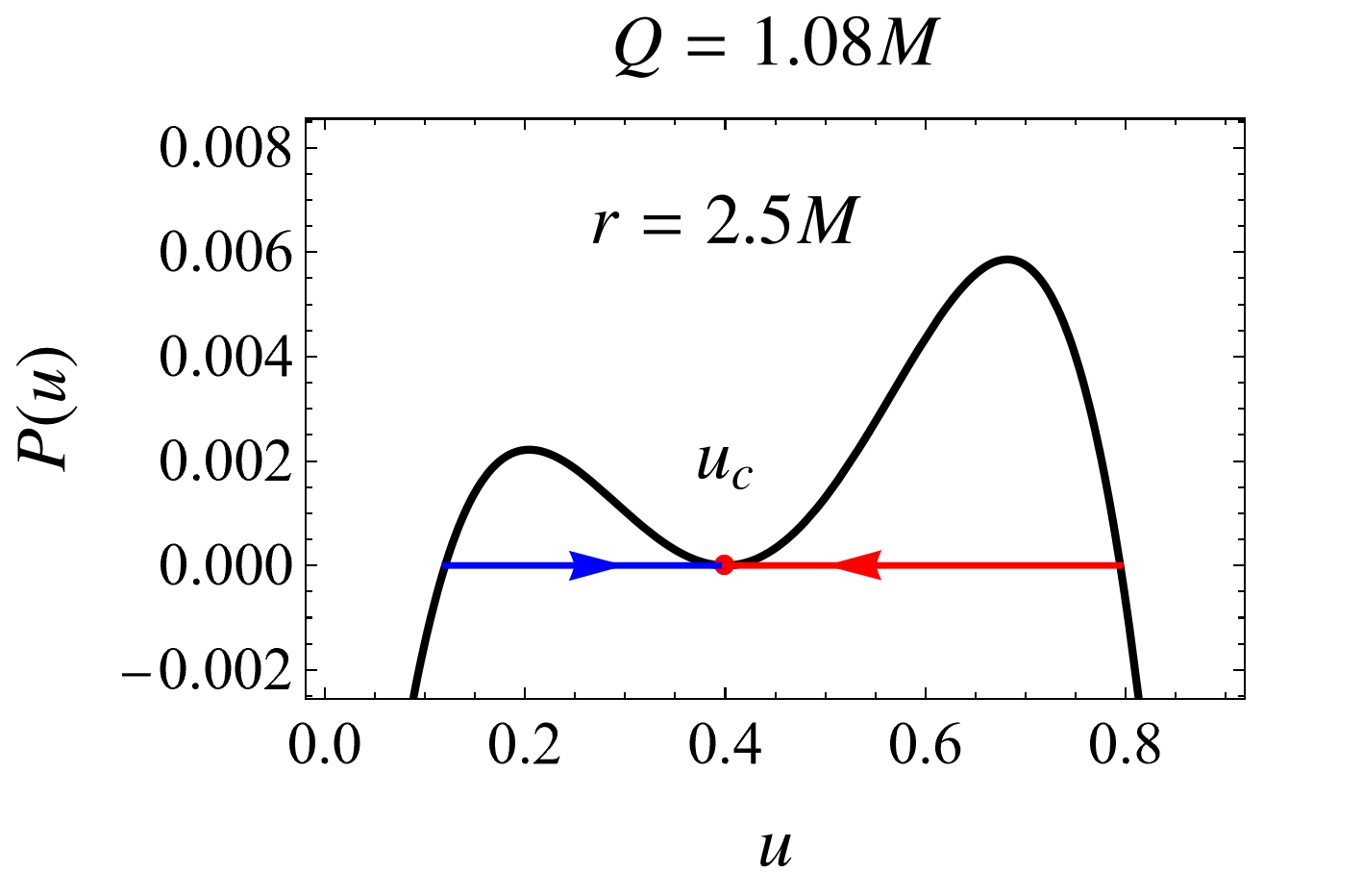}}
\enspace
\subfloat{\includegraphics[scale=.64]{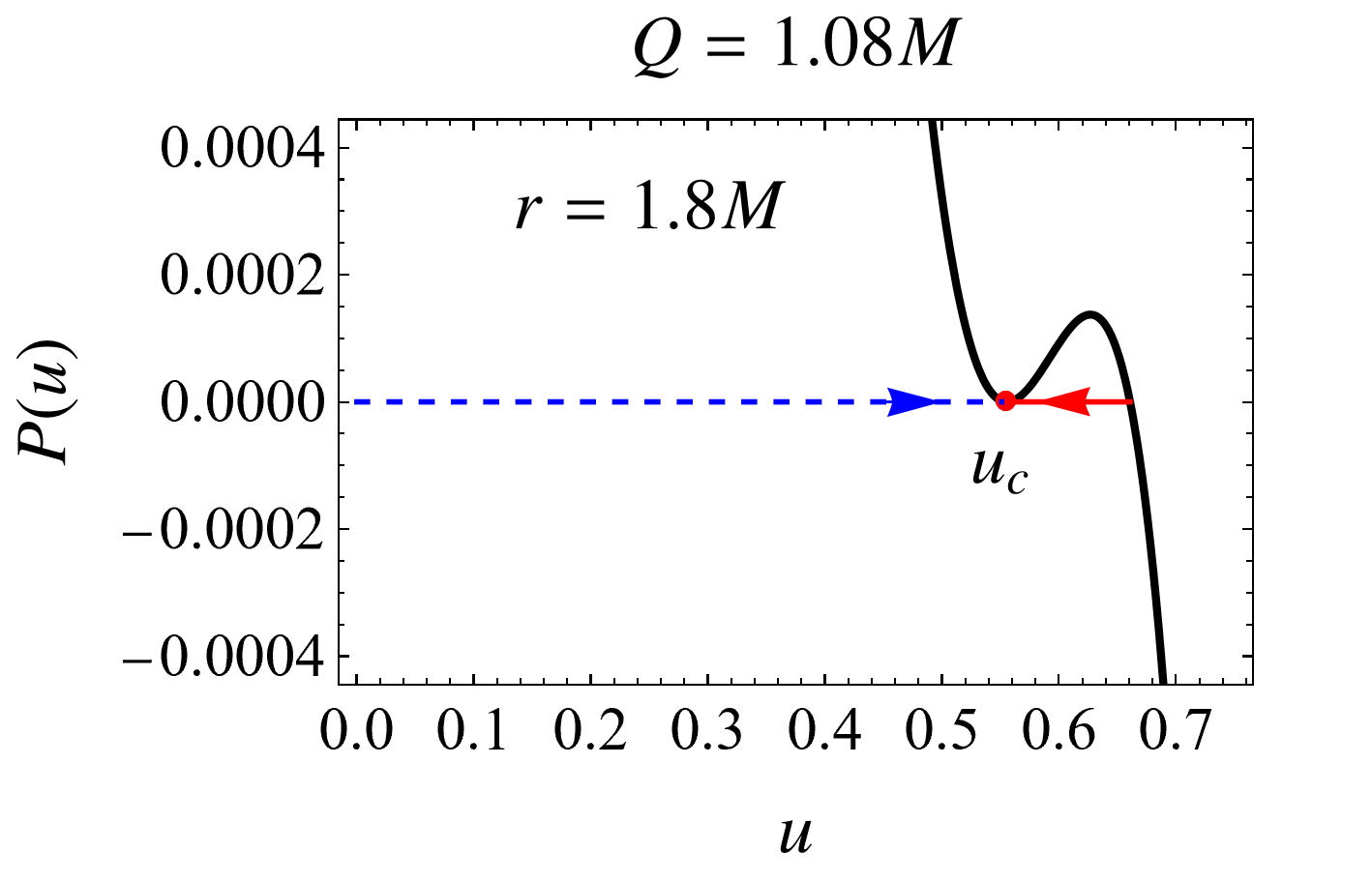}}
\end{minipage}\par\smallskip
\begin{minipage}{\linewidth}
\centering
\subfloat{\includegraphics[scale=.7]{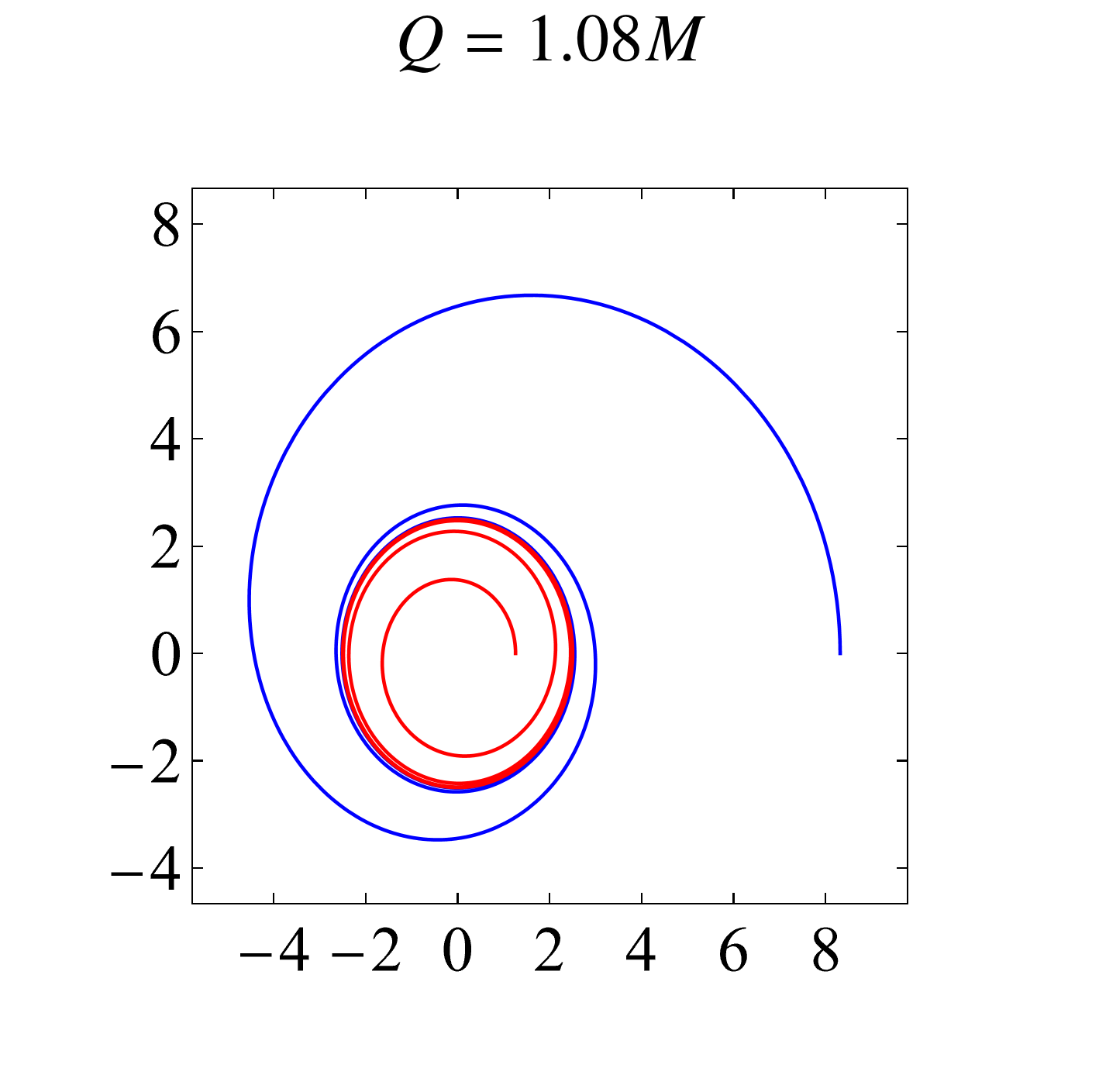}}
\end{minipage}%
\caption{$P(u)$ plots at unstable $\rc$ values with $E<1$ in the top left figure and $E\geq1$ in 
the bottom figure. Top right: The pair of homoclinic orbits at $\rc=2.5M$ of $Q=1.08M$ plotted with 
Eqs.(\ref{EF4realoutap}) and (\ref{EFinnerhomo}) accordingly.}
\end{figure}\label{p(u)108homo}

For Schwarzschild black holes \cite{yklimzcy24,ChandrashekarBook}, perturbing an unstable 
cirular orbit causes the resulting trajectory to become either an escaping orbit or a plunging orbit based on the 
direction of perturbation. This is since the closed domain in $P(u)$ describing bound orbits 
cease to exist whenever two roots coalesce. However, for RN spacetime, we still observe closed domains in $P(u)$ plots
for the entire unstable circular orbits segment (Fig.\hyperref[p(u)108homo]{10}).
This suggests that the oscillatory motion could reoccur if the middle $\uc$ root is perturbed towards the 
direction of another non-negative real root, in other words, the emerging trajectory of particles will 
remained bound geometrically. This is indeed the case and in fact, it exhibit 
chaotic motion \cite{bombelliChaos,levinhomo2}. Trajectories that arised from perturbations 
start in a similiar fashion at small $\Delta\phi$ ($\approx15\pi$ in Fig.\hyperref[Q108homopert]{11}) before evolving in 
radically different manner as $\Delta\phi$ increases.

Eq.(\ref{EF4realinap}) can be applied to motions caused by inward perturbation but for outward 
perturbation, we require a different expression and not Eq.(\ref{EF4realoutap}). This is the 
condition in 253.00, pg.107 \cite{brydEI} that describes trajectories starting from the periastron. Then, 
we have $\psi=\arcsin{\sqrt{\dfrac{(b-d)(c-u)}{(c-d)(b-u)}}}$ and with (\ref{ginvk}),
\begin{equation}\label{EFpertout}
r(\phi)_{\,\RN{5}}=\dfrac{(d-b)+(c-d)\,\sn^2(g^{-1}\,\phi,k)}
                         {(d-b)\,c+b\,(c-d)\,\sn^2(g^{-1}\,\phi,k)}
\end{equation}

\begin{figure}[h!]
    \centering
        \begin{minipage}{\linewidth}
        \centering
        \subfloat{\includegraphics[scale=.35]{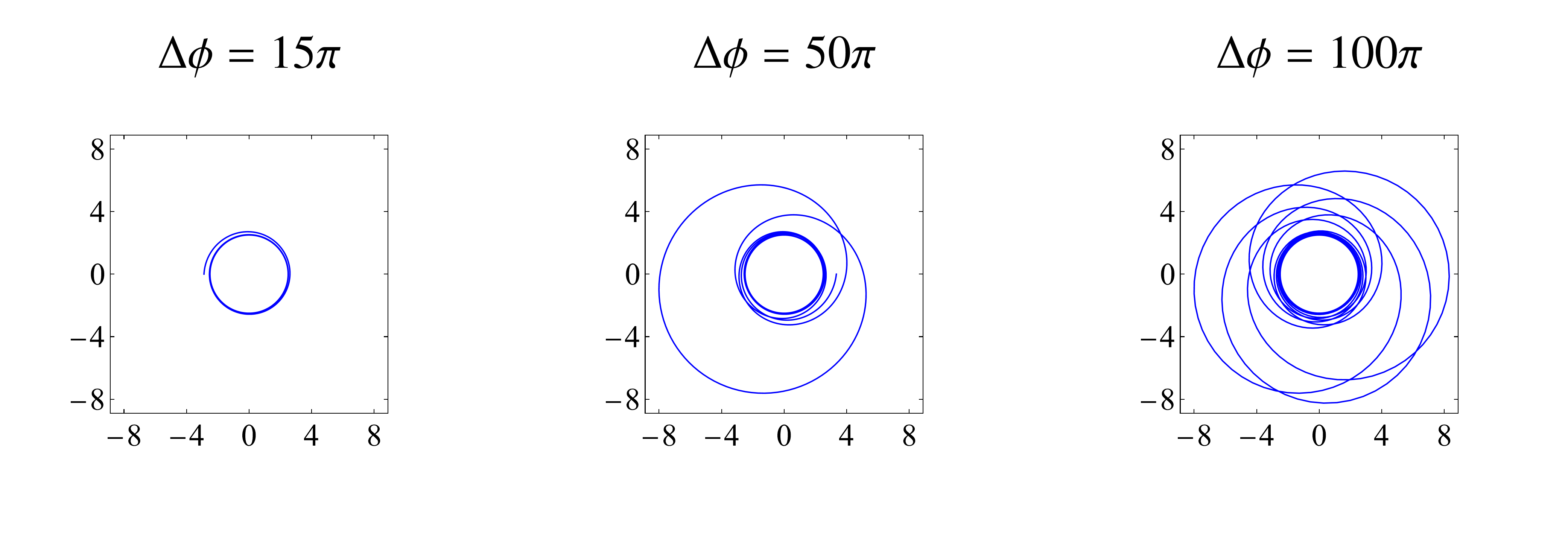}}
        \end{minipage}\par\smallskip
        \begin{minipage}{\linewidth}
        \centering
        \subfloat{\includegraphics[scale=.35]{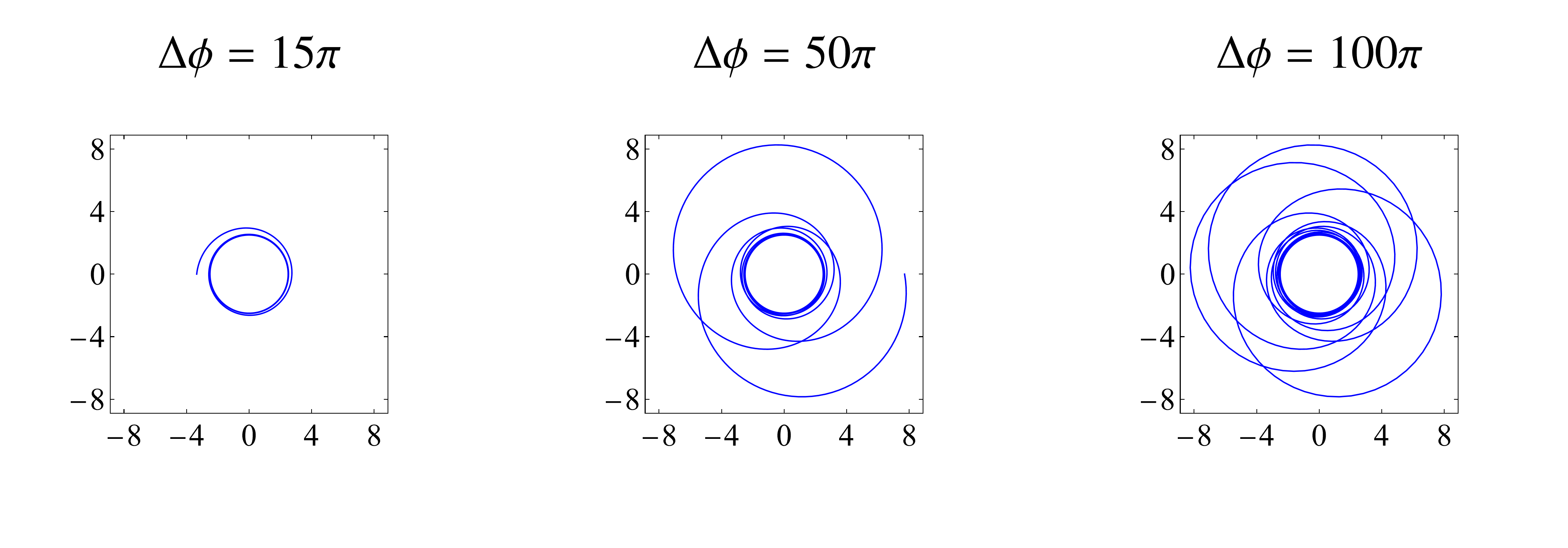}}
        \end{minipage}\par\smallskip
        \begin{minipage}{\linewidth}
        \centering
        \subfloat{\includegraphics[scale=.35]{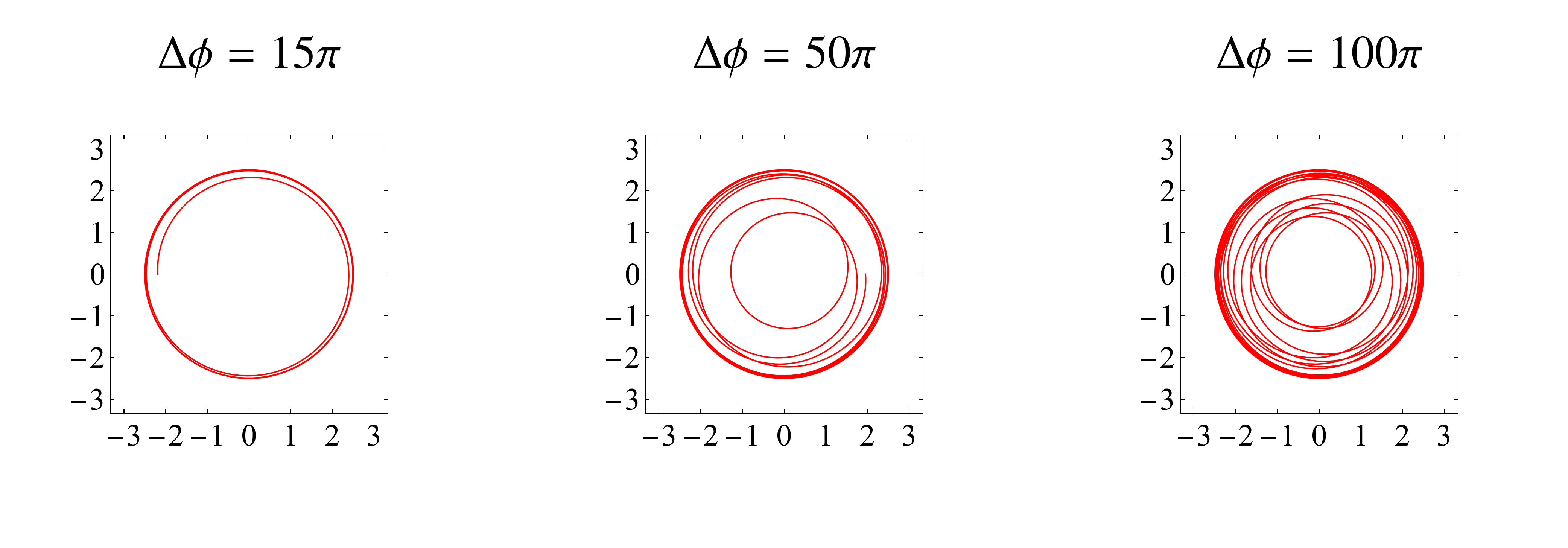}}
        \end{minipage}\par\smallskip
        \begin{minipage}{\linewidth}
        \centering
        \subfloat{\includegraphics[scale=.35]{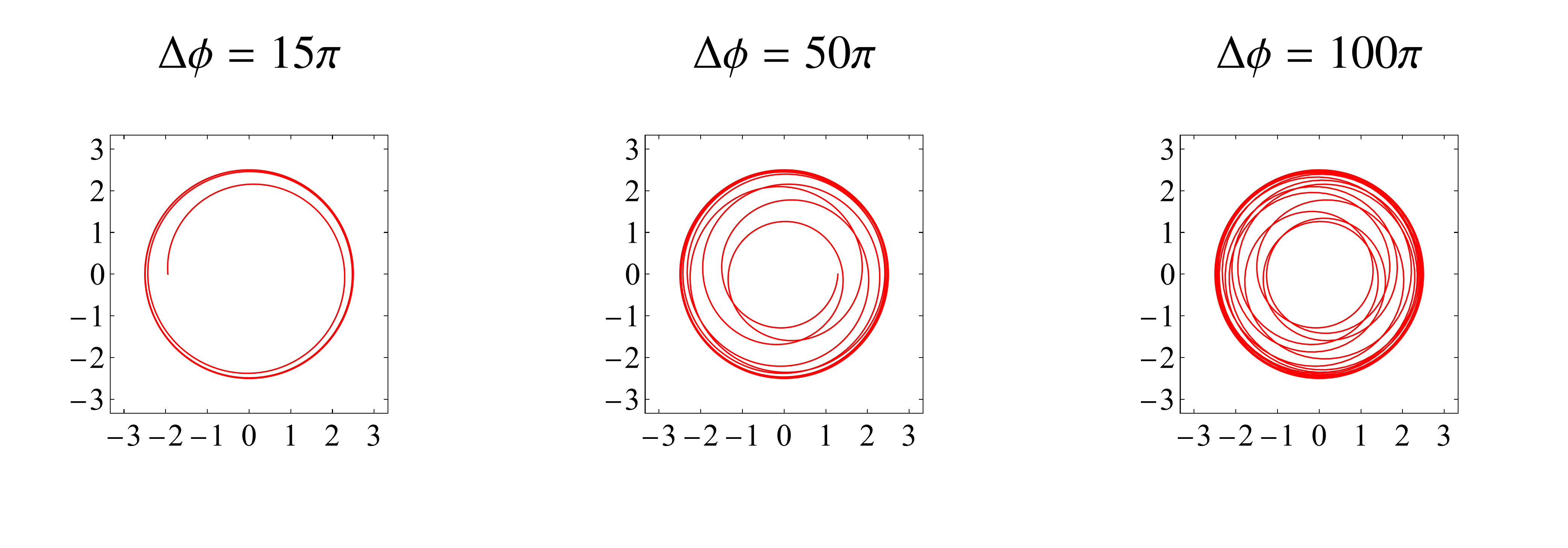}}
        \end{minipage}
        \caption{The chaotic nature stems from the sensitivity to 
        small perturbations at $\rc=2.5M$ of $Q=1.08M$.  
        The blue and red orbits are arise from outward and inward perturbations 
        respectively.}
\end{figure}\label{Q108homopert}

At the critical points in all of Case 1 and 2 naked singularities and also black holes, another root, 
$b$ coalesece with $\uc$ at the ISCO. Moreover, root $a$ coalesces with $\uc$ at the OSCO of Case 2. Hence, 
the critical points contain a triply degenerate root, leaving only two distinct root values overall and a 
single closed domain in $P(u)$. The solution for homoclinic orbits at these critical points cannot take the general form 
(\ref{EF4realgen}) introduced previously since the threefold repeated root made all $A_1,A_2,A_3,A_4$
terms vanish. Hence, a $P(u)$ expression with the degenerate root $u_c$ factored out is required, as in 
pg. 95 of \cite{szelakova}. A simple rearrangement gives
\begin{subequations}\label{tripdeghomo}
\begin{align}
P(u)&=(u-u_c)^3(2M-3Q^2u_c-Q^2u_c)\nonumber\\
\Rightarrow r(\phi)_{\,\RN{6}}&=\dfrac{\phi^2\,m(Q,\uc)^2+Q^2}
                               {2\,m(Q,\uc)+\uc(\phi^2\,m(Q,\uc)^2+Q^2)}\,,\\
m(Q,\uc)&=(M-2Q^2\uc)
\end{align}    
\end{subequations}
Fig.\hyperref[3deghomoorb]{12} homoclinic orbits are plotted with Eq.(\ref{tripdeghomo}).
By observing the location of the closed domain of the ISCOs, we deduce that an inner time-like 
homoclinic orbit asymptotes to $r_{ISCO}$ and only chaotic motion from inward perturbation is possible. 
The homoclinic orbit at Case 2A OSCOs is formed by trajectories starting from inifinity and asymptote to 
the null circular orbit radius. Although the sole null circular orbit at $r=1.5M$ for $Q=\sqrt{9/8}M$ is not 
strictly a solution to (\ref{cuspcond}), it exhibits a triply degenerate root like every other OSCO of Case 2 
and hence can be treated as one. This explains why we consider $Q=\sqrt{\frac{9}{8}}\,M$ as part of Case 2. 

Finally, we shall briefly mention some trivial $\rc$ values. At the intersection of the two 
stable circular orbits segments in $(L,E)$-space of Case 1 and 2 (Fig.\hyperref[Fig:2]{2}), 
there is a pair of degenerate roots, $c=d$ and $a=b$. So, there could only be a pair of circular orbits, 
each with radius corresponding to the degenerate roots 
(Fig.\hyperref[3deghomoorb]{12}). At $r=2.5 M$ of $Q=\frac{\sqrt{5}}{2}\,M$ we find the only possible case 
where all four roots degenerate to the same value, $a=b=c=d$, i.e. the point where the 
two critical points from Case 2 coalesce upon transitioning to Case 3. Both of these 
types of circular orbits do not change shape when perturbed,
thus we can name them as \textit{hyperstable circular orbits}.

\begin{figure}[t!]
    \centering
    \begin{minipage}{\linewidth}
    \centering
    \subfloat{\includegraphics[scale=.62]{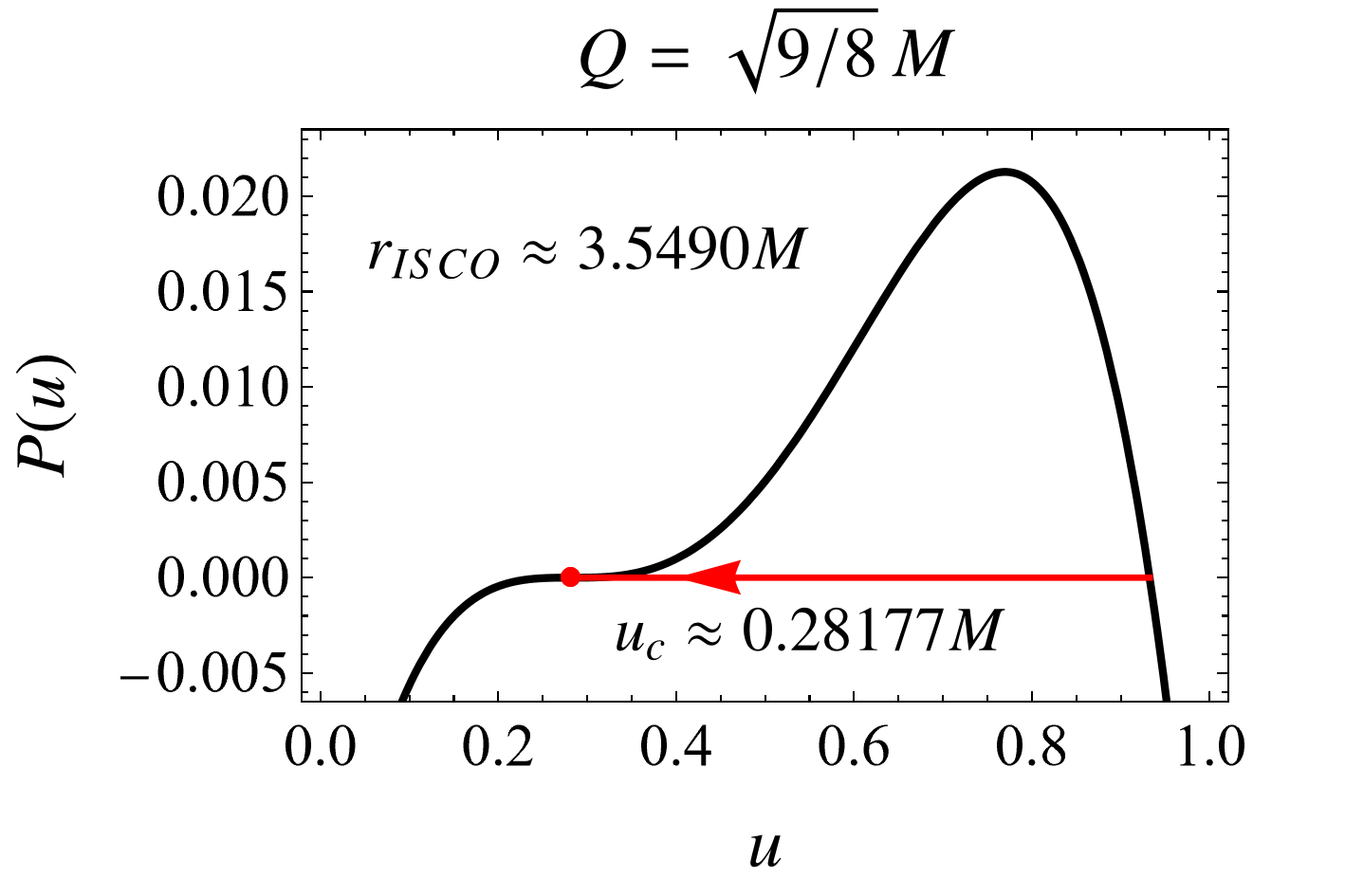}}
    \enspace
    \subfloat{\includegraphics[scale=.62]{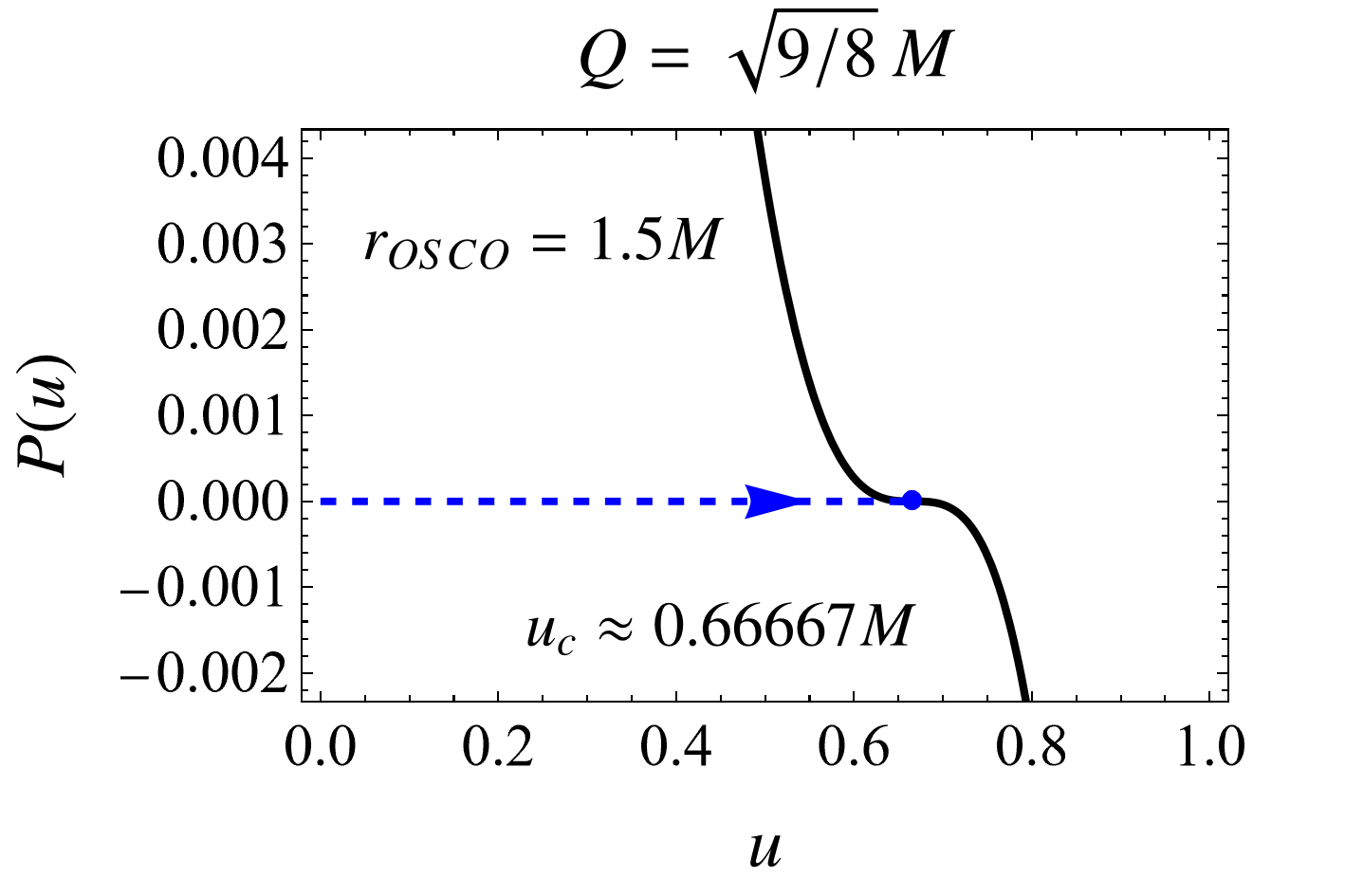}}
    \end{minipage}\par\smallskip
    \begin{minipage}{\linewidth}
    \centering
    \subfloat{\includegraphics[scale=.42]{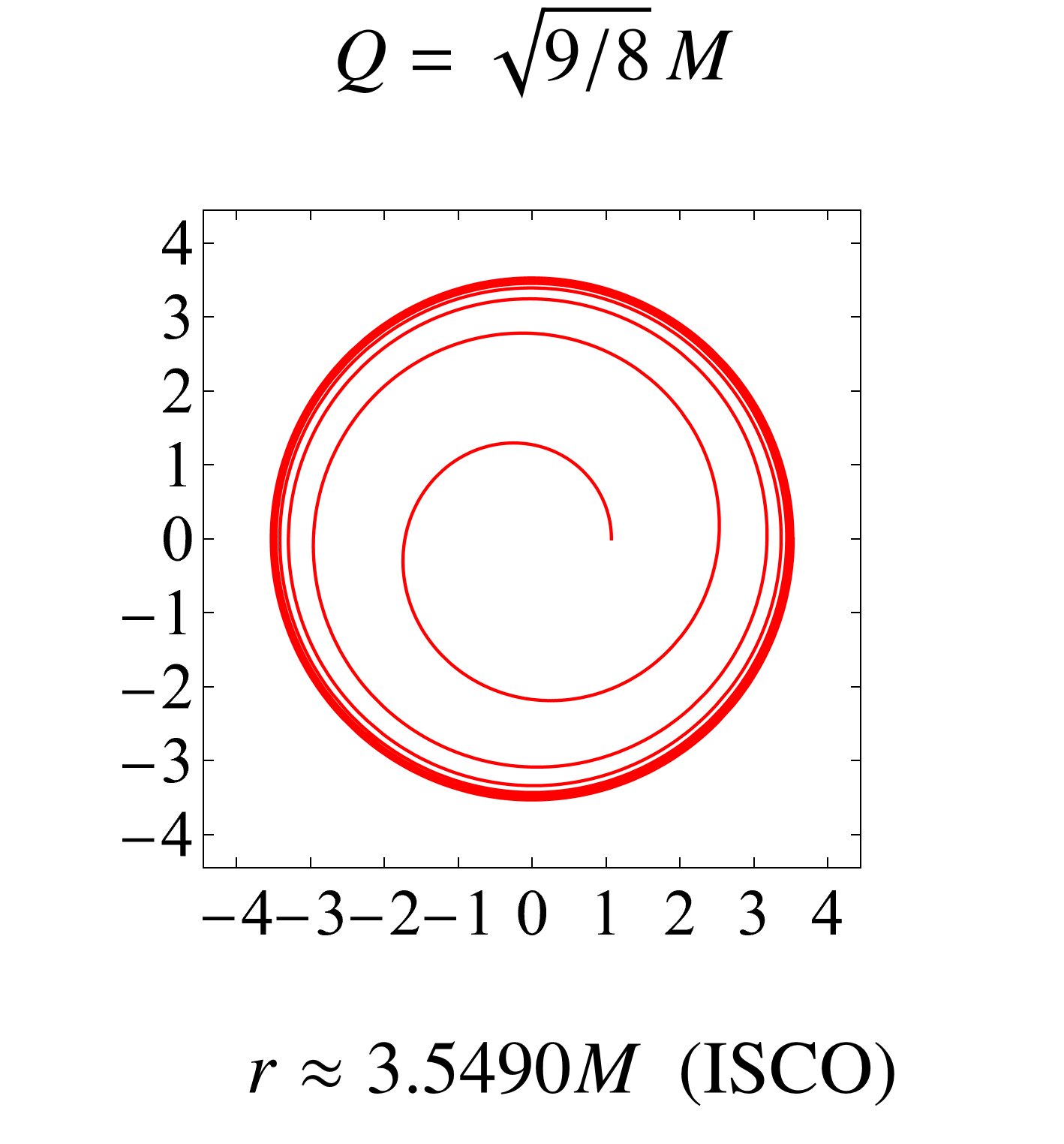}}
    \enspace
    \subfloat{\includegraphics[scale=.64]{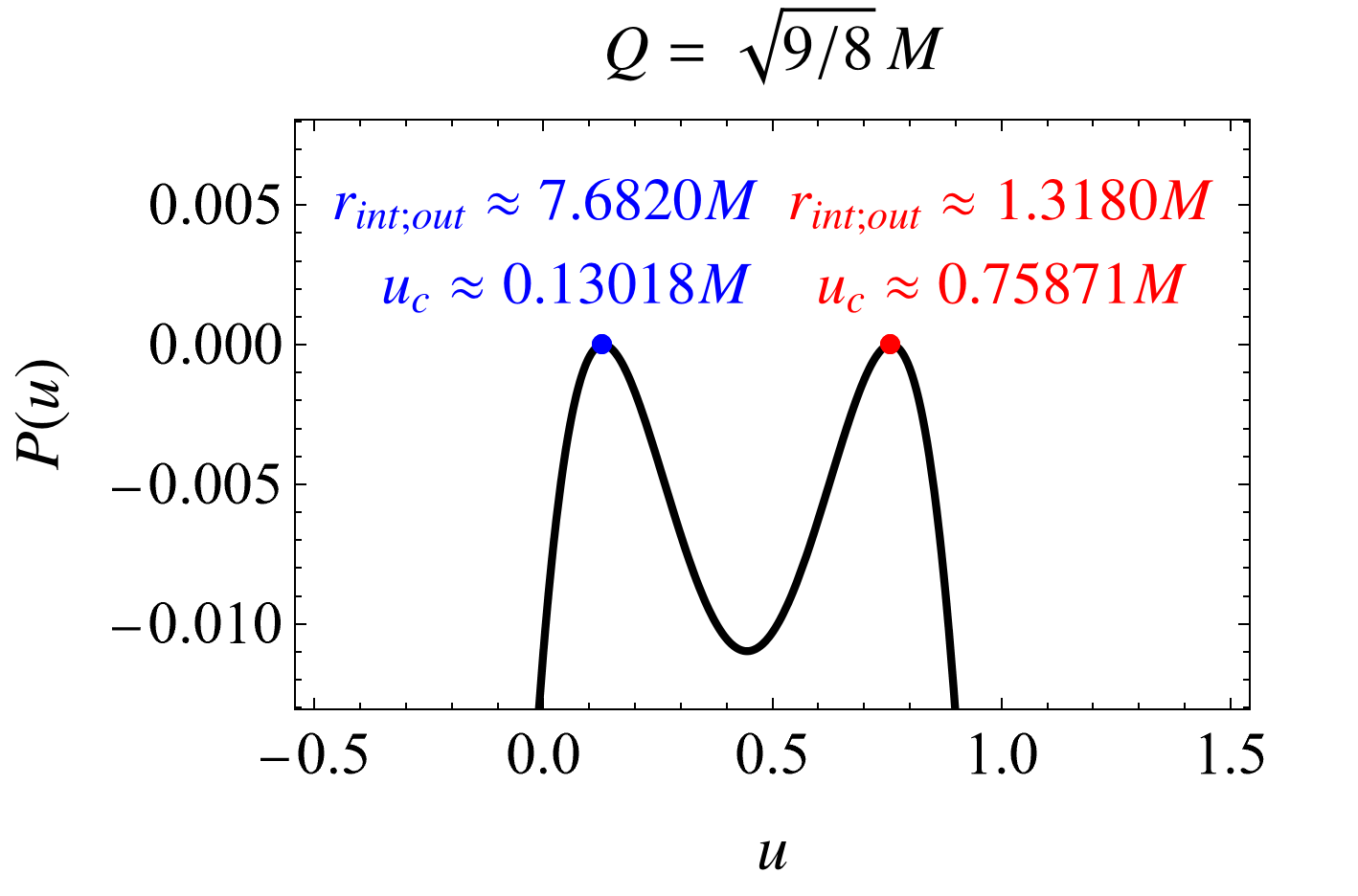}}
    \enspace
    \subfloat{\includegraphics[scale=.42]{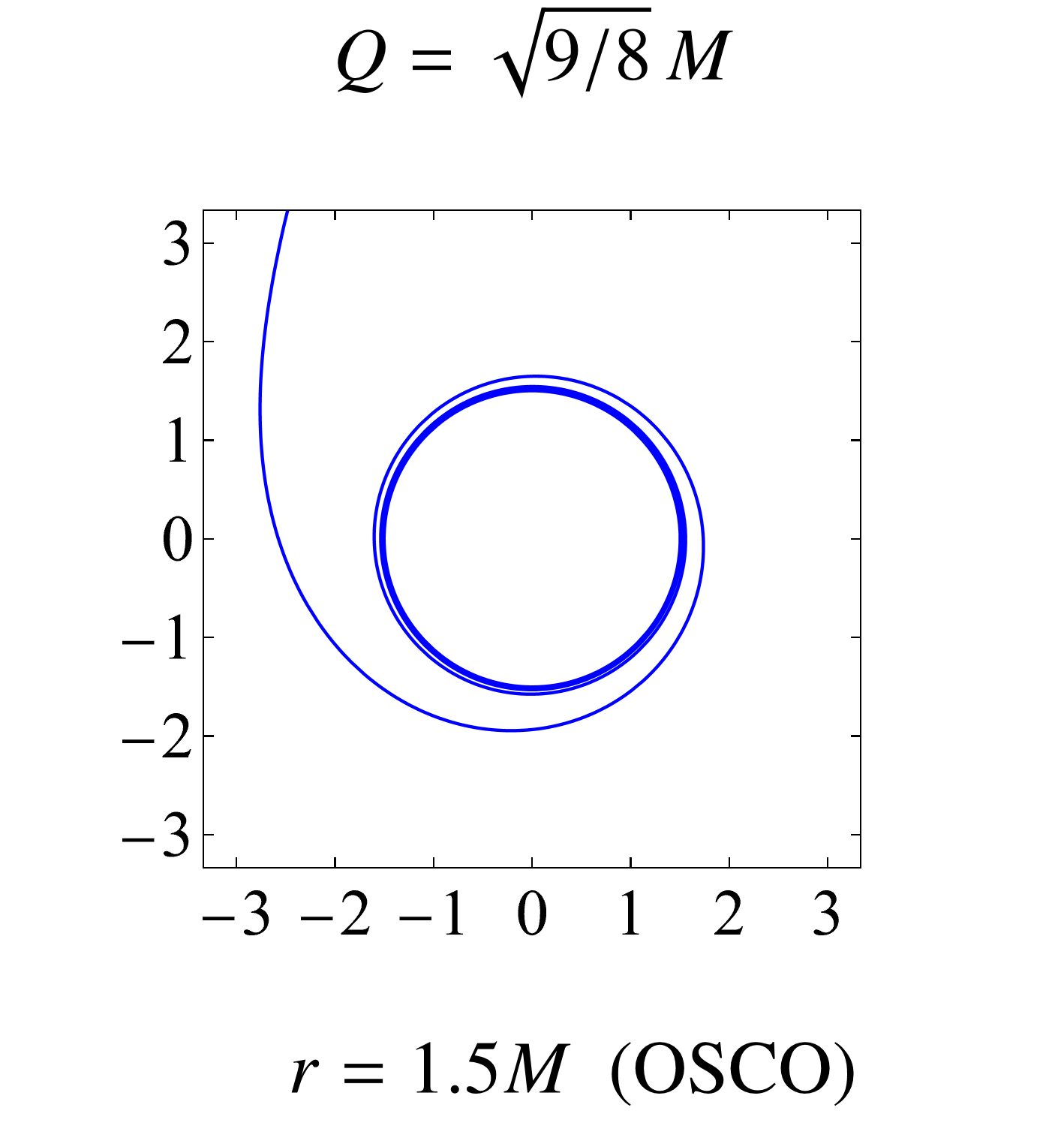}}
    \end{minipage}%
    \caption{$P(u)$ plots at each critical points of $Q=\sqrt{9/8}M$ in the top row 
    and the resulting homoclinic orbits in the bottom row. Bottom middle figure shows 
    the $P(u)$ at the intersection point $r_{int}$ of the two stable $\rc$ segments.}
    \end{figure}\label{3deghomoorb} 

The $P(u)$ roots structure of any $\rc$ values from $r_{IUCO}\leq\rc\leq r_{\gamma^{-}}$ (Case 1) or 
$r_{IUCO}\leq\rc\leq r_{OSCO}$ (Case 2A) on the inner stable circular orbits segment clearly indicate an
inner circular orbit associated with an unbound outer orbit, but now, values of both roots
for the outer region are smaller than $\uc$ of the inner region. The periastron 
value $c$ of $r_{IUCO}$ is the smallest possible radial distance away from the singularity where trajectories
from the outer region could reach. Therefore, there is always an outer unbound orbit with 
$r_{IBCO}\leq\rc\leq r_{IUCO}$ possessing the same $L$ and $E$ values with a type of inner bound 
orbit; either periodic, circular or chaotic, depending on the value of $\rc$.

\section{Generalization of domains and naked singularities \texorpdfstring{$q$}{TEXT}-branch distributions}\label{sec_4}

\subsection{Domains \texorpdfstring{$\dom_k$}{TEXT} where periodic orbits are located in \texorpdfstring{$(L,E)-$}{TEXT}space}\label{sec_4.1}
Gathering relevant results from the previous sections, we can outline the domains $\dom_k$ in 
$(L,E)$-space where periodic orbits can or cannot be found. As in \cite{yklimzcy24},
domains in this section are defined as the sets of $(L,E)$ values confined within a 
specific region in the parameter space that give rise to either one or two particular type of orbits.

We first bring up Case 1 and the two variations of Case 2. In Sec. \ref{sec_3.2}, we observed 
that $(L,E)$ values with four real roots lie in the narrow area confined by the three distinct 
segments of the $\rc$ curve representing the two stable regions and the unstable region. This area includes 
all points that lie exactly on both the stable circular segments of the curve but exclude $r_{int}$
and the critical points. Therefore, we define $\dom_1$ to be the part of the area lying entirely below 
the $E=1$ line and $\dom_2$ to be the part of the area above and including the $E=1$ line. $\dom_1$ is the 
only region whose $(L,E)$ values are capable of producing periodic orbit pair of two different size. 
Furthermore, $\dom_2$ is where we find a pair of an outer escaping orbit and a low eccentricity inner periodic 
orbit that is remarkably unbound energetically and bound geometrically. Also, $\dom_2$ does not exist for Case 2B 
since the whole $\rc$ graph lie entirely below the $E=1$ line.     

Then in Sec. \ref{sec_3.3}, we uncovered areas with two real and two complex roots. We label $\dom_3$
as the area entirely to the right of inner stable $\rc$ segment, below the $E=1$ line, above and including the outer stable 
$\rc$ segment. $\dom_4$ will be the area confined to the right of the $L=0\,$ axis, left of both the 
unstable $\rc$ segment plus the inner stable $\rc$ segment, including all points on the latter, and entirely 
below both the outer stable $\rc$ segment and the $E=1$ line. Again, $r_{int}$ and critical points are not part 
of both of these domains. The complex roots prohibit a second bound orbit from forming in both $\dom_3$ and $\dom_4$.
Both of these domains typically contain $(z,0,v)$ orbits with large $z$ and relatively low $v$. The entire
$E\geq1$ region minus both $\dom_2$ and the unstable $\rc$ segment also have this root configuration.
This will be our $\dom_5$.

Finally, in Case 3, there are definitely at least two complex roots present in the entire parameter space. 
As such, the domains here are relatively simple to define. Unbound orbits always appear in the $E\geq1$
region. So it is also $\dom_5$ defined the same way as the preceeding paragraph. Case 3 bound orbits will
lie in $\dom_6$, the entire region confined below $E=1$ and above the $\rc$ curve. $\dom_6$ share a 
resemblance with black holes graphs, in that there is only a single continous region describing stable bound 
orbits. Let $\dom_7$ be the lower right quadrant-like part of the parameter space that yield four complex 
roots in all three cases. This means no orbit possessing $(L,E)$ values from $\dom_7$ are physically possible.
For completeness, we will still list a domain $\dom_8$ where we can find the chaotic orbits from perturbations 
as shown in Sec. \ref{sec_3.3} since its inception homoclinic orbits was established to fit into the taxonomy. 

Table \hyperref[Domtable]{1} classifies all $\dom_k$ together with choices of analytical solutions of 
periodic orbits or any bound orbits in general. Fig.\hyperref[LE-domain]{13} illustrate $\dom_k$ locations 
in $(L,E)$-space graphically.

\begin{table}\label{Domtable}
\caption{Summary of the essential periodic orbits description for domains $\dom_k$ in $(L,E)$-space}
\begin{tabularx}{\textwidth}{@{}CCLL@{}}
\toprule
Domains & Roots configuration & Orbit types & Periodic orbit solution\footnotemark[8]\\
\midrule
$\dom_{1}$ & Four real & Outer bound, Inner bound & Outer: $r(\phi)_{\,\RN{1}}$, 
                 Inner: $r(\phi)_{\,\RN{2}}$ \\
$\dom_{2}$ & Four real including one negative & Outer unbound, Inner bound & Inner: $r(\phi)_{\,\RN{2}}$ \\
$\dom_{3}$ & Two real and two complex & Only outer bound & $r(\phi)_{\,\RN{3}}$ \\
$\dom_{4}$ & Two real and two complex & Only inner bound & $r(\phi)_{\,\RN{3}}$ \\
$\dom_{5}$ & Two real including one negative and two complex & Only unbound & - \\
$\dom_{6}$ & Two real and two complex & One bound &  $r(\phi)_{\,\RN{3}}$ \\ 
$\dom_{7}$ & Four complex & No orbits possible & - \\ [0.3cm]
$\dom_{8}$ ($E<1$) $r_{ISCO}<\rc<r_{IBCO}$ & Three real including one degenerate & 
           Homoclinic orbit pairs, perturb gives chaos bound & 
           Outer homo: $r(\phi)_{\,\RN{1}}$ Inner homo: $r(\phi)_{\,\RN{4}}$
           Outer chaos: $r(\phi)_{\,\RN{5}}$ Inner chaos: $r(\phi)_{\,\RN{2}}$\\ [0.3cm]
$\dom_{8}$ ($E\geq1$) Case 1: $r_{IBCO}\leq\rc\leq r_{\gamma^{+}}$ 
           Case 2: $r_{IBCO}\leq\rc\leq r_{OSCO}$ & Three real including one negative and one degenerate
           & Outer unbound, Inner homoclinic and chaos 
           & Inner homo: $r(\phi)_{\,\RN{4}}$ Inner chaos: $r(\phi)_{\,\RN{2}}$\\ [0.3cm]    
$\dom_{8}$ (Critical points) & Two real including one triply degenerate & 
           ISCO: Inner homoclinic, OSCO: Outer homoclinic & $r(\phi)_{\,\RN{6}}$ \\
\bottomrule
\end{tabularx}
\footnotetext[8]{We selected our formulas for $r(\phi)_{\,\RN{1}-\RN{5}}$ from Ref.\cite{brydEI}. 
For those preferring Ref.\cite{gradshteyn2014table}, the equivalent formulas can be found in 3.147, 
pg.275-276. For four reals, they are are no. 2,6,7,3 for 
$r(\phi)_{\,\RN{1}}$,$\,r(\phi)_{\,\RN{2}}$,$\,r(\phi)_{\,\RN{4}}$,$\,r(\phi)_{\,\RN{5}}$ respectively. 
With complex roots, the alternate form for $r(\phi)_{\,\RN{3}}$ is 3.145, no.2, pg.274.}
\end{table}

\begin{figure}[h!]
    \centering
        \begin{minipage}{\linewidth}
        \centering
        \subfloat[Case 1]{\includegraphics[scale=.54]{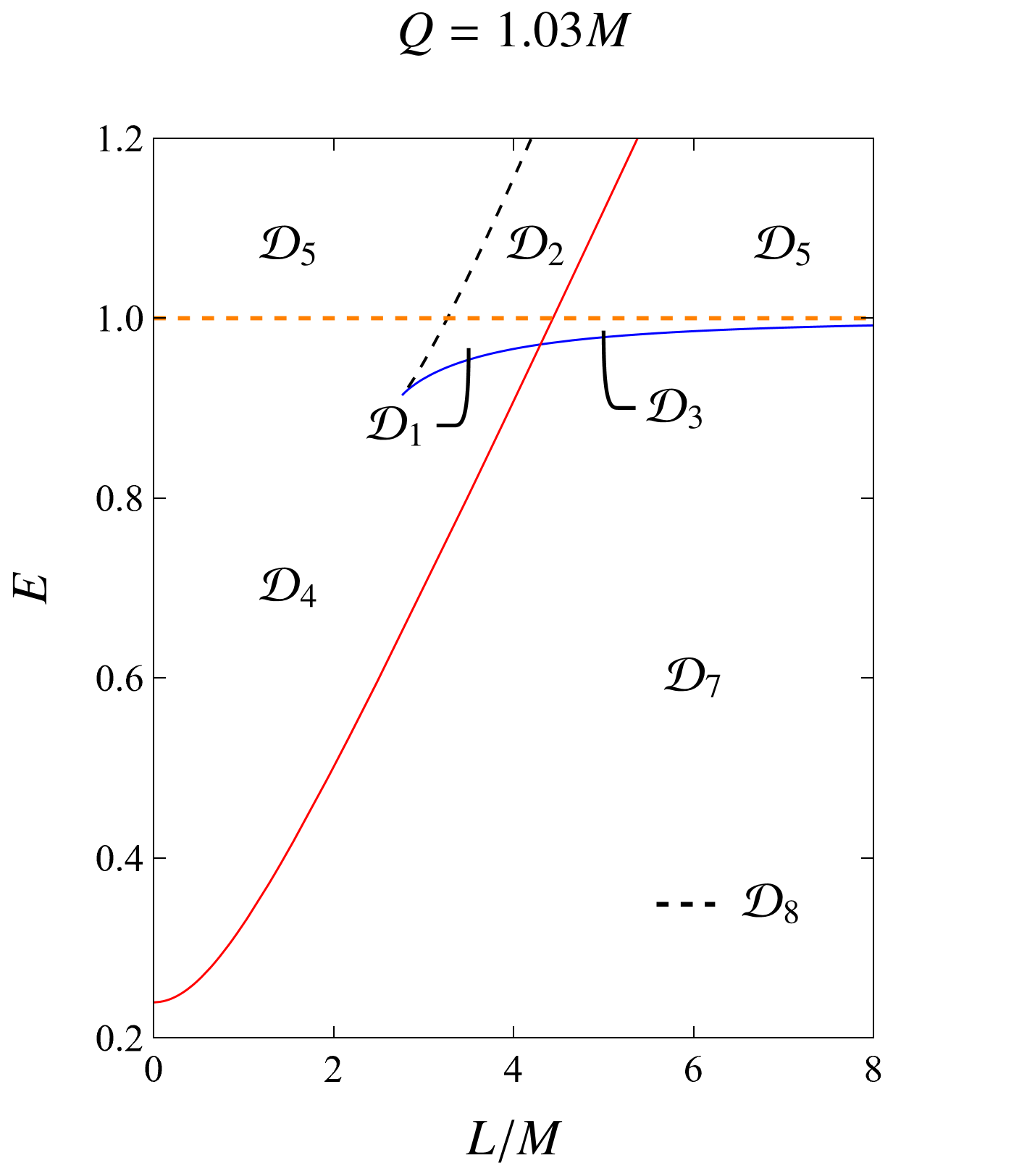}}
        \enspace
        \subfloat[Case 2A]{\includegraphics[scale=.58]{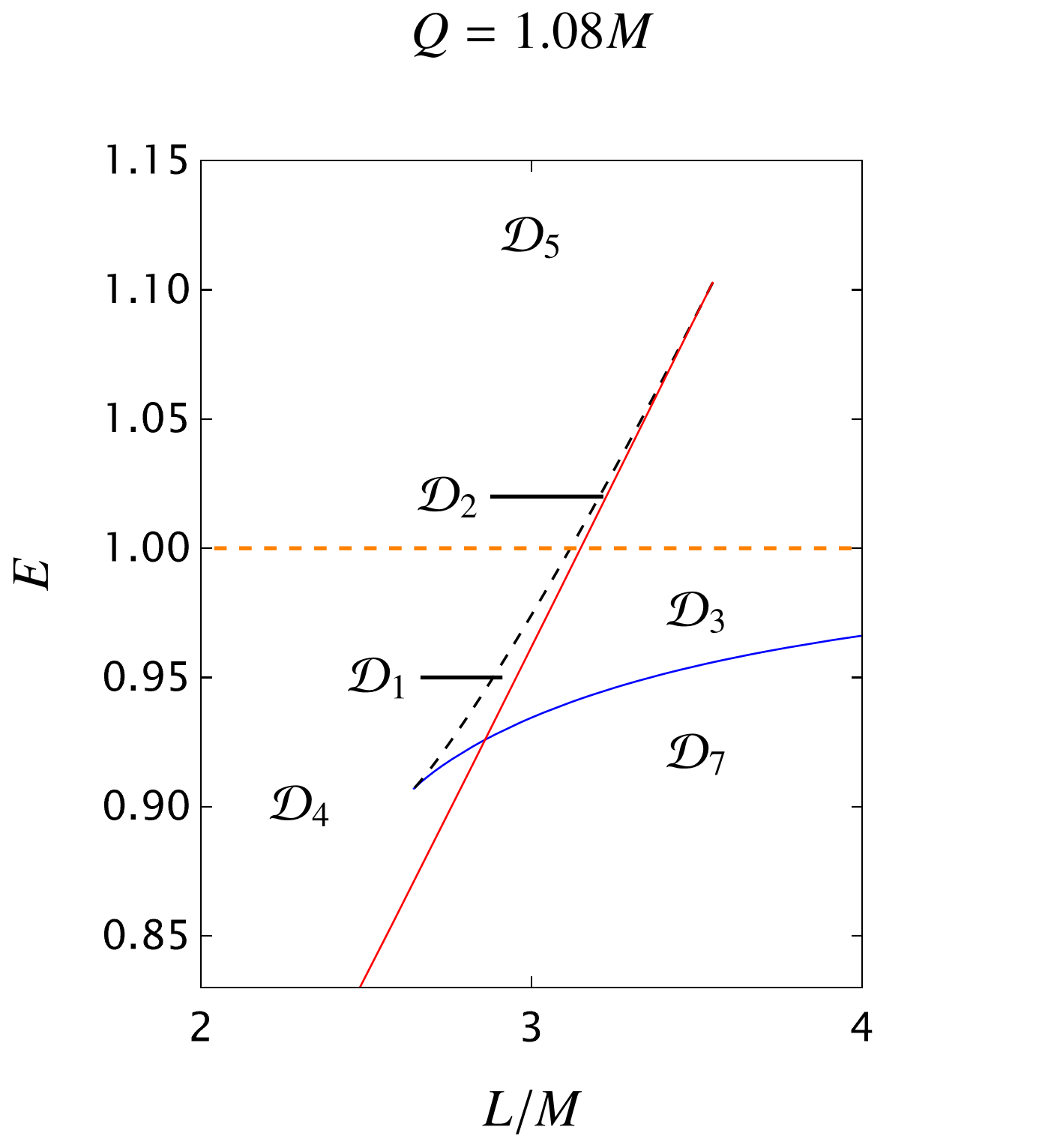}}
        \end{minipage}\par\medskip
        \begin{minipage}{\linewidth}
        \centering
        \subfloat[Case 2B]{\includegraphics[scale=.6]{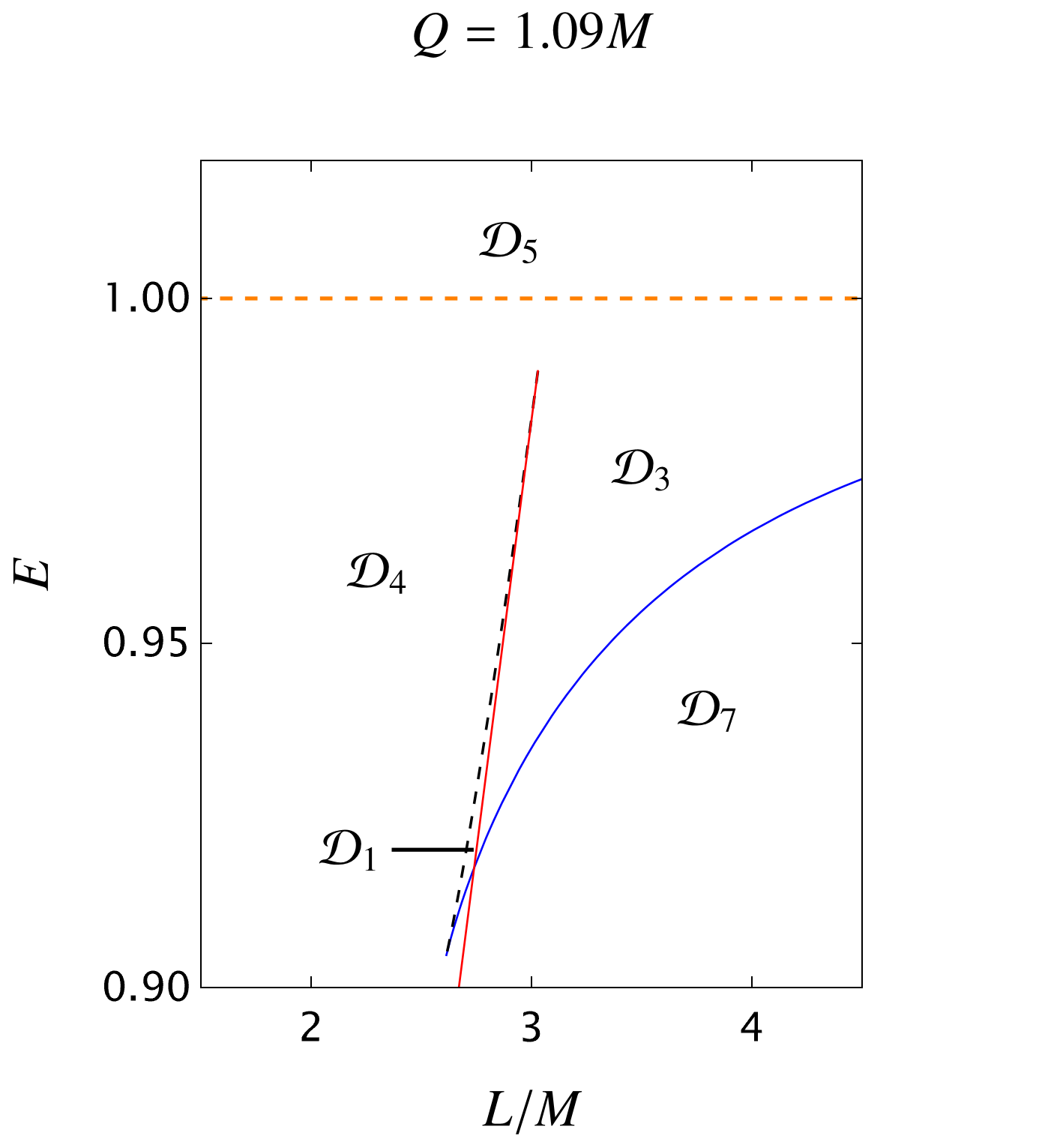}}
        \enspace
        \subfloat[Case 3]{\includegraphics[scale=.54]{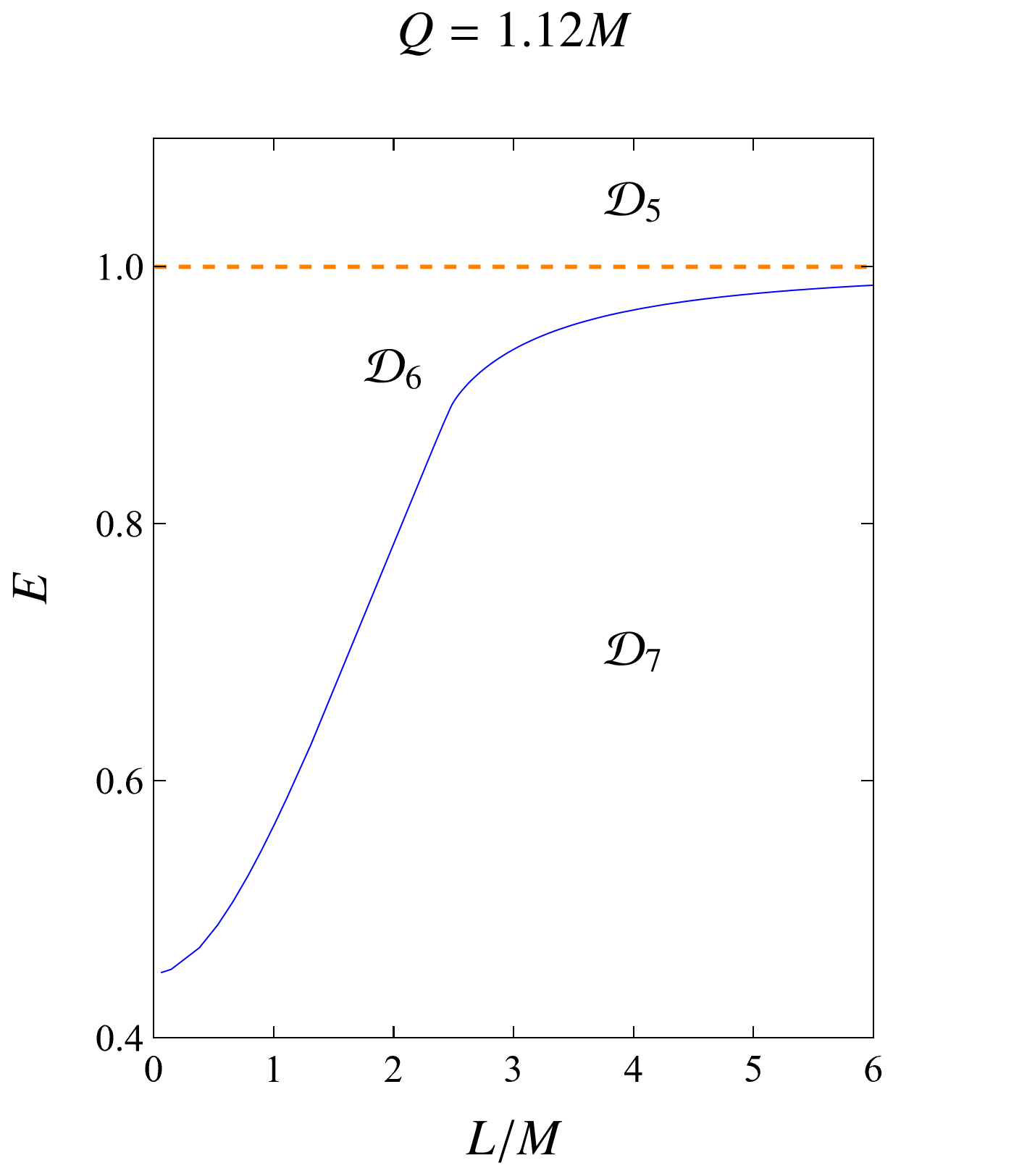}}
        \end{minipage}%
\caption{The four naked singularity cases and their 
$\dom_k$ locations in $(L,E)$-space. The entire 
unstable circular orbits segment (dashed black lines) is $\dom_8$.}
\end{figure}\label{LE-domain}

\subsection{General distribution pattern of \texorpdfstring{$q$}{TEXT}-branches for Cases 1,2 and 3}\label{sec_4.2}

Now that we know the domains in $(L,E)$-space to find periodic orbits, we can explore properties of the
$q$-branch distribution for both regions in each case. Beginning with
Case 1 and 2 again, the distribution for the outer region basically follow the same rules as black holes  
in Sec. \hyperref[sec_2.3]{2.3} (Fig.\hyperref[BHqbranch]{4}). The main difference is that now
there are two different domains, $\dom_1$ and $\dom_3$ where the outer $q$-branch can reside.
This situation was highlighted in Fig.\hyperref[Q101branchB]{8} earlier where we have to use 
different analytical solutions for each separate $e$ interval, akin to a piecewise function.   
On the same note, this type of $q$-branch seem to be inherent for all inner orbits
$q$-branches that emanate from $\dom_2$. Those cross three separate domains in order from $\dom_2$ 
to $\dom_1$ to $\dom_4$.

There are other notable differences of $q$-branches for the outer and inner region. One is the 
difference in direction of emanation for increasing $w,v$ values, which is equivalent to increasing
$q$ overall since $q=w+\frac{v}{z}$ \cite{levinpg08}. Points of increasing $q$ goes from right to 
left for the outer region, whereas for the inner region, it goes upwards along the slope of the inner 
stable $\rc$ segment. The infinite $z$ limit branch for the inner region is visible in a compact 
$(L,E)$-space map. It lie in $\dom_4$ close to the $L=0$ axis, acting as the leftmost rational 
$q$-branch for the inner region (Figs.\hyperref[Q101branchB]{8},\hyperref[Case2qbranch]{14},\hyperref[Case3qbranch]{15}). 
Only quasiperiodic orbits exist to the left of this limit branch.      

Strangely, we find a peculiar $q$-branch trend when increasing $q$ values as we get to larger
charge $Q$ from Case 2 onwards. Starting from Case 2A, inner $q$-branches with 
$w\geq1$ emanating from some point close to $r_{int}$ begin to abruptly end at some $e<1$ without 
reaching close to the $E=1$ line. Further increasing $q$ in direction up the slope shrinks the branch 
more untill rationals with large $w$ (typically $w\geq3$) super close to the unstable $\rc$ segment do not emanate 
anymore branches (Fig.\hyperref[Case2qbranch]{14}). A similar trend applies to the outer $q$-branch for 
Case 2B but this time, large $w$ branch could still appear and those emanating from the stable $\rc$ segment 
within $\dom_1$ could not cross the inner stable $\rc$ segment.

\begin{figure}[t!]
\centering
\begin{minipage}[t]{\linewidth}
\centering
\subfloat[]{\includegraphics[scale=.65]{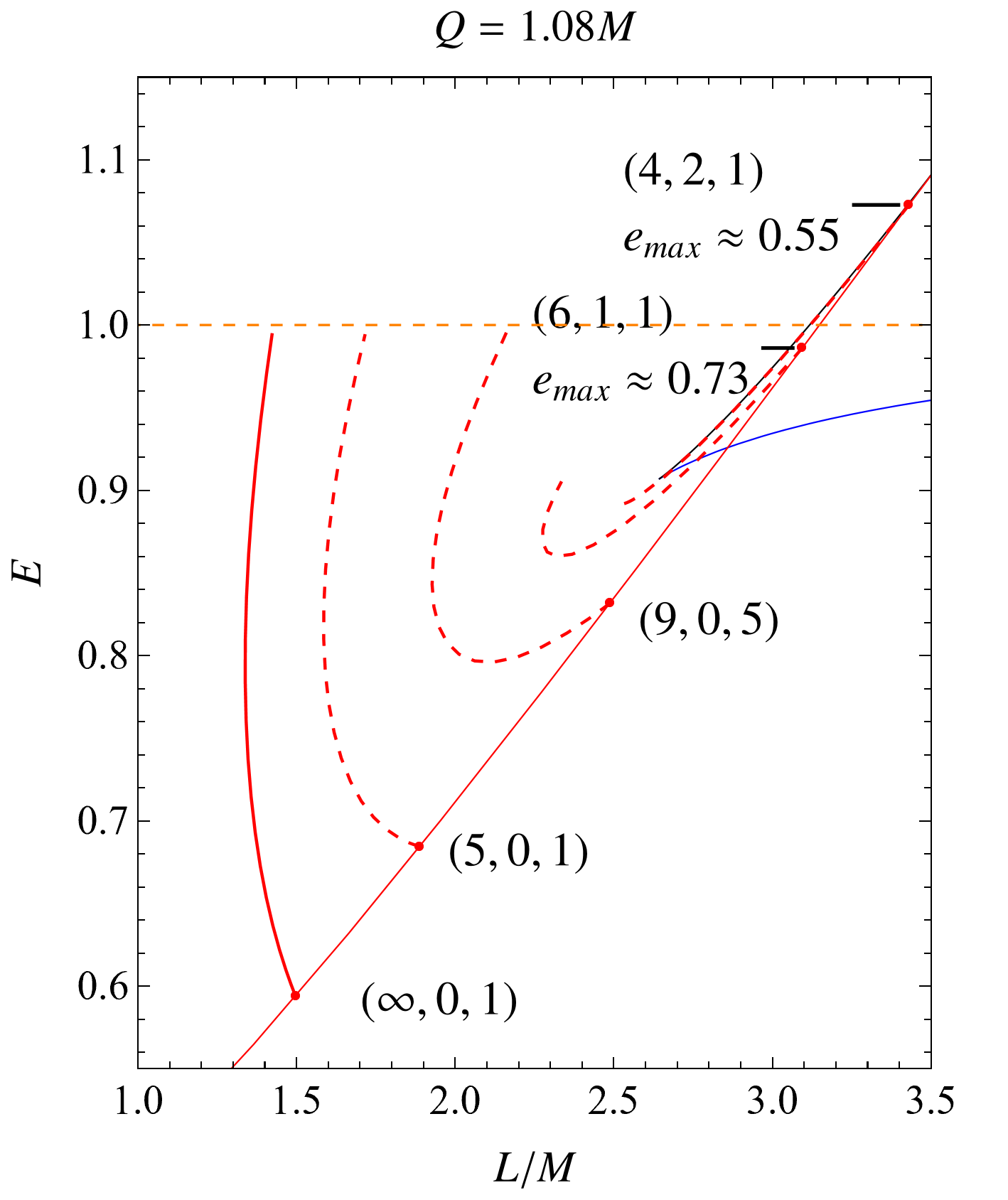}}
\end{minipage}\par\smallskip
\begin{minipage}[t]{\linewidth}
\centering
\subfloat[]{\includegraphics[scale=.54]{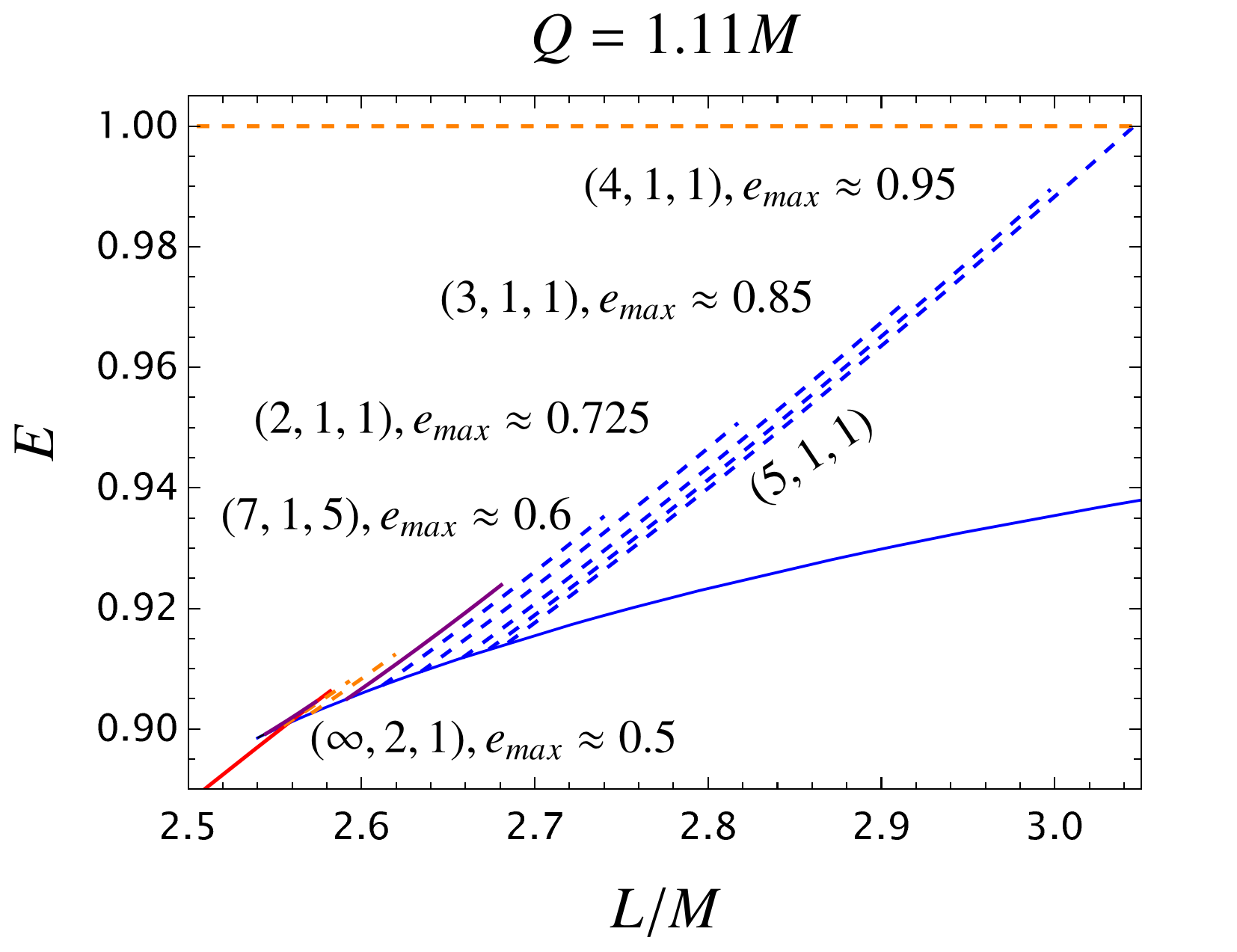}}
\enspace
\subfloat[]{\includegraphics[scale=.54]{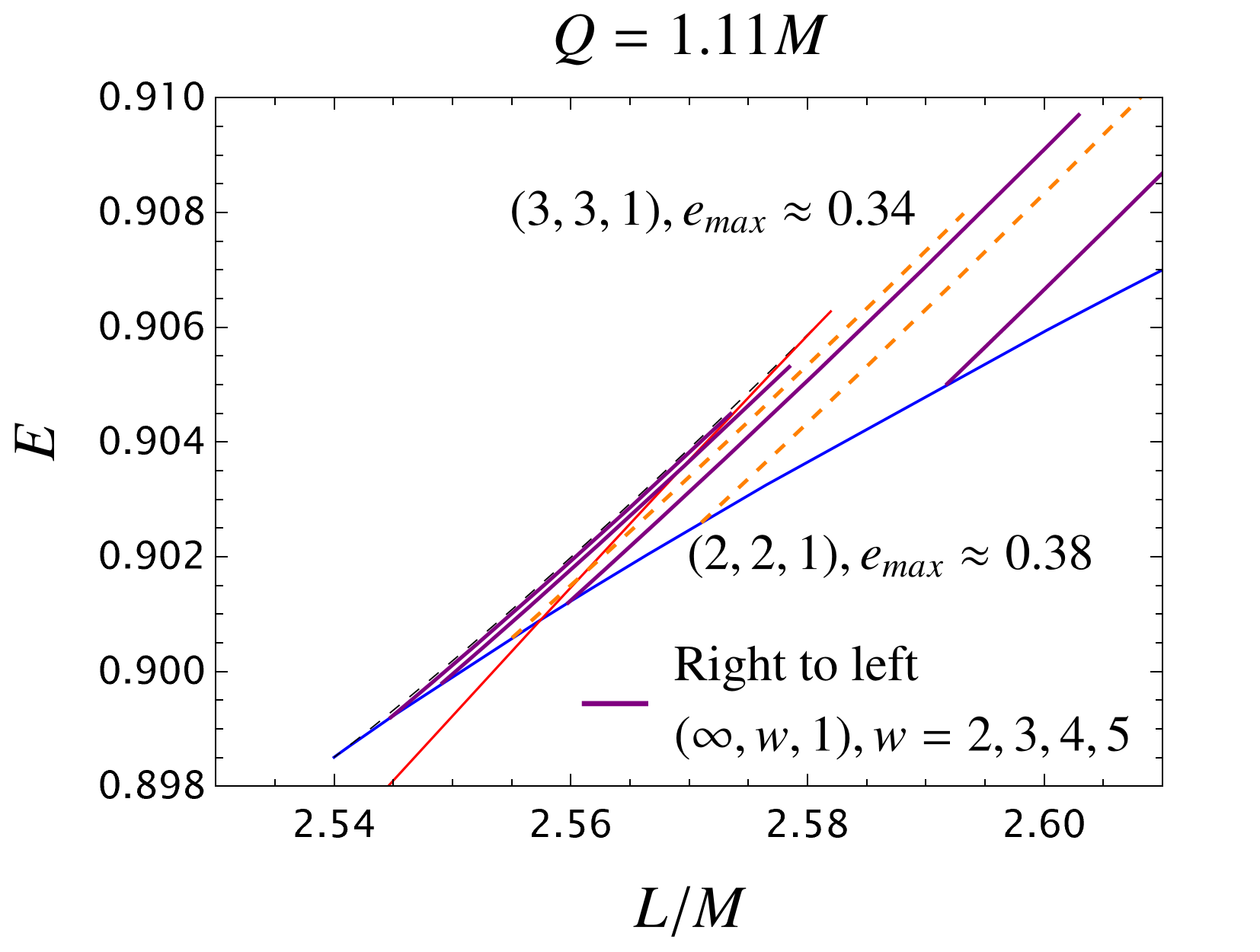}}
\end{minipage}%
\caption{The odd trend of $q$-branches terminating approximately close to the $e$ values shown 
for Case 2.} 
\end{figure}\label{Case2qbranch}

A possible mathematical reasoning we found for this `shrinking branch' trend, especially for $q$-branch 
lying in domains containing complex roots, $\dom_3$ and $\dom_4$, is that both sides of 
Eq.(\ref{LRnumericalrel}) yield complex values. For large rationals, after reaching a certain $e$ value, 
only complex $\lr$ and $q$ solution remain. In $(L,E)$-space, taking some arbitary points slightly above 
the tip of the shorter rational $q$-branches return an irrational $q$ value that represent a completely different branch 
arbitarily close to the former. However, this does not account for the short rational $q$-branches lying in the 
four real root domain in Case 2B, where the trend seemingly persists into that domain. Physically, it might indicate 
that periodic orbits with a large number of near-center whirls are harder to form proximately close to both kinds of 
marginally stable circular orbits (ISCO and OSCO).   

Coming to Case 3, $Q=\frac{\sqrt{5}}{2}\,M$ is the transition charge between Case 2 
and 3. The disappearance of the second stable $\rc$ segment starting from this $Q$ cause the 
$q$-branch to distribute in a single region manner. Some distribution properties from Case 2 carry 
over. For one, this appear to be the last $Q$ to have an infinite $w$ limit (at the $r=2.5M$ point)
and thus homoclinic orbits are not possible in Case 3. The infinite $z$ limit branch lie towards 
the left side like the ones for the inner region in Case 1 and 2. Direction of increasing $q$ is the
reverse to that of the outer region, from left to right coupled with the shrinking branch length trend. 
As such, $q$-branches can now only emanate from the left portion of the $\rc$ curve, particularly from 
the steep slope (Fig.\hyperref[Case3qbranch]{15}). 

Increasing $Q$ further in Case 3 seems to impose a lower limit to the integer value that $z$ could take.
This restricts the number of allowed periodic orbits and thus the range of the
$q$-branch distribution (Fig.\hyperref[Case3qbranch]{15}). Eventually, the exquisite looking small $q$ value periodic orbits are no 
longer possible for sufficiently large $Q$ and what remain are the generic Keplerian and unbound 
orbits. This may likely be caused by the repulsive `anti-gravity' effect of the naked singularity 
which increases with $Q$ \cite{cohen79}. The range of $E$ values for Case 3 $\rc$ curves shrinks as 
$Q\rightarrow\infty$ till the point where we need $E\simeq1$ values to maintain bound 
orbits at really large $Q$.

\begin{figure}[t!]
    \centering
        \begin{minipage}{\linewidth}
        \centering
        \subfloat[]{\includegraphics[scale=.61]{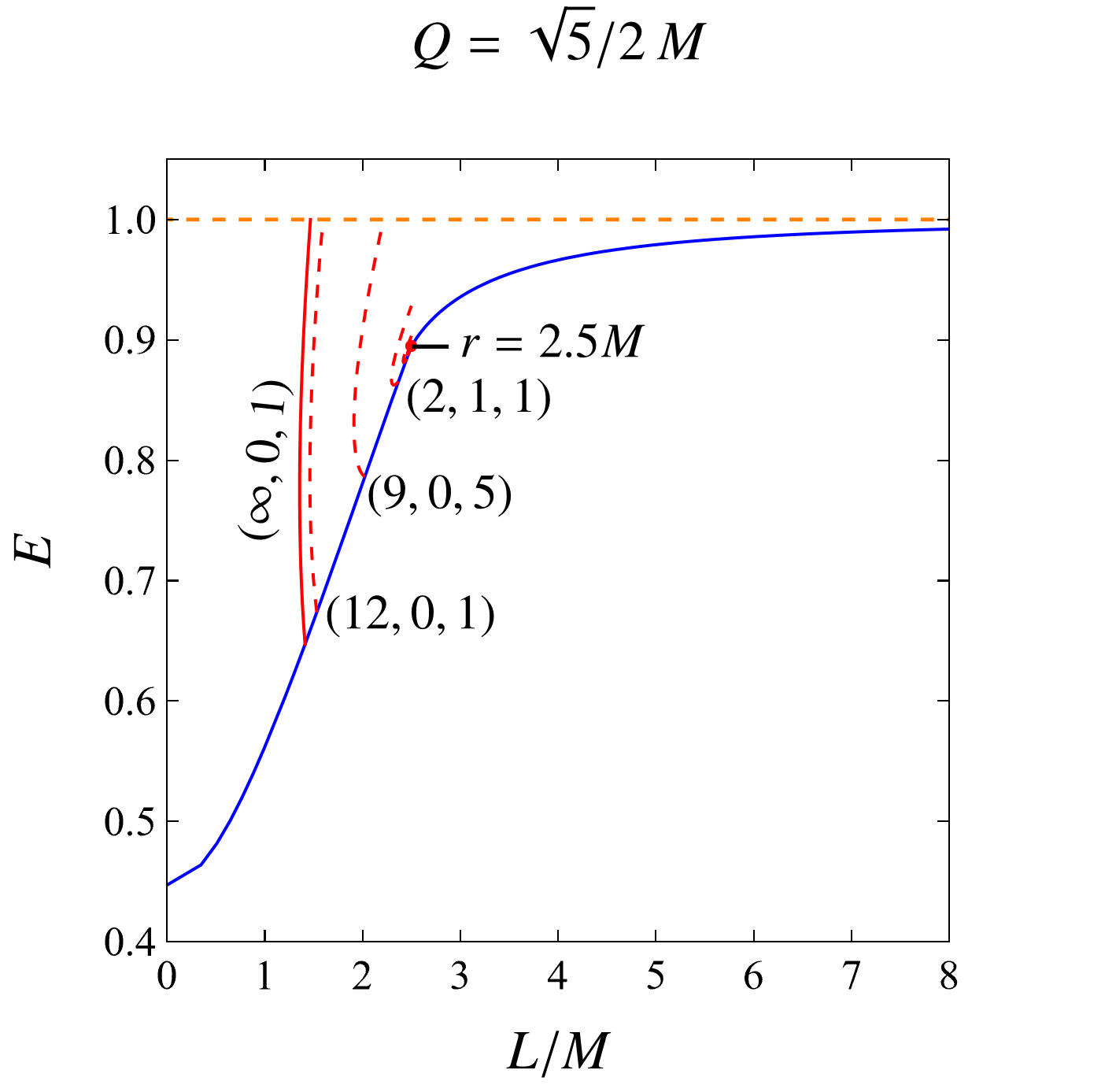}}
        \enspace
        \subfloat[]{\includegraphics[scale=.58]{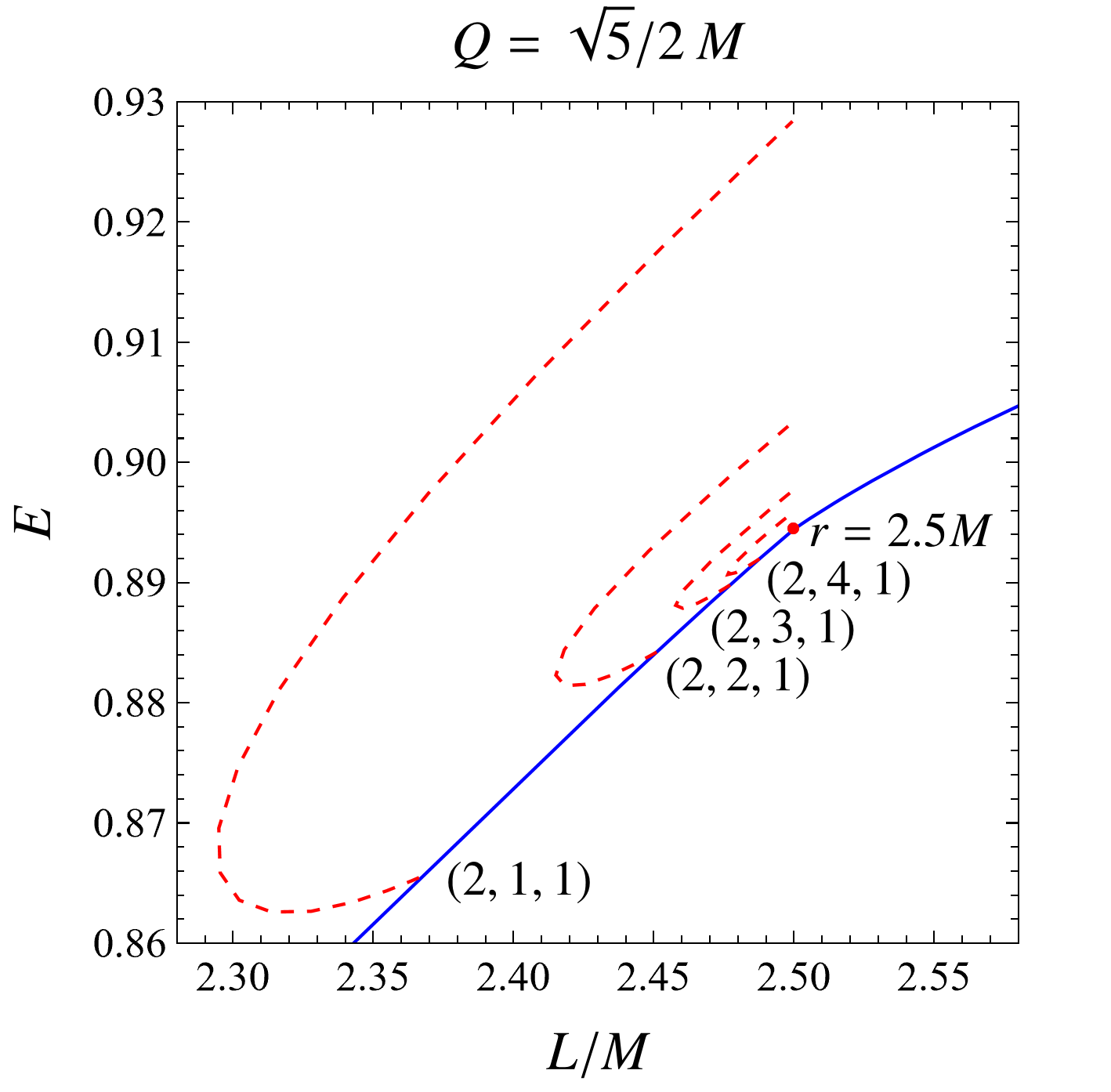}}
        \end{minipage}\par\smallskip
        \begin{minipage}{\linewidth}
        \centering
        \subfloat[]{\includegraphics[scale=.57]{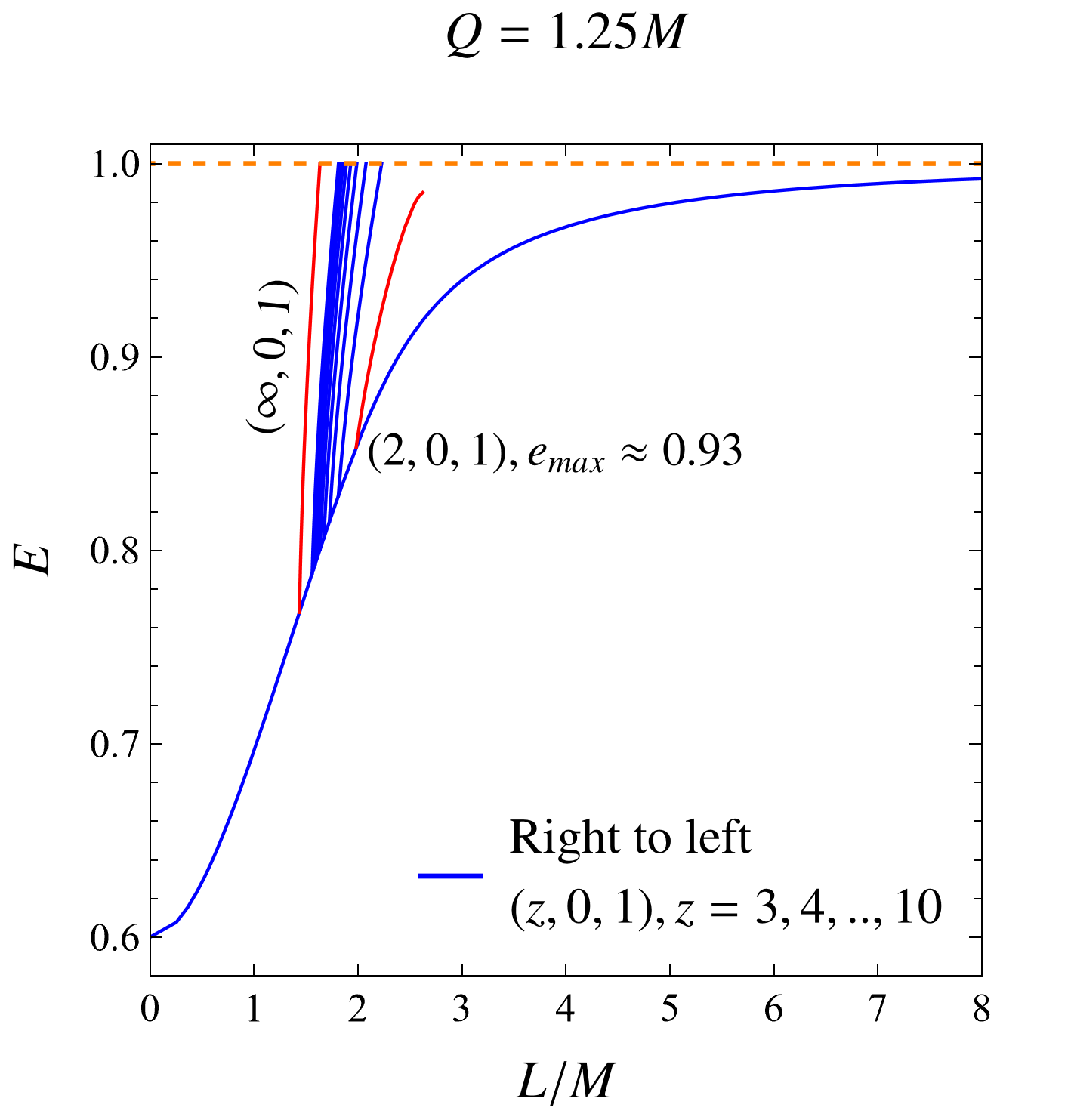}}
        \enspace
        \subfloat[]{\includegraphics[scale=.58]{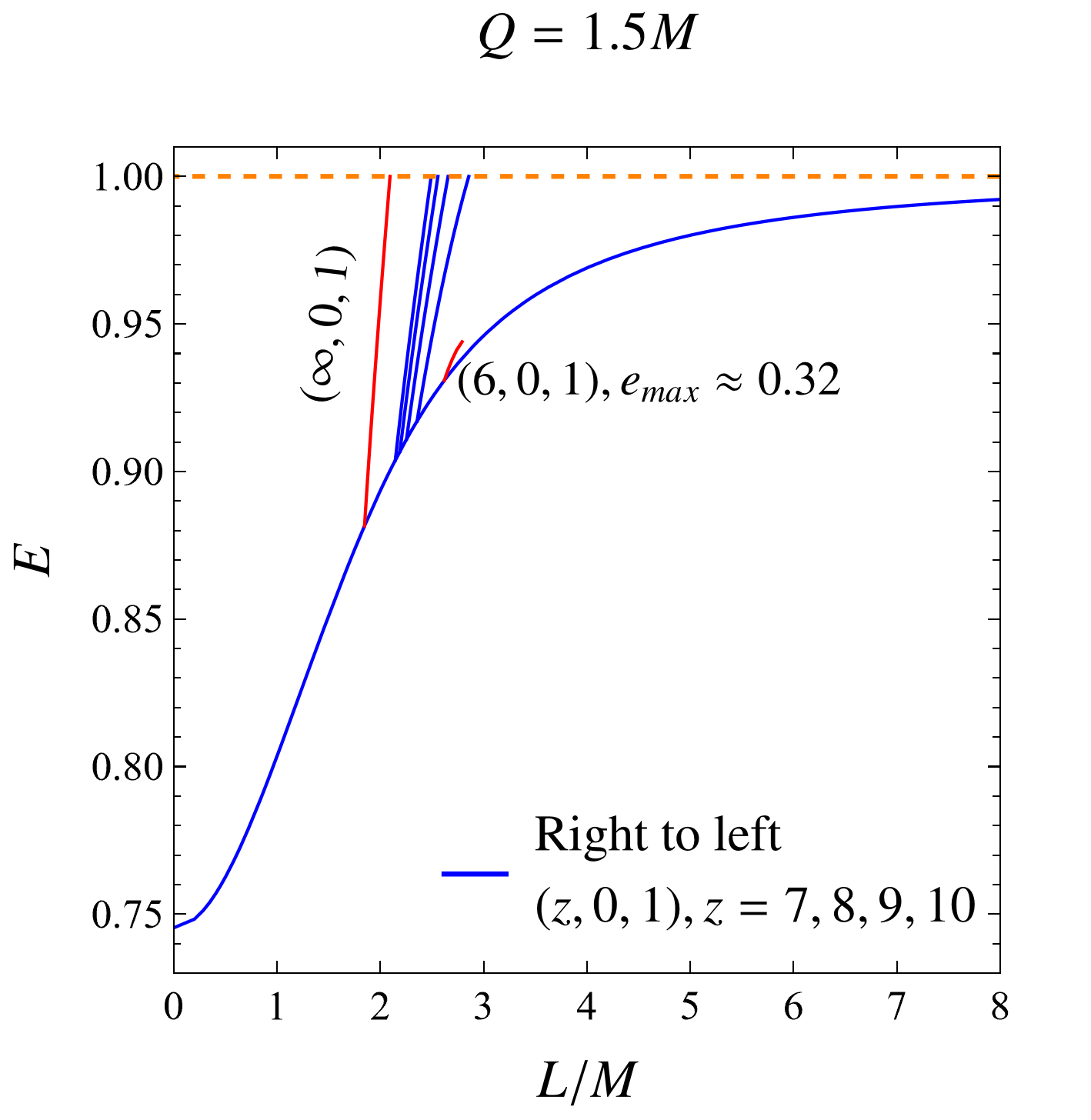}}
        \end{minipage}%
        \caption{The $q$-branch distribution for Case 3.}
\end{figure}\label{Case3qbranch}

\section{Conclusion}\label{sec_5}
\begin{parindent}15pt
\noindent In this paper, we have shown how it is possible for time-like neutral test particles to have 
a pair of periodic orbits with the same $(z,w,v)$ and $(L,E)$ values but different size revolving around 
a RN naked singularity. We explained the importance of understanding the root configuration of the quartic 
inverse radial coordinate polynomial $P(u)$ at different parts in $(L,E)$ parameter space. The root configuration
indicates whether any given $(L,E)$ values correspond to an associative orbit pair from a possible combination of 
periodic, circular, homoclinic or unbound. Relatively simple analytical solutions in terms of Jacobi elliptic 
functions and all four roots of $P(u)$ are used to plot the first three aforementioned orbits, providing computational 
advantages. The solution of homoclinic orbits at the critical points is the exception to this.    

We have analysed the three possible scenarios of a RN naked singularity, reasoning which domains
in $(L,E)$-space could contain physical parameters for periodic orbits. We parametrised several expressions 
with geometric parameters $e$ and $\lr$, where these parameters are encoded in $q$-branch that emanate 
from both stable circular orbits segments of the circular orbits curve. We inferred possible physical properties
from looking at the different kinds of shape of the branches for each scenarios. 
We found out how tweaking $z,w,v$ values change the location of a given $q$-branch and then illustrate 
the general distribution in all scenarios, revealing some novel trends. 
The analysis and procedures in this paper might be useful for those intending to 
provide periodic orbit descriptions to other spherically symmetric spacetime or naked singularity 
background possessing a radial polynomial with degree 4. 

The strange trend of decreasing $q$-branch length for large rationals in both Case 2 and 3, 
especially those with $w\geq1$ suggest it is harder for periodic orbits with more whirls to zoom
out far from its near-circular centre. We had yet to pinpoint whether it is purely a mathematical 
consequence from the underlying nature of the polynomial roots or there are inherent physical properties in 
the different regions in naked singularity background that restrict the potential 
extent of large rationals orbits. In any case, this trend may help in probing the exact size of each 
region and to constrain the dynamics of non-circular bound orbits surrounding a naked singularity,
combined with a more robust mathematical relationship that link rational $q$ with complex numbers.       
\end{parindent}

\section*{Acknowledgments}
Y.-K. L is supported by Xiamen University Malaysia Research Fund (Grant No. XMUMRF/
2021-C8/IPHY/0001).

\begin{appendices}
\section*{Appendix A: Viet\'{e} Theorem}\label{app_A}
    Here, we demonstrate how to derive expressions for $L,E$ and roots $a,b$ parametrised with
    $e$ and $\lr$. We fix the expressions of roots $c,d=\frac{1\pm e}{\lr}$ to conveniently follow 
    Kepler's first law. This means we need to determine the formulas for 
    the two other roots. First, apply Vi\`{e}te theorem \cite{viete} to compare the coefficient 
    of terms in (\ref{Pu1}) and the arbitary form of roots $a,b,c,d$ in (\ref{Pu2}). It is written out as,
    \begin{align}
    \dfrac{1-E^2}{Q^2L^2}&=abcd\,, \settag{A1} \label{A1}\\
    \dfrac{2M}{Q^2L^2}&=abc+abd+acd+bcd\,, \settag{A2} \label{A2}\\
    \dfrac{1}{Q^2}+\dfrac{1}{L^2}&=ab+ac+ad+bc+bd+cd\,, \settag{A3} \label{A3}\\
    \dfrac{2M}{Q^2}&=a+b+c+d \settag{A4} \label{A4}
    \end{align}
    In our case, $cd=\frac{1-e^2}{\lr^2}$ and $c+d=\frac{2}{\lr}$. Substituting these into 
    (\hyperref[A1]{A1}) and (\hyperref[A4]{A4}) gives
    $ab=\brac{\frac{1-E^2}{Q^2L^2}}\brac{\frac{\lr^2}{1-e^2}}$ and 
    $a+b=\frac{2M}{Q^2}-\frac{2}{\lr}$ respectively. We recommend deriving $L$ and $E$ without finding 
    the explicit expression of $a$ and $b$ beforehand. Observe that the RHS of (\hyperref[A2]{A2}) and 
    (\hyperref[A3]{A3}) can be factored into $ab\,(c+d)+cd\,(a+b)$ and $ab+cd+(c+d)(a+b)$ respectively. 
    From here, treat $ab,\,cd,\,a+b$ and $c+d$ as variables, do relevant substitutions and then 
    solve (\hyperref[A2]{A2}) and (\hyperref[A3]{A3}) simultaneuosly to obtain the  
    expressions for $L$ and $E$ as given in Eqs.(\ref{LEellipse}).
    
    Then, solving (\hyperref[A1]{A1}) and (\hyperref[A4]{A4}) leads to the exact form of roots 
    $a,b$. These roots naturally takes the form $a,b=\frac{A\pm\sqrt{B}}{C}$ where $A,B,C$ are 
    functions involving $L,E$ terms. Substituting the $L,E$ expressions (\ref{LEellipse}) derived earlier 
    and further simplifications should yield Eqs.(\ref{rootpaireqs1}). This is why we recommend seeking the $L,E$ 
    expressions first. In general, this procedure can be use for finding explicit expressions of two 
    quartic roots given the form of the other two roots are fixed.

\section*{Appendix B: Circular Orbits Expressions}\label{app_B}
Here, we list down some useful circular orbits formulas for naked singularities given in
\cite{pugliese11n} that aid our periodic orbit search.
Solving the conditions for critical points, $V_{\mathrm{eff}}''=0$ for $Q>M$ produce three real solutions. 
These solutions can be expressed in terms of trigonometric functions. In order of increasing magnitude, 
$r_1<r_2<r_3$, they are
    \begin{align}
    r_1&=2M-2\sqrt{4M^2-3Q^2}\,\times\sin\sbrac{\dfrac{\pi}{6}+\dfrac{1}{3}\arccos\brac{B(Q)}},
                 \settag{B0} \label{B0}\\
    r_2&=2M-2\sqrt{4M^2-3Q^2}\,\times\sin\sbrac{\dfrac{1}{3}\arcsin\brac{B(Q)}},
                 \settag{B1} \label{B1}\\
    r_3&=2M+2\sqrt{4M^2-3Q^2}\,\times\cos\sbrac{\dfrac{1}{3}\arccos\brac{B(Q)}},
                 \settag{B2} \label{B2}\\
    &B(Q)=\dfrac{8M^4-9M^2Q^2+2Q^4}{M(4M^2-3Q^2)^{3/2}}\nonumber    
    \end{align}

The smallest solution, (\hyperref[B0]{B0}) is unphysical as it lie within the $r_*$ equilibrium sphere 
and should be ignored. The largest value (\hyperref[B2]{B2}) represents the ISCO of the outer region surrounding 
a naked singularity for Cases 1 and 2 and also black holes. (\hyperref[B1]{B1}) gives the OSCO of the 
inner region solely for Case 2. Real solutions terminate at Case 3 (except for $Q=\frac{\sqrt{5}}{2}\,M$ where 
$\mbox{B}1=\mbox{B}2=2.5M$) since no critical points exist for this case. 
    
The null circular orbits radii in Case 1 are $r_{\gamma^{\pm}}=\half\brac{3M\pm\sqrt{9M^2-8Q^2}}$ and the minimum radius is simply 
$r_*=Q^2/M$. Time-like circular orbits and hence emanation points for periodic orbits from
the inner region cannot exist in the following range; $r_{\gamma^{-}}\leq\rc\leq r_{\gamma^{+}}$ 
and $\rc\leq r_*$. This is to ensure that the values of $E$ and $L$ for 
time-like particles are well-defined by having the denominator in
Eqs.(\ref{LEcircular}) obey $\rc^2-3M\rc+2Q^2>0$ and the numerator of 
$L(\rc,Q)$ obey $M\rc-Q^2>0$. 
    
The intersection point $r_{int}$ of the two stable circular orbits segments in Case 1 and 2 is 
best solved graphically.
\end{appendices}

\bibliographystyle{sn-mathphys-num}

\end{document}